\documentclass[12pt]{report}
\usepackage{amsmath}
\usepackage{amssymb}
\usepackage{epsfig}
\begin{document}
\numberwithin{equation}{chapter}
\numberwithin{table}{chapter}
\numberwithin{figure}{chapter}

\begin{table}[t]
  \begin{flushright}
     OUTP-04-07P    \\
     hep-th/0403003 \\
     February 2004
  \end{flushright}
\end{table}

\title{$ SO(10) $ heterotic M-theory vacua}

\author{Richard S. Garavuso  
\\
Wolfson College
\\
\\
Theoretical Physics Department
\\
University of Oxford
\\
\\
\\
\\
\\
\\
\\
\\
\\
Thesis submitted in partial fulfillment of the requirements 
\\
for the Degree of Doctor of Philosophy at the
\\
University of Oxford
}

\date{February 2004}

\maketitle

\begin{abstract}

The $ SO(10) $ embedding of the Standard Model spectrum is supported by
evidence for neutrino masses.
This thesis adapts the available formalism to study a class of  
heterotic M-theory vacua with $ SO(10) $ grand unification group.
Compactification to four dimensions with  $ \mathcal{N} = 1 $
supersymmetry is achieved on a torus fibered Calabi-Yau 3-fold   
$ \mathbf{Z} = \mathbf{X} / \tau_{\mathbf{X}} $ with first homotopy group
$ \pi_{1}(\mathbf{Z}) = \mathbb{Z}_{2} $.
Here $ \mathbf{X} $ is an elliptically fibered Calabi-Yau 3-fold which
admits two global sections and $ \tau_{\mathbf{X}} $ is a freely acting
involution on $ \mathbf{X} $. The vacua in this class have net number of
three generations of chiral fermions in the observable sector and
may contain M5-branes in the bulk space which wrap holomorphic
curves in $ \mathbf{Z} $.  Vacua with nonvanishing and vanishing instanton
charges in the observable sector are considered.  The latter case
corresponds to potentially viable matter Yukawa couplings.  Since
$ \pi_{1}(\mathbf{Z}) = \mathbb{Z}_{2} $, the grand unification group can
be broken with $ \mathbb{Z}_{2} $ Wilson lines.

Realistic free-fermionic models preserve the $ SO(10) $ embedding of the 
Standard Model spectrum. These models have a stage in their construction
which corresponds to $ \mathbb{Z}_{2} \times \mathbb{Z}_{2} $ orbifold
compactification of the weakly coupled 10-dimensional heterotic string.
This correspondence identifies associated Calabi-Yau 3-folds which
possess the structure of the above $ \mathbf{Z} $ and $ \mathbf{X} $.
This, in turn, allows the above formalism to be used to study heterotic
M-theory vacua associated with realistic free-fermionic models.
It is argued how the top quark Yukawa coupling in these models can be
reproduced in the heterotic M-theory limit.

\end{abstract}

\pagebreak

\pagenumbering{roman}

\tableofcontents

\chapter*{Acknowledgements}
\addcontentsline{toc}{chapter}{Acknowledgements}
\label{chapter:acknow}
I would like to thank my thesis supervisor Alon E. Faraggi, and
Jose M. Isidro.

\section*{Publications}                  
The new results presented in this thesis have been published by
Faraggi, Garavuso and Isidro \cite{FarGarIsi:Non} and
Faraggi and Garavuso \cite{FarGar:Yuk}.  Chapter \ref{HetVac} presents
rules for constructing the heterotic M-theory vacua of 
\cite{FarGarIsi:Non} and \cite{FarGar:Yuk}. These vacua are
discussed in Chapters \ref{NoYukVac} and \ref{YukVac}, respectively.  
Chapter \ref{Free} explains the geometrical overlap \cite{FarGarIsi:Non}
with realistic free-fermionic models, and argues how the top quark Yukawa
coupling in these models can be reproduced \cite{FarGar:Yuk} in the
heterotic M-theory limit.

\chapter{\label{Intro}Introduction} 
\pagenumbering{arabic}
\setcounter{page}{1}

When the \emph{11-dimensional supergravity} \cite{CreJulSch:Sup}
limit of \emph{M-theory} \cite{M-theory} is compactified
on an orbifold $ \mathbf{S}^{1} / \mathbb{Z}_{2} $, cancellation of gauge
and gravitational anomalies requires the presence of a chiral 
$ \mathcal{N} = 1 $, $ E_{8} $ vector supermultiplet on each of the two
orbifold fixed planes.  If the gauginos in these supermultiplets have
\emph{opposite} chirality, then one obtains a theory which has no
spacetime supersymmetry \cite{FabHor:Cas}. On the other hand, if the
gauginos have the \emph{same} chirality, then one obtains a theory with $
\mathcal{N} = 1 $ supersymmetry.  This supersymmetric theory is known as 
\emph{Ho\v{r}ava-Witten theory} \cite{HorWit:Het,HorWit:Ele}; it gives the
low-energy strongly coupled limit of the $ E_{8} \times E_{8} $ 
heterotic string.  

Compactifications of Ho\v{r}ava-Witten theory leading to
unbroken $ \mathcal{N} = 1 $ supersymmetry in four dimensions
\cite{Wit:Str} are (to lowest order) based on the spacetime structure  
\begin{equation}
\mathbf{M}^{11} = \mathbf{M}^{4} \times \mathbf{S}^{1}/\mathbb{Z}_{2} 
\times \boldsymbol{\mathcal{X}},
\end{equation}
where $ \mathbf{M}^{4} $ is 4-dimensional Minkowski space and
$ \boldsymbol{\mathcal{X}} $ is a Calabi-Yau 3-fold.
Generally, on a given fixed plane, some subgroup $ G $ of the $ E_{8} $
symmetry survives on the Calabi-Yau
3-fold. $ E_{8} $ is broken to $ H $, where the grand unification
group $ H $ is the commutant subgroup of $ G $ in $ E_{8} $. 
Table \ref{commutant} lists the commutant subroup of $ G = SU(n) $ in 
$ E_{8} $ for the cases $ n = 3,4,5 $.
\begin{table}[t]
$$ \begin{array}{c|c}
G & H
\\
\hline
SU(3) & E_{6}
\\
SU(4) & SO(10)
\\
SU(5) & SU(5)
\end{array} $$
\caption{Commutant subgroup $ H $ of $ G = SU(n) $ in $ E_{8} $ for the
cases $ n = 3,4,5 $.}
\label{commutant}
\end{table} 
The `standard embedding', in which the spin connection of the Calabi-Yau
3-fold is embedded in a $ G = SU(3) $ subgroup of one of the $ E_{8} $ 
gauge groups, corresponds to $ H = E_{6} $.  
Vacua with nonstandard embeddings may contain M5-branes 
\cite{Wit:Str,LukOvrWal:Non} in the bulk space. One refers to
Ho\v{r}ava-Witten theory compactified to lower dimensions with arbitrary
gauge vacua as \emph{heterotic M-theory}.

Powerful techniques in algebraic geometry have been developed which
allow the study of a large class of the $ G = SU(n) $ vacua discussed
above. Consider a Calabi-Yau 3-fold $ \boldsymbol{\mathcal{X}} $. The
gauge fields associated with $ G $ `live' on $ \boldsymbol{\mathcal{X}} $,
and hence $ (3 + 1)$-dimensional Poincar\'{e} invariance is left unbroken.
The requirement of unbroken $ \mathcal{N} = 1 $ supersymmetry implies that
the corresponding field strengths must satisfy the \emph{Hermitian
Yang-Mills constraints} 
\begin{equation}
F_{ab} = F_{\overline{a} \overline{b}} 
         = g^{a \overline{b}} F_{a \overline{b}} = 0.
\end{equation}  
Donaldson \cite{Don} and Uhlenbeck and Yau \cite{UhlYau} prove that each
solution to the 6-dimensional \emph{Yang-Mills equations}
\begin{equation}
D^{A}F_{AB} = 0
\end{equation}
satisfying the Hermitian Yang-Mills constraints
corresponds to a semistable holomorphic vector bundle 
$ V_{ \boldsymbol{\mathcal{X}} } $
over $ \boldsymbol{\mathcal{X}} $ with structure group being the
complexification $ G_{\mathbb{C}} $ of the group $ G $, and conversely.  
Whereas solving the above Yang-Mills equations may be untenable in
practice, some methods for constructing semistable holomorphic vector
bundles are known.

Semistable holomorphic vector bundles $ V_{\mathbf{X}} $ with structure
groups
\begin{equation} 
G_{\mathbb{C}} = SU(n)_{\mathbb{C}} \subset E_{8 \mathbb{C}} 
\end{equation}
can be explicitly constructed over an \emph{elliptically} fibered Calabi-Yau
3-fold $ \mathbf{X} $ using the \emph{spectral cover method}
\cite{FriMorWit:Vec,Don:Pri,BerJohPanSad:On}. That
is, $ \mathbf{X} $ is a \emph{torus} fibered Calabi-Yau 3-fold which
admits a global section. It consists of a complex base 2-surface 
$ \mathbf{B} $ and elliptic curves $ \mathbf{E}_{b} $ fibered over each
point $ b \in \mathbf{B} $.  The Calabi-Yau condition 
\begin{equation}
c_{1}(\mathbf{TX}) = 0
\end{equation}
restricts the base $ \mathbf{B} $ \cite{Gra,MorVaf:ComII} 
to be a del Pezzo $ (\mathbf{dP}_{r}, r = 0,\dots,8) $, rational elliptic
$ (\mathbf{dP}_{9}) $, Hirzebruch $ (\mathbb{F}_{r}, r \geq 0) $, blown-up
Hirzebruch, or an Enriques surface. 
Donagi, Lukas, Ovrut and Waldram \cite{DonLukOvrWal:Non,DonLukOvrWal:Hol} 
used the above construction and the results of \cite{Cur:Chi,And:On} to
present a class of heterotic M-theory vacua. The vacua in this class have
net number of generations $ N_{\textrm{gen}} = 3 $ of chiral fermions in
the observable sector with grand unification groups such as $ H = E_{6} $, 
$ SO(10) $, and  $ SU(5) $. The global section restricts the fundamental
group of $ \mathbf{X} $ to be trivial and hence Wilson lines cannot be
used to break $ H $ to the standard model gauge group.  The exception is 
$ \mathbf{X} $ fibered over an Enriques base which, however
\cite{DonLukOvrWal:Hol}, is not consistent with the requirement 
$ N_{\textrm{gen}} = 3 $.  The vacua with nonstandard embeddings
generically contain M5-branes in the bulk space at specific points in the 
$ \mathbf{S}^{1} / \mathbb{Z}_{2} $ orbifold direction.  These M5-branes
are required to span the 4-dimensional uncompactified space (to
preserve $ (3 + 1) $-dimensional Poincar\'{e} invariance) and wrap
holomorphic curves in $ \mathbf{X} $ (to preserve $ \mathcal{N} = 1 $
supersymmetry in four dimensions). 

Considering instead torus fibered Calabi-Yau 3-folds which 
\emph{do not} admit a global section, one expects to find nontrivial first
homotopy groups.  Such a 3-fold $ \mathbf{Z} $ can be constructed from an
elliptically fibered Calabi-Yau 3-fold $ \mathbf{X} $ by modding out by a
discrete group of freely acting symmetries $ \Gamma $. The smooth 3-fold 
$ \mathbf{Z} = \mathbf{X} / \Gamma $ has first homotopy group 
$ \pi_{1} (\mathbf{Z}) = \Gamma $. To construct semistable holomorphic
vector bundles on $ \mathbf{Z} $, one finds those bundles on 
$ \mathbf{X} $ which are invariant under $ \Gamma $. These then descend to
bundles on $ \mathbf{Z} $. Donagi, Ovrut, Pantev and Waldram
\cite{DonOvrPanWal:Sta,DonOvrPanWal:Non} 
used these results to construct $ N_{\textrm{gen}} = 3 $ vacua with grand
unification group $ H = SU(5) $ which is broken to the standard model gauge
group with a $ \mathbb{Z}_{2} $ Wilson line.  This was done by taking  
$ \Gamma = \mathbb{Z}_{2} $ and constructing an elliptically fibered
Calabi-Yau 3-fold $ \mathbf{X} $ which admits \emph{two} global sections
and a freely acting involution
\begin{equation}
\tau_{ \mathbf{X} }: \mathbf{X} \rightarrow \mathbf{X}.
\end{equation} 
The torus fibered Calabi-Yau 3-fold
\begin{equation}
\mathbf{Z}  = \frac{ \mathbf{X} }{ \tau_{\mathbf{X}} }
\end{equation}
has first homotopy group
\begin{equation}
\pi_{1}(\mathbf{Z}) = \mathbb{Z}_{2}.
\end{equation}

This thesis extends the above work by Donagi, Ovrut, Pantev and Waldram by
considering the $ H = SO(10) $ case.  Hence the title `$ SO(10) $
heterotic M-theory vacua'.  The $ SO(10) $ embedding of the Standard Model
spectrum is supported by experimental evidence \cite{neutrino} for
neutrino masses.  A class of string models which preserves this embedding
are the
\emph{realistic free-fermionic models} \cite{(FSU5),(PS),(SLM),(LRS)}. 
These $ N_{\textrm{gen}} = 3 $, 4-dimensional string models are
constructed using the free-fermionic formulation \cite{free} of the weakly
coupled heterotic string. They have a stage in their construction which
corresponds \cite{correspond} to $ \mathbb{Z}_{2} \times \mathbb{Z}_{2} $
orbifold compactification of the weakly coupled 10-dimensional heterotic
string. This correspondence identifies associated Calabi-Yau 3-folds which
possess the structure of the above $ \mathbf{Z} $ and $ \mathbf{X} $.
This, in turn, allows the above formalism to be used to study heterotic
M-theory vacua associated with the realistic free-fermionic models.

A free-fermionic model is generated by a suitable choice of boundary
condition basis vectors (which encode the spin structure of the worldsheet
fermions) and generalized GSO projection coefficients.  The boundary
condition basis vectors associated with the realistic free-fermionic
models are constructed in two stages.  The first stage constructs the NAHE
set \cite{FarNan:Nat} of five basis vectors denoted by 
$\{\mathbf{1},\mathbf{S},\mathbf{b}_{1},\mathbf{b}_{2},\mathbf{b}_{3}\}$.
After generalized GSO projections over the NAHE set, the residual gauge
group is 
\begin{equation*}
SO(10) \times SO(6)^{3} \times E_{8}.  
\end{equation*}
NAHE set models have
$ \mathcal{N} = 1 $ spacetime supersymmetry
and 48 chiral generations in the $ \mathbf{16} $
representation of $ SO(10) $ (16 from each of the sectors 
$ \mathbf{b}_{1} $, $ \mathbf{b}_{2} $, and $ \mathbf{b}_{3} $).
The sectors $ \mathbf{b}_{1} $, $ \mathbf{b}_{2} $, and $ \mathbf{b}_{3} $
correspond to the three twisted sectors of the associated 
$ \mathbb{Z}_{2} \times \mathbb{Z}_{2} $ orbifold.  The second stage of
the construction adds three (or four) basis vectors, typically denoted by
$\{ \boldsymbol{\alpha},\boldsymbol{\beta},\boldsymbol{\gamma},\ldots \}$, 
which correspond to Wilson lines in the associated 
$ \mathbb{Z}_{2} \times \mathbb{Z}_{2} $ orbifold formulation.  These
basis vectors break the $ SO(10) \times SO(6)^{3} \times E_{8} $ gauge
group
and reduce the number of chiral generations from 48 to 3 (one from each of
the sectors
$ \mathbf{b}_{1} $, $ \mathbf{b}_{2} $, and $ \mathbf{b}_{3} $).
The $ SO(10) $ symmetry is broken to one of its subgroups. The
flipped $ SU(5) $ \cite{(FSU5)},
Pati-Salam \cite{(PS)},
Standard-like \cite{(SLM)}, and
left-right symmetric \cite{(LRS)}
$ SO(10) $ breaking patterns are shown in Table \ref{SO(10)breaking}.
\begin{table}[t]
$$ \begin{array}{cl}
\textrm{flipped} \ SU(5) & SO(10) \rightarrow SU(5) \times U(1)
\\
\textrm{Pati-Salam}      & SO(10) \rightarrow SO(6) \times SO(4)
\\
\textrm{Standard-like}   &
SO(10) \rightarrow SU(3) \times SU(2) \times U(1)^{2}
\\
\textrm{left-right symmetric} & 
  SO(10) \rightarrow SU(3) \times SU(2)_{L} \times SU(2)_{R} \times U(1)
\end{array} $$
\caption{$ SO(10) $ breaking patterns in realistic free-fermionic models.}
\label{SO(10)breaking}
\end{table}
In the former two cases, an additional 
$ \mathbf{16} $ and $ \mathbf{\overline{16}} $
representation of $ SO(10) $ is obtained from the set
$\{\boldsymbol{\alpha},\boldsymbol{\beta},\boldsymbol{\gamma},\ldots\}$.
Similarly, the hidden $ E_{8} $ is broken to one of its
subgroups.  The flavor $ SO(6) $ symmetries are broken to flavor $ U(1) $
symmetries. Three such symmetries arise from the subgroup of the
observable $ E_{8} $ which is orthogonal to $ SO(10) $. Additional 
$ U(1) $ symmetries arise from the pairing of real fermions.  The final
observable gauge group depends on the number of such pairings.

The precise geometrical realization of the full $ N_{\textrm{gen}} = 3 $
realistic free-fermionic models is not yet known.  However, the extended 
NAHE set
$ \{ \mathbf{1}, \mathbf{S}, \\ \mathbf{b}_{1}, \mathbf{b}_{2},
     \mathbf{b}_{3}, \boldsymbol{\xi}_{1} \} $,
or equivalently
$ \{ \mathbf{1}, \mathbf{S}, \boldsymbol{\xi}_{1}, \boldsymbol{\xi}_{2},
\mathbf{b}_{1}, \mathbf{b}_{2} \} $, has been shown \cite{correspond} 
to yield the same data as the $ \mathbb{Z}_{2} \times \mathbb{Z}_{2} $
orbifold of a toroidal Narain model \cite{Nar:New} with nontrivial background 
fields \cite{NarSarWit:A_No}.  This 
$ \mathbb{Z}_{2} \times \mathbb{Z}_{2} $
orbifold is denoted by $ \mathbf{Z}_{+} $ or $ \mathbf{Z}_{-} $, depending
on the choice of sign for the GSO projection coefficient 
$ C (^{ \boldsymbol{\xi_{1}} }_{ \boldsymbol{\xi_{2}} } ) $.
Each of the three twisted sectors produces eight chiral generations in the  
$ \mathbf{27} $ representation of $ E_{6} $ in the case of 
$ \mathbf{Z}_{+} $, or $ \mathbf{16} $ representation of $ SO(10) $ in
the case of $ \mathbf{Z}_{-} $.  The untwisted sector produces an
additional three $ \mathbf{27} $ and $ \mathbf{ \overline{27} } $ or
$ \mathbf{16} $ and $ \mathbf{ \overline{16} } $
repesentations of $ E_{6} $ or $ SO(10) $, respectively, yielding 
$ N_{\textrm{gen}} = 24 $.  As the NAHE set is common to all realistic
free-fermionic models, $ \mathbf{Z}_{+} $ and $ \mathbf{Z}_{-} $ are at
their core.  A freely acting shift $ \gamma $ relates the
$ (h^{(1,1)},h^{(2,1)}) = (27,3) $ orbifold $ \mathbf{Z}_{+} $
($ \mathbf{Z}_{-} $) to the $ (h^{(1,1)},h^{(2,1)}) = (51,3) $ orbifold
$ \mathbf{X}_{+} $ ($ \mathbf{X}_{-} $).  As will be discussed, the
Calabi-Yau 3-folds associated with the (27,3) and (51,3) orbifolds
have the structure of the above $ \mathbf{Z} $ and $ \mathbf{X} $. 
 
\begin{figure}[t]
\epsfig{file=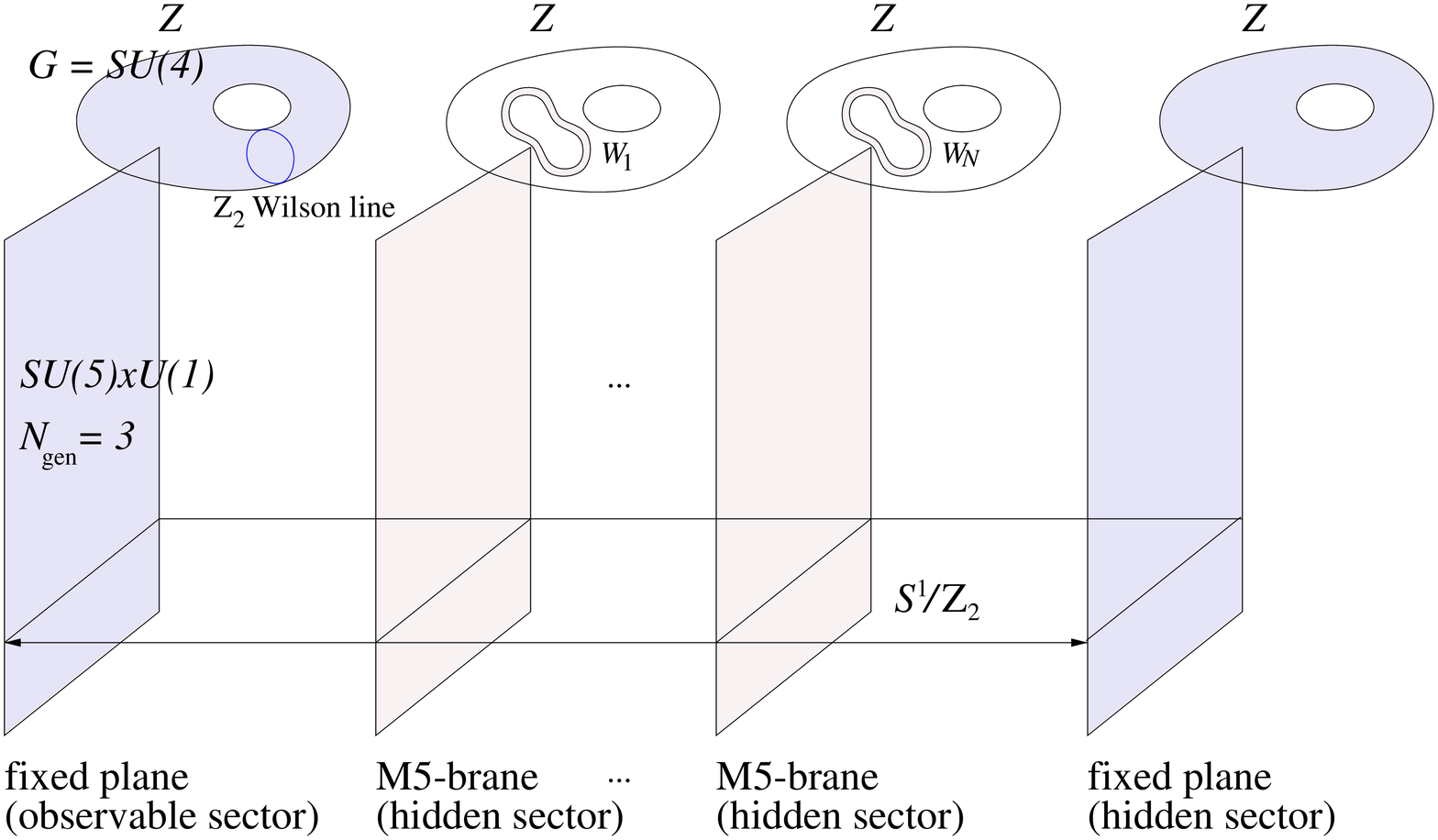, width=13cm}
\caption{Heterotic M-theory brane-world. A $ \mathbb{Z}_{2} $ Wilson line
breaks the $ SO(10) $ grand unification group to $ SU(5) \times  U(1) $
with $ N_{\textrm{gen}} = 3 $ net chiral generations in the observable
sector.
There are $ N $ M5-branes wrapping holomorphic curves $
W^{(n)}_{\mathbf{Z}} $ $ (n = 1,\ldots,N) $ in the torus fibered
Calabi-Yau 3-fold $ \mathbf{Z} $.}
\label{HetBrane}
\end{figure}

The new results presented in this thesis have been published by 
Faraggi, Garavuso and Isidro \cite{FarGarIsi:Non} and 
Faraggi and Garavuso \cite{FarGar:Yuk}.
In the former work, it is shown that a class of $ N_{\textrm{gen}} = 3 $,
$ H= SO(10) $ vacua is admitted by the $ \mathbf{B} = \mathbb{F}_{2} $
surface, while the $ \mathbf{B} = \mathbf{dP}_{3} $ surface does not admit
such a class.  Furthermore, 
$ \mathbf{B} = \mathbb{F}_{r} $ $ (r \ \textrm{even} \geq 4) $ admits this
class when certain conditions are satisfied, while 
$ \mathbf{B} = \mathbb{F}_{0} $ and 
$ \mathbf{B} = \mathbb{F}_{r} $ $ (r \ \textrm{odd} \geq 1) $ do not admit
this class.  The $ SO(10) $ symmetry is broken to 
$ SU(5) \times U(1) $ with a $ \mathbb{Z}_{2} $ Wilson line. 
Such vacua are concrete realizations of the heterotic M-theory
brane-world shown in Figure \ref{HetBrane}. 
Finally, the former work utilizes the 
$ \mathbb{Z}_{2} \times \mathbb{Z}_{2} $
correspondence to connect realistic free-fermionic models to the above
formalism.  In the latter work, vacua with potentially viable matter
Yukawa couplings are searched for.  Arnowitt and Dutta \cite{ArnDut:Yuk}
argue that such vacua can be obtained by requiring vanishing instanton
charges in the observable sector.  This restriction rules out the
solutions found in \cite{FarGarIsi:Non}.  However, by replacing the
previously considered sufficient (but not necessary) constraints on the 
vector bundles with more general constraints, it is shown that a class of 
$ N_{\textrm{gen}} = 3 $, $ H = SO(10) $ vacua with potentially viable
matter Yukawa couplings is admitted by
$ \mathbf{B} = \mathbf{dP}_{7} $ when certain conditions are satisfied,
while
$ \mathbf{B} = \mathbf{dP}_{r} $ $ ( r = 0,\ldots,6,8 ) $
and $ \mathbf{B} = \mathbb{F}_{r} $ $ (r \geq 0) $ do not admit such a
class. It is then argued how the top quark Yukawa coupling \cite{top}
in realistic free-fermionic models can be reproduced in the heterotic 
M-theory limit.  Although the precise geometrical realization of the full
$  N_{\textrm{gen}} = 3 $ models is unknown, this coupling can be computed
as a $ \mathbf{27}^{3} $ $ E_{6} $ or $ \mathbf{16} \cdot \mathbf{16}
\cdot \mathbf{10} $ $ SO(10) $ coupling with $ N_{\textrm{gen}} = 24 $.
This leads to the presentation of rules for constructing heterotic
M-theory vacua allowing $ H = E_{6} $ and $ SO(10) $ grand unification
groups with arbitrary $ N_{\textrm{gen}} $.

The presentation is organized as follows:  Chapter \ref{HetM-theory}
reviews Ho\v{r}ava-Witten theory, its compactification to four dimensions
with $ \mathcal{N} = 1 $ supersymmetry, and the associated
4-dimensional low energy effective action.  Chapter \ref{HetVac}
discusses rules for constructing heterotic M-theory vacua allowing grand
unification groups such as $ H = E_{6} $, $ SO(10) $, and $ SU(5) $ with
arbitrary $ N_{\textrm{gen}} $. The vacua
appearing in \cite{FarGarIsi:Non} and \cite{FarGar:Yuk} are presented in 
Chapters \ref{NoYukVac} and \ref{YukVac}, respectively. Chapter \ref{Free}
reviews realistic free-fermionic models and their 
$ \mathbb{Z}_{2} \times \mathbb{Z}_{2} $ correspondence in more detail,
explains how this correspondence identifies associated Calabi-Yau 3-folds
which possess the structure of the above $ \mathbf{Z} $ and
$ \mathbf{X} $, and argues how the top quark Yukawa coupling in 
realistic free-fermionic models can be reproduced in the heterotic
M-theory limit.  Chapter \ref{Summary} summarizes the new results
presented in this thesis. Appendix \ref{Chern} reviews Chern classes.  The
spectral cover method is reviewed in Appendix \ref{Spectral}.

\chapter{\label{HetM-theory}Heterotic M-theory}

M-theory on the orbifold $ \mathbf{S}^{1} / \mathbb{Z}_{2} $ is believed
to describe the strong coupling limit of the $ E_{8} \times E_{8} $
heterotic string \cite{HorWit:Het}.  At low energy, this theory is
described by 11-dimensional supergravity \cite{CreJulSch:Sup} coupled to
one 10-dimensional $ E_{8} $ Yang-Mills supermultiplet on each of the two
orbifold fixed planes. This low energy description, known as
\emph{Ho\v{r}ava-Witten theory} \cite{HorWit:Ele}, will be summarized in
Section \ref{HorWit}.  As discussed in Chapter \ref{Intro},
Ho\v{r}ava-Witten theory compactified to lower dimensions with arbitrary
gauge vacua is referred to as \emph{heterotic M-theory}. Compactification
on a Calabi-Yau 3-fold to four dimensions with unbroken 
$ \mathcal{N} = 1 $ supersymmetry \cite{Wit:Str} is reviewed in Section
\ref{Comp}, and the corresponding 4-dimensional effective theory is
discussed in Section \ref{4DLow}.

The following conventions will be employed.  The 11-dimensional space
$ \mathbf{M}^{11} $ is parameterized by coordinates
$ x^{I} $ with indices 
\begin{equation}
I,J,K,\ldots = 0,\ldots,9,11. 
\end{equation}
The metric 
\begin{equation}
g_{IJ} = \eta_{mn} e^{m}_{I} e^{n}_{J} 
\end{equation}
has Lorentz signature 
$ (-,+,\ldots,+) $.  The $ 32 \times 32 $ real gamma matrices satisfy 
\begin{align}
\{ \Gamma_{I}, \Gamma_{J} \} & = 2 g_{IJ} 
\\
\Gamma_{0} \ldots \Gamma_{9} \Gamma_{11} & = 1,
\end{align}
and
\begin{equation}
\Gamma^{ I_{1} \ldots I_{n} } \equiv 
  \Gamma^{[I_{1}} \ldots \Gamma^{I_{n}]}.
\end{equation}  
The orbifold $ \mathbf{S}^{1} / \mathbb{Z}_{2} $
is chosen to be tangent to  $ x^{11} $, with $ \mathbb{Z}_{2} $ acting as
\begin{equation}
\mathbb{Z}_{2}: x^{11} \rightarrow - x^{11}.  
\end{equation}
Barred indices
\begin{equation}
\bar{I},\bar{J},\bar{K},\ldots = 0,\ldots,9
\end{equation}
are used to
label the coordinates of the 10-dimensional space orthogonal to the
orbifold. In the \emph{upstairs} picture, 
$ x^{11} \in [-\pi \rho, \pi \rho ] $ with the endpoints
identified. There are two 10-dimensional hyperplanes, 
$ \mathbf{M}^{10}_{ \textrm{\tiny{($i$)}} } \ (i = 1,2) $, 
locally specified by 
$ x^{11} = 0 $ and $ x^{11} = \pi \rho $, 
which are fixed under
$ \mathbb{Z}_{2} $ action. In other words, the orbifold is a circle 
$ \mathbf{S}^{1} $ of radius $ \rho $ with fixed points at $ x^{11} = 0 $
and  $ x^{11} = \pi \rho $.  In the \emph{downstairs} picture, the
orbifold is an interval $ x^{11} \in [0, \pi \rho] $ with
$ \mathbf{M}^{10}_{ \textrm{\tiny{($i$)}} } $ $ (i = 1,2) $ forming
boundaries to 
$ \mathbf{M}^{11} $. The fields are required to have a definite behavior
under the $ \mathbb{Z}_{2} $ action.  A bosonic field $ \Phi $ must be
even or odd; that is, 
\begin{equation}
\Phi(x^{11}) = \pm \Phi(-x^{11}). 
\end{equation}
For an 11-dimensional Majorana spinor $ \Psi $, the condition is
\begin{equation}
\Psi(x^{11}) = \pm \Gamma_{11} \Phi(-x^{11})  
\end{equation}
so that the projection to
an orbifold fixed plane yields a 10-dimensional Majorana-Weyl spinor.
The 11-dimensional supergravity multiplet consists of the metric
$ g_{IJ} $, a 3-form potential $ C_{IJK} $ with field strength
$ G_{IJKL} = d_{I} C_{JKL} = 4! \, \partial_{ [I}C_{JKL] } $,
and the gravitino $ \psi_{I \alpha} $ ($ \alpha $ is a 32-component
Majorana spinor index).  For the bosonic fields,
$ g_{ \bar{I} \bar{J} } $, $ g_{11,11} $ and
$ C_{\bar{I} \bar{J} 11} $ must be even under
$ \mathbb{Z}_{2} $, while $ g_{\bar{I} 11} $ and
$ C_{ \bar{I} \bar{J} \bar{K} } $ must be odd.  For the 11-dimensional
gravitino, the condition is
\begin{equation}
\label{gravitino}
\psi_{\bar{I}}(x^{11}) = \Gamma_{11} \psi_{\bar{I}}(-x^{11}), \quad
\psi_{11}(x^{11}) = - \Gamma_{11} \psi_{11}(-x^{11}).
\end{equation}
The 11-dimensional supergravity multiplet is coupled to one $ E_{8} $
Yang-Mills supermultiplet 
$ (A^{ \textrm{\tiny{($i$)}} a }_{\bar{I}},
   \chi^{ \textrm{\tiny{($i$)}} a }) $ 
on each orbifold fixed plane 
$ \mathbf{M}^{10}_{ \textrm{\tiny{($i$)}} } $ 
$ (i = 1,2) $. Here, $ a = 1,\ldots,248 $ labels the adjoint
representation of $ E_{8} $. The $ E_{8} $ gauge field 
$ A^{ \textrm{\tiny{($i$)}} }_{\bar{I}} $ has field strength
\begin{equation}
F^{ \textrm{\tiny{($i$)}} }_{\bar{I}\bar{J}} 
  =   \partial_{\bar{I}} A^{ \textrm{\tiny{($i$)}} }_{\bar{J}}
    - \partial_{\bar{J}} A^{ \textrm{\tiny{($i$)}} }_{\bar{I}}
    + [A^{ \textrm{\tiny{($i$)}} }_{\bar{I}},
       A^{ \textrm{\tiny{($i$)}} }_{\bar{J}}],
\end{equation} 
and the gaugino $ \chi^{ \textrm{\tiny{($i$)}} } $ satisfies 
\begin{equation}
\Gamma_{11} \chi^{ \textrm{\tiny{($i$)}} } 
  = \chi^{ \textrm{\tiny{($i$)}} }.
\end{equation}  
An inner product is defined by 
\begin{equation}
X^{a} X^{a} = \textrm{tr}(X^{2}) = \frac{1}{30} \textrm{Tr}(X^{2}),
\end{equation}
with `Tr' the trace in the adjoint representation.  The spin connection
\begin{equation}
\Omega_{IJK} \equiv \Omega^{mn}_{I} e_{Jm}e_{Kn} 
\end{equation}
is understood to be
given by the solution of the field equation that results from varying it
as an independent field.  The Riemann tensor is the field strength
constructed from $ \Omega $. When the theory is further compactified on a
Calabi-Yau manifold in Section \ref{Comp}, indices 
\begin{equation}
A,B,C,\ldots = 4,\ldots,9  
\end{equation}
label the Calabi-Yau coordinates.
Holomorphic and
antiholomorphic indices on the Calabi-Yau space are denoted by $
a,b,c,\ldots $ and $ \bar{a},\bar{b},\bar{c},\ldots $, respectively.
Coordinates of the 4-dimensional uncompactified space $ \mathbf{M}^{4} $
are labeled by indices 
\begin{equation}
\mu, \nu, \ldots = 0,\ldots,3.  
\end{equation}

\section{\label{HorWit}Ho\v{r}ava-Witten theory}

Ho\v{r}ava-Witten theory can be formulated as an expansion \cite{HorWit:Ele} 
in the 11-dimensional gravitational coupling $ \kappa $. To lowest order in 
this expansion, Ho\v{r}ava-Witten theory is 11-dimensional supergravity
(which is of order $ \kappa^{-2} $), with the fields restricted under the
$ \mathbb{Z}_{2} $ action as described above. In the upstairs picture, the
action is

\begin{multline}
\label{S_SG}
S_{\textrm{SG}} = - \frac{1}{ \kappa^{2} } \int_{\mathbf{M}^{11}} d^{11}x
                    \sqrt{g}
         \Biggl[ \frac{1}{2} R
         + \frac{1}{2} \overline{\psi}_{I} \Gamma^{IJK}
            D_{J}(\Omega) \psi_{K}
         + \frac{1}{48} G_{IJKL}G^{IJKL}     
         \Biggr.    
\\
         + \frac{ \sqrt{2} }{192}
           \left( \overline{\psi}_{I} \Gamma^{IJKLMN} \psi_{N}
                  + 12 \overline{\psi}^{J} \Gamma^{KL} \psi^{M}
           \right) G_{JKLM}                            
\\
         \left. 
         + \frac{ \sqrt{2} }{3456} \epsilon^{ I_{1} \ldots I_{11} }
           C_{ I_{1} I_{2} I_{3} } G_{ I_{4} \ldots I_{7} }
           G_{ I_{8} \ldots I_{11} }
         + \textrm{(Fermi)}^{4}
         \right].
\end{multline}  
The terms which are quartic in the gravitino can be absorbed into the
definition of supercovariant objects.  The condition (\ref{gravitino})
means that the gravitino is chiral from a 10-dimensional perspective, and
so the theory has a gravitational anomaly localized on the fixed planes.
$ S_{\textrm{SG}} $ is invariant under the local supersymmetry
transformations
\begin{align}
\delta e_{I}^{m} &= \frac{1}{2} \bar{\eta} \Gamma^{m} \psi_{I} 
\\
\label{superC}
\delta C_{IJK} &= - \frac{\sqrt{2}}{8} \bar{\eta} 
                     \Gamma_{ [IJ } \psi_{ K] } 
\\
\delta \psi_{I} &= D_{I}(\Omega) \eta + \frac{ \sqrt{2} }{288}
   (\Gamma_{IJKLM} - 8 g_{IJ} \Gamma_{KLM}) G^{JKLM} \eta 
 + (\textrm{Fermi})^{2}
\end{align}
whose infinitesimal spacetime dependent Grassmann parameter $ \eta $
(which transforms as a Majorana spinor) satisfies the orbifold condition
 \begin{equation}          
\eta(x^{11}) = \Gamma_{11} \eta(-x^{11}).
\end{equation}
This condition means that the theory has 32 supersymmetries in the bulk,
but only 16 (chiral) supersymmetries on the orbifold fixed planes.  At
this order in $ \kappa $, the 4-form field strength $ G_{IJKL} $ satisfies
the boundary conditions
\begin{align}
G_{ \bar{I} \bar{J} \bar{K} \bar{L} } |_{x^{11}=0} &= 0
\\
G_{ \bar{I} \bar{J} \bar{K} \bar{L} } |_{x^{11}=\pi \rho} &= 0,
\end{align}
the equation of motion
\begin{equation}
D_{I}(\Omega) G^{IJKL} =  0,
\end{equation}
and the Bianchi identity
\begin{equation}
(dG)_{IJKLM} = 0.
\end{equation}
To this order, it is consistent to set $ G_{IJKL} = 0 $.

Cancellation of the gravitational anomaly requires
the introduction of one $ E_{8} $ Yang-Mills supermultiplet 
$ (A^{ \textrm{\tiny{($i$)}}a}_{\bar{I}},\chi^{\textrm{\tiny{($i$)}}a}) $ 
on each orbifold fixed plane 
$ \mathbf{M}^{10}_{ \textrm{\tiny{($i$)}} } $ $ (i = 1,2) $.  
The minimal Yang-Mills action is
\begin{equation}
S_{\textrm{YM}} = - \frac{1}{\lambda^{2}} \sum_{i=1}^{2} 
                    \int_{ \mathbf{M}^{10}_{ \textrm{\tiny{($i$)}} } }
                    d^{10}x \sqrt{g} \ \textrm{tr}
 \left( 
        \frac{1}{4} F^{ \textrm{\tiny{($i$)}} }_{\bar{I} \bar{J}}
                    F^{ \textrm{\tiny{($i$)}} \bar{I} \bar{J} } 
       +\frac{1}{2} \bar{\chi}^{ \textrm{\tiny{($i$)}}  } 
                    \Gamma^{\bar{I}} D_{\bar{J}}(\Omega)
                    \chi^{ \textrm{\tiny{($i$)}}  }
 \right)
\end{equation}
where $ \lambda $ is the 10-dimensional gauge coupling. This action is
invariant under the global supersymmetry transformations
\begin{align}
\delta A^{ \textrm{\tiny{($i$)}} a }_{\bar{I}} 
  &= \frac{1}{2} \bar{\eta} \Gamma_{ \bar{I} } 
                 \chi^{ \textrm{\tiny{($i$)}} a }
\\
\delta \chi^{ \textrm{\tiny{($i$)}} a } 
  &= - \frac{1}{4} \Gamma^{\bar{I}\bar{J}}
                   F^{ \textrm{\tiny{($i$)}} a }_{ \bar{I} \bar{J} }
                   \eta.
\end{align}
The challenge is then to add interactions and modify the supersymmetry
transformation laws so that 
$ S = S_{\textrm{SG}} + S_{\textrm{YM}} + \cdots $ is locally
supersymmetric.  This involves coupling the gravitino to the Yang-Mills
supercurrent.  However, since the gravitino lives in the 11-dimensional
bulk, while the Yang-Mills supermultiplets live on the 10-dimensional
fixed planes, a locally supersymmetric theory cannot be achieved simply by
adding interactions on the fixed planes.  To achieve local supersymmetry,
the Bianchi identity must be modified to read
\begin{multline}
\label{Bianchimod}
(dG)_{11 \bar{I} \bar{J} \bar{K} \bar{L}} =  
  8 \pi^{2} \sqrt{2} \ \frac{ \kappa^{2} }{ \lambda^{2} }
  \Biggl\{   J^{(0)} \delta(x^{11}) 
           + J^{(N+1)} \delta (x^{11} - \pi \rho)
  \Biggr.           
\\
  \Biggl.  + \frac{1}{2} \sum_{n=1}^{N} J^{(n)}
             \left[ \delta (x^{11} - x_{n}) + \delta (x^{11} + x_{n})
             \right]
  \Biggr\}_{\bar{I} \bar{J} \bar{K} \bar{L}}
\end{multline}
where 
\begin{align}
J^{(0)} 
 &= - \frac{1}{16 \pi^{2}}  
      \left[  
              \textrm{tr} 
              \left( 
                F^{ \textrm{\tiny{(1)}} } \wedge F^{ \textrm{\tiny{(1)}} } 
              \right) 
             -\frac{1}{2} \, \textrm{tr} \, (R \wedge R)
      \right]_{x^{11} = 0}
\\[10pt]
J^{(N+1)}    
 &= - \frac{1}{16 \pi^{2}}  
      \left[  
              \textrm{tr} 
              \left( 
                F^{ \textrm{\tiny{(2)}} } \wedge F^{ \textrm{\tiny{(2)}} } 
              \right)
             -\frac{1}{2} \, \textrm{tr} \, (R \wedge R)
      \right]_{x^{11} = \pi \rho} 
\end{align} 
are the sources on the fixed planes at $ x^{11} = x_{0} \equiv 0 $
and $ x^{11} = x_{N+1} \equiv \pi \rho $, respectively, and 
$ J^{(n)} $ $ (n = 1,\ldots,N) $ are the M5-brane sources located at
$ x^{11} = x_{1},\ldots,x_{N} $ 
$ (0 \leq x_{1} \leq \ldots \leq x_{N} \leq \pi \rho) $.
Each M5-brane at $ x = x_{n} $ is paired with a mirror M5-brane at 
$ x = - x_{n} $ with the same source since the Bianchi identity must be
even under the $ \mathbb{Z}_{2} $ action.

With the modified Bianchi identity (\ref{Bianchimod}), 
$ S = S_{\textrm{SG}} + S_{\textrm{YM}} + \ldots $ can be made locally
supersymmetric.  However, having gained supersymmetry, Yang Mills
gauge invariance has been lost.  The modified Bianchi identity implies
that $ G_{ 11 \bar{I} \bar{J} \bar{K} } $ is invariant under the
infinitesimal gauge transformations 
\begin{equation}
\delta A^{ \textrm{\tiny{($i$)}} a }_{\bar{I}} 
  = D_{\bar{I}}(\Omega) \epsilon^{ \textrm{\tiny{($i$)}} a } 
\end{equation}
if $ C_{ 11 \bar{I} \bar{J} } $ transforms as
\begin{equation}
\delta C_{ 11 \bar{I} \bar{J} } =
 - \frac{ \kappa^{2} }{6 \sqrt{2} \lambda^{2}}
   \left[
      \delta (x^{11})  \ \textrm{tr} 
      \left( 
        \epsilon^{ \textrm{\tiny{(1)}} } 
        F^{ \textrm{\tiny{(1)}} }_{\bar{I} \bar{J}} 
      \right)
    + \delta (x^{11} - \pi \rho)  \ \textrm{tr} 
      \left( 
        \epsilon^{ \textrm{\tiny{(2)}} } 
        F^{ \textrm{\tiny{(2)}} }_{\bar{I} \bar{J}} 
      \right)
   \right].
\end{equation}
This implies that the $ CGG $ Chern-Simons interaction is not gauge
invariant.  Thus, the classical theory is not gauge invariant, and a
consistent classical theory does not exist.  At the quantum level, there
is in addition the 10-dimensional Majorana-Weyl anomaly which cancels the
gauge anomaly of the CGG interaction provided
\begin{equation}
\frac{1}{\lambda^{2}} = \frac{1}{2 \pi \kappa^{2}} 
                        \left( \frac{\kappa}{4 \pi} \right)^{2/3}. 
\end{equation}
The quantum theory is anomaly free; gauge, gravitational and mixed
anomalies are cancelled with a refinement \cite{HorWit:Het,HorWit:Ele} of
the standard \emph{Green-Schwarz mechanism} \cite{GreSch:Ano}.

To order $ \kappa^{-2 + (2/3)} $, the Ho\v{r}ava-Witten action is
\begin{multline}
\label{HorWitAct}
S = - \frac{1}{ \kappa^{2} } \int_{\mathbf{M}^{11}} d^{11}x
                    \sqrt{g}
         \Biggl[ \frac{1}{2} R
         + \frac{1}{2} \overline{\psi}_{I} \Gamma^{IJK}
            D_{J} \left( \textrm{\scriptsize $ \frac{1}{2} $} 
                         ( \Omega + \hat{\Omega} )
                  \right) \psi_{K}
         \Biggr.   
\\
         + \frac{ \sqrt{2} }{384}
           \left( \overline{\psi}_{I} \Gamma^{IJKLMN} \psi_{N}
                  + 12 \overline{\psi}^{J} \Gamma^{KL} \psi^{M}
           \right) ( G_{JKLM} + \hat{G}_{JKLM} )                           
\\
         \left.
         + \frac{1}{48} G_{IJKL}G^{IJKL}
         + \frac{ \sqrt{2} }{3456} \epsilon^{ I_{1} \ldots I_{11} }
           C_{ I_{1} I_{2} I_{3} } G_{ I_{4} \ldots I_{7} }
           G_{ I_{8} \ldots I_{11} }
         \right]
\\
    - \frac{1}{2 \pi \kappa^{2}}
      \left( \frac{\kappa}{4 \pi} \right)^{2/3}
      \sum_{i=1}^{2} \int_{ \mathbf{M}^{10}_{ \textrm{\tiny{($i$)}} } }
      d^{10}x \sqrt{g}
      \Biggl[
         \frac{1}{4} F^{ \textrm{\tiny{($i$)}} a }_{\bar{I} \bar{J}} 
                     F^{ \textrm{\tiny{($i$)}} a \bar{I} \bar{J} }
       + \frac{1}{2} \bar{\chi}^{ \textrm{\tiny{($i$)}} a } 
                     \Gamma^{\bar{I}} D_{\bar{J}}(\hat{\Omega}) 
                     \chi^{ \textrm{\tiny{($i$)}} a }
      \Biggr.
\\
      \left.
       + \frac{1}{8} \bar{\psi}_{I} \Gamma^{\bar{J}\bar{K}} 
         \Gamma^{\bar{I}} 
         \left(   F^{ \textrm{\tiny{($i$)}} a }_{\bar{J} \bar{K}} 
                + \hat{F}^{ \textrm{\tiny{($i$)}} a }_{\bar{J} \bar{K}}
         \right) \chi^{ \textrm{\tiny{($i$)}} a }
       - \frac{\sqrt{2}}{48} \bar{\chi}^{ \textrm{\tiny{($i$)}} a } 
         \Gamma^{\bar{I} \bar{J} \bar{K}} \chi^{ \textrm{\tiny{($i$)}}a } 
         \hat{G}_{\bar{I} \bar{J} \bar{K} 11}  
      \right].
\end{multline}
where the quartic Fermi terms are absorbed into the supercovariant objects
\begin{align}
\hat{\Omega}_{IJK} & = \Omega_{IJK} 
                     + \frac{1}{8} \bar{\psi}^{L} \Gamma_{LIJKM} \psi^{M} 
\\[10pt]
\hat{G}_{IJKL} &=   G_{IJKL} 
                  + \frac{3 \sqrt{2}}{4} \bar{\psi}_{[I} 
                    \Gamma_{JK} \psi_{L]}
\\[10pt]
\hat{F}^{ \textrm{\tiny{($i$)}} a }_{\bar{I} \bar{J}} 
    &=   F^{ \textrm{\tiny{($i$)}} a }_{\bar{I} \bar{J}} 
       - \bar{\psi}_{[\bar{I}} \Gamma_{\bar{J} \, ]} 
         \chi^{ \textrm{\tiny{($i$)}} a }.
\end{align}
This action is invariant under the local supersymmetry transformations
\begin{align}
\delta e_{I}^{m} &= \frac{1}{2} \bar{\eta} \Gamma^{m} \psi_{I}
\\[10pt]  
\delta C_{IJK} \ &\textrm{is given by (\ref{superC}) with correction} 
               \ \delta' C_{11\bar{J}\bar{K}}  
\\[10pt]
\delta \psi_{\bar{I}} 
  &=   D_{\bar{I}}(\hat{\Omega}) \eta 
     + \frac{\sqrt{2}}{288}
       (\Gamma_{\bar{I}JKLM} - 8 g_{\bar{I}J} \Gamma_{KLM}) 
       \hat{G}^{JKLM} \eta
     + \delta' \psi_{\bar{I}}
\\[10pt]
\delta \psi_{11}
  &=   D_{11}(\hat{\Omega}) \eta
     + \frac{\sqrt{2}}{288}
       (\Gamma_{11JKLM} - 8 g_{11J} \Gamma_{KLM})
       \hat{G}^{JKLM} \eta
     + \delta' \psi_{11}
\\[10pt]
\delta A^{ \textrm{\tiny{($i$)}} a }_{\bar{I}} 
  &= \frac{1}{2} \bar{\eta} \Gamma_{ \bar{I} } 
     \chi^{ \textrm{\tiny{($i$)}} a }
\\[10pt]
\delta \chi^{ \textrm{\tiny{($i$)}} a } 
  &= - \frac{1}{4} \Gamma^{\bar{I}\bar{J}}
       F^{ \textrm{\tiny{($i$)}} a }_{ \bar{I} \bar{J} } \eta
     + \delta' \chi^{ \textrm{\tiny{($i$)}} a }
\end{align}
where the supersymmetry transformation law corrections are
\begin{align}
\delta' C_{11 \bar{J} \bar{K}}
  &= \frac{1}{6 \sqrt{2}} \,
       \frac{1}{2 \pi} \left( \frac{\kappa}{4 \pi} \right)^{2/3}
     \sum_{i=1}^{2} \textrm{tr} 
     \left( 
         A^{ \textrm{\tiny{($i$)}} }_{\bar{J}} 
         \delta A^{ \textrm{\tiny{($i$)}} }_{\bar{K}}  
       - A^{ \textrm{\tiny{($i$)}} }_{\bar{K}} 
         \delta A^{ \textrm{\tiny{($i$)}} }_{\bar{J}}
     \right) 
\\[10pt]
\delta' \psi_{\bar{I}}
  &= - \frac{1}{576 \pi} \left( \frac{\kappa}{4 \pi} \right)^{2/3}
       \left[
          \delta (x^{11}) \, \bar{\chi}^{ \textrm{\tiny{(1)}} a } 
          \Gamma^{\bar{J} \bar{K} \bar{L}} 
          \chi^{ \textrm{\tiny{(1)}} a } 
       \right. 
\nonumber
\\
   & \left. \quad \quad \quad
     + \, \delta (x^{11} - \pi \rho) \, 
          \bar{\chi}^{ \textrm{\tiny{(2)}} a }
          \Gamma^{\bar{J} \bar{K} \bar{L}} 
          \chi^{ \textrm{\tiny{(2)}} a }
      \right] 
      \left( 
              \Gamma_{\bar{J} \bar{K} \bar{L}} 
             - 6 g_{\bar{I}\bar{J}} \Gamma_{\bar{K}\bar{L}} 
      \right) \eta
\\[10pt]
\delta' \psi_{11}
  &= + \frac{1}{576 \pi} \left( \frac{\kappa}{4 \pi} \right)^{2/3}
       \left[
          \delta (x^{11}) \, 
          \bar{\chi}^{ \textrm{\tiny{(1)}} a }
          \Gamma^{JKL} \chi^{ \textrm{\tiny{(1)}}a }
       \right.
\nonumber
\\
  & \left. \quad \quad \quad \quad \quad \quad \quad
     + \, \delta (x^{11} - \pi \rho) \, 
          \bar{\chi}^{ \textrm{\tiny{(2)}} a }
          \Gamma^{JKL} \chi^{ \textrm{\tiny{(2)}} a }
      \right] \Gamma_{JKL} \, \eta
\\[10pt]
\delta' \chi^{ \textrm{\tiny{($i$)}} a } 
   &= \frac{1}{4} \bar{\psi}_{\bar{I}} \Gamma_{\bar{J}} 
      \chi^{ \textrm{\tiny{($i$)}} a }
      \Gamma^{\bar{I} \bar{J}} \eta.                        
\end{align}
At this order in $ \kappa $, the 4-form field strength $ G_{IJKL} $
satisfies the boundary conditions

\begin{align}   
G_{ \bar{I} \bar{J} \bar{K} \bar{L} } |_{x^{11}=0} 
  &= - \frac{3}{\sqrt{2}} \,  
         \frac{1}{2 \pi} \left( \frac{\kappa}{4 \pi} \right)^{2/3}  
       \left(
          F^{ \textrm{\tiny{(1)}} a }_{[\bar{I} \bar{J}} 
          F^{ \textrm{\tiny{(1)}} a }_{\bar{K} \bar{L}]}
        - \frac{1}{2} \,
          \textrm{tr} \, R_{[\bar{I} \bar{J}} R_{\bar{K} \bar{L}]}
       \right)
\\[10pt]
G_{ \bar{I} \bar{J} \bar{K} \bar{L} } |_{x^{11}=\pi \rho} 
  &= + \frac{3}{\sqrt{2}} \,                                           
         \frac{1}{2 \pi} \left( \frac{\kappa}{4 \pi} \right)^{2/3}
       \left(
          F^{ \textrm{\tiny{(2)}} a }_{[\bar{I} \bar{J}} 
          F^{ \textrm{\tiny{(2)}} a }_{\bar{K} \bar{L}]}
        - \frac{1}{2} \,
          \textrm{tr} \, R_{[\bar{I} \bar{J}} R_{\bar{K} \bar{L}]}
       \right),
\end{align}
the equation of motion
\begin{equation}
D_{I}(\Omega) G^{IJKL} =  0,
\end{equation}
and the Bianchi identity
\begin{multline}
(dG)_{11 \bar{I} \bar{J} \bar{K} \bar{L}} =
  4 \sqrt{2} \pi \left( \frac{\kappa}{4 \pi} \right)^{2/3}
  \Biggl\{   J^{(0)} \delta(x^{11})
           + J^{(N+1)} \delta (x^{11} - \pi \rho)
  \Biggr.
\\
  \Biggl.  + \frac{1}{2} \sum_{n=1}^{N} J^{(n)}
             \left[ \delta (x^{11} - x_{n}) + \delta (x^{11} + x_{n})
             \right]
  \Biggr\}_{\bar{I} \bar{J} \bar{K} \bar{L}}.
\end{multline}
Note that the $ \textrm{tr} \, (R \wedge R) $ terms appearing in the above
boundary
conditions and Bianchi identity are not required by the low energy
theory, but (since they are needed for anomaly cancellation, given the 
structure of the one-loop chiral anomalies) must be present in the full
quantum M-theory. 

\section{\label{Comp}Compactification to $ D = 4 $
with $ \mathcal{N} = 1 $}

The compactification of Ho\v{r}ava-Witten theory to four dimensions with
unbroken $ \mathcal{N} = 1 $ supersymmetry is discussed in \cite{Wit:Str}.  
The procedure starts with the spacetime structure
\begin{equation}
\label{spacetime}
\mathbf{M}^{11} = \mathbf{M}^{4} \times \mathbf{Z} \times
\mathbf{S}^{1}/\mathbb{Z}_{2},
\end{equation}
where $ \mathbf{M}^{4} $ is 4-dimensional Minkowski space and 
$ \mathbf{Z} $ is a Calabi-Yau 3-fold.  M5-branes can be included in the
bulk space at points throughout the orbifold interval.  These M5-branes
are required to span $ \mathbf{M}^{4} $ (to preserve 
$ (3 + 1) $-dimensional Poincar\'{e} invariance) and wrap holomorphic
curves in $ \mathbf{Z} $ (to preserve $ \mathcal{N} = 1 $ supersymmetry in 
four dimensions).

The correction to the background (\ref{spacetime}) is computed
perturbatively. The set of equations to be solved consists of the Killing
spinor equation
\begin{equation}
\label{Killing}
\delta \psi_{I} = D_{I} \eta + \frac{ \sqrt{2} }{288}
   (\Gamma_{IJKLM} - 8 g_{IJ} \Gamma_{KLM}) G^{JKLM} \eta = 0,
\end{equation}
the equation of motion   
\begin{equation}
\label{motion}
D_{I} G^{IJKL} = 0,
\end{equation} 
and the Bianchi identity
\begin{multline}
\label{modBianchi}
(dG)_{11 \bar{I} \bar{J} \bar{K} \bar{L}} 
    = 4 \sqrt{2} \pi 
      \left( \frac{\kappa}{4 \pi} \right)^{2/3}
  \Biggl\{ 
     J^{(0)} \delta(x^{11}) + J^{(N+1)} \delta (x^{11} - \pi \rho)
  \Biggr.           
\\  
  \Biggl.  
   + \frac{1}{2} \sum_{n=1}^{N} J^{(n)}
       \left[ 
         \delta (x^{11} - x_{n}) + \delta (x^{11} + x_{n})
       \right]
     \Biggr\}_{\bar{I} \bar{J} \bar{K} \bar{L}}.
\end{multline}

The Bianchi identity (\ref{modBianchi}) can be viewed as an expansion in
powers of $ \kappa^{2/3} $.  To linear order in $ \kappa^{2/3} $,
the solution to the Killing spinor equation, equation of motion, and
Bianchi identity takes the form
\begin{gather}
(ds)^{2}  = (1 + b)\eta_{\mu \nu} dx^{\mu} dx^{\nu}
            + (g^{ \textrm{\tiny{(CY)}} }_{AB} + h_{AB}) dx^{A} dx^{B}
            + (1 + \gamma) \left( dx^{11} \right)^{2}    \\
\begin{align}
G_{ABCD}  &= 0 + G'_{ABCD}                              \\
G_{ABC11} &= 0 + G'_{ABC11}                             \\
\eta      &= (1 + \psi) \eta^{ \textrm{\tiny{(CY)}} }.
\end{align}
\end{gather}
with all other components of $ G_{IJKL} $ vanishing.  
$ g^{ \textrm{\tiny{(CY)}} }_{AB} $ and 
$ \eta^{ \textrm{\tiny{(CY)}} } $
are the Ricci-flat metric and the covariantly constant spinor on the
Calabi-Yau 3-fold.

As discussed in \cite{LukOvrWal:On}, the first order corrections
$ b $, $ h_{AB} $, $ \gamma $, $ G' $ and $ \psi $ can be expressed
in terms of a single $ (1,1) $-form $ \mathcal{B}_{a \bar{b}} $ on
the Calabi-Yau 3-fold.  The results are
\begin{align}
b    
  &= \frac{\sqrt{2}}{6} \beta   
\\
h_{a \bar{b}}
  &= \sqrt{2} i 
     \left(   
         \beta_{a \bar{b}} 
       - \frac{1}{3} \omega_{a \bar{b}} \beta 
     \right)
\\
\gamma 
  &= - \frac{\sqrt{2}}{3} \beta
\\
G'_{ABCD} 
  &= \frac{1}{2} \epsilon_{ABCDEF} \partial_{11} \beta^{EF}  
\\
G'_{ABC11}
  &= \frac{1}{2} \epsilon_{ABCDEF} \partial^{D} \beta^{EF}
\\
\psi 
  &= - \frac{\sqrt{2}}{24} \beta
\end{align}
where $ \beta = \omega^{AB} \beta_{AB} $ and 
$ \omega_{a \bar{b}} = -i g^{ \textrm{\tiny{(CY)}} }_{a \bar{b}} $ 
is the K\"{a}hler form. All that remains then is to determine 
$ \mathcal{B}_{a \overline{b}} $,
which can be expanded in terms of eigenmodes of the Laplacian on the
Calabi-Yau 3-fold.  For the purpose of computing low energy effective
actions, it is sufficient to keep only the zero eigenvalue or `massless'
terms in this expansion; that is, the terms proportional to the harmonic 
$ (1,1) $ forms of the Calabi-Yau 3-fold.
Let $ \{ \omega_{ia \bar{b}} \} $ be a basis for these
harmonic $ (1,1) $-forms, where $ i = 1, \ldots, h^{(1,1)} $.
Then
\begin{equation}
\mathcal{B}_{a \bar{b}} = \sum_{i} b_{i} \omega^{i}_{a \bar{b}} 
         + (\textrm{massive terms}).
\end{equation}
The $ \omega_{ia \bar{b}} $ are Poincar\'{e} dual to the 4-cycles
$ \mathcal{C}_{4i} $, and one can define the integer charges
\begin{equation}
\beta^{(n)}_{i} = \int_{\mathcal{C}_{4i}} J^{(n)},
\quad n = 0,1,\ldots,N, N+1.
\end{equation}
$ \beta^{(0)}_{i} $ and $ \beta^{(N+1)}_{i} $ are the instanton charges on
the orbifold fixed planes and $ \beta^{(n)}_{i} $, $ n = 1, \ldots, N $   
are the the magnetic charges of the M5-branes.  The expansion coefficients
$ b_{i} $ are found in \cite{LukOvrWal:Non} in terms of these charges,
the normalized orbifold coordinates
\begin{equation}
\label{normorb} 
z = \frac{x^{11}}{\pi \rho},  \quad z_{n} = \frac{x_{n}}{\pi \rho} \quad
(n = 1, \ldots N), \quad z_{0} = 0, \quad z_{1} = 1
\end{equation}
and the expansion parameter
\begin{equation}
\epsilon = \left( \frac{\kappa}{4 \pi} \right)^{2/3} \frac{2 \pi^{2}
\rho}{\mathcal{V}^{2/3}},
\end{equation}
where 
\begin{equation}
 \mathcal{V} = \int_{ \mathbf{Z} } d^{6}x \sqrt{g^{ \textrm{\tiny{(CY)}}
}} 
\end{equation}
is the Calabi-Yau volume.  The result is
\begin{equation}
b_{i} = \frac{\epsilon}{\sqrt{2}}
        \left[
            \sum_{m=0}^{n} \beta^{(m)}_{i} ( |z| - z_{m} )
          - \frac{1}{2} \sum_{m=0}^{N=1} ( 1 - z_{m} )^{2} \beta^{(m)}_{i}
        \right]
\end{equation}
in the interval
\begin{equation}
z_{n} \leq |z| \leq z_{n+1}
\end{equation}
for fixed $ n $, where $ n = 0,\dots,N $.

A cohomological constraint on the Calabi-Yau 3-fold $ \mathbf{Z} $, the
gauge bundles $ V^{ \textrm{\tiny{($i$)}} }_{\mathbf{Z}} $ on the orbifold
fixed planes 
$ \mathbf{M}^{10}_{ \textrm{\tiny{($i$)}} } $ $ (i = 1,2) $, and the
M5-branes can be found by
integrating the Bianchi identity over a 5-cycle which spans the orbifold
interval together with an arbitrary 4-cycle $ \mathcal{C}_{4} $ in the
Calabi-Yau 3-fold. Since $ dG $ is exact, this integral must vanish.
Physically, this is the statement that there can be no net charge in a
compact space, since there is nowhere for the flux to `escape'.
Performing the integral over the orbifold interval yields   
\begin{equation}
\sum_{n=0}^{N+1} \int_{ \mathcal{C}_{4} } J^{(n)} = 0.
\end{equation} 
Since this is true for an arbitrary 4-cycle $ \mathcal{C}_{4} $ in the
Calabi-Yau 3-fold, it follows that the sum of the sources must be
cohomologically trivial.  That is
\begin{align}
\label{CohomTriv}
0 &= \sum_{n=0}^{N+1} \left[ J^{(n)} \right] 
   = - \frac{1}{16 \pi^{2}}
      \left\{ 
         \sum_{i=1}^{2} 
           \left[
              \textrm{tr} 
                \left( 
                  F^{ \textrm{\tiny{($i$)}} } \wedge 
                  F^{ \textrm{\tiny{($i$)}} } 
                \right)
           \right]
        -  \left[
              \textrm{tr} \, ( R \wedge R )
           \right] 
      \right\}
    + \sum_{n=1}^{N} \left[ J^{(n)} \right].
\end{align}
From Appendix \ref{Chern}, one obtains for the second Chern classes of 
$ V^{ \textrm{\tiny{($i$)}} }_{\mathbf{Z}} $ and the
tangent bundle $ \mathbf{TZ} $ 
\begin{align}
c_{2} \! \left( V^{ \textrm{\tiny{($i$)}} }_{\mathbf{Z}} \right)
  &= - \frac{1}{16 \pi^{2}}
       \left[  
         \textrm{tr} \!
         \left( 
           F^{ \textrm{\tiny{($i$)}} } \wedge 
           F^{ \textrm{\tiny{($i$)}} } 
         \right)
       \right]
\\
c_{2} (\mathbf{TZ})
  &= - \frac{1}{16 \pi^{2}} 
       \left[
         \textrm{tr} \,
         ( R \wedge R )
       \right].
\end{align}
It follows that (\ref{CohomTriv}) can be written as
\begin{equation}
[W_{\mathbf{Z}}] =   c_{2}(\mathbf{TZ}) 
                   - c_{2} \! 
                     \left( 
                       V^{ \textrm{\tiny{(1)}} }_{\mathbf{Z}}
                     \right) 
                   - c_{2} \! 
                     \left( 
                       V^{ \textrm{\tiny{(2)}} }_{\mathbf{Z}} 
                     \right)
\end{equation}
where
\begin{equation}
[W_{\mathbf{Z}}] = \sum_{n=1}^{N} \left[ J^{(n)} \right]
\end{equation}
is the 4-form cohomology class associated with the M5-branes. 
Note that if $ V^{ \textrm{\tiny{(2)}} }_{\mathbf{Z}} $ is chosen to be
trivial
(so that $ c_{2} ( V^{ \textrm{\tiny{(2)}} }_{\mathbf{Z}} ) $ vanishes),
then (\ref{CohomTriv}) shows that vacua with standard embeddings 
(for which 
$   \textrm{tr} \left( F^{ \textrm{\tiny{(1)}} } \wedge 
                       F^{ \textrm{\tiny{(1)}} } \right)
  - \textrm{tr} \, ( R \wedge R ) = 0 $) do not contain M5-branes.

\section{\label{4DLow}$ D = 4 $ low energy effective action}

The 4-dimensional low energy effective action is obtained 
\cite{LukOvrWal:Fiv} by reducing the 11-dimensional action
(\ref{HorWitAct}) on the background solution discussed in Section
\ref{Comp}.  The results obtained are summarized below.

The field content of the action splits into $ N+2 $ sectors coupled to one
another only gravitationally.  Two of these sectors arise from the $ E_{8}
$ gauge supermultiplets on the orbifold fixed planes, while the other $ N
$ originate from the degrees of freedom on the M5-brane worldvolume
theories.  All of these sectors have $ \mathcal{N} = 1 $ supersymmetry.

Generally, the $ E_{8} $ gauge groups on the observable and hidden
orbifold fixed planes are broken to $ H_{(1)} $ and
$ H_{(2)} $, respectively.  There are chiral matter fields
in those representations $ \mathcal{R} $ of $ H_{(i)} $ that appear in the
decomposition 
\begin{equation}
\mathbf{248}_{E_{8}} \rightarrow \oplus_{\mathcal{S},\mathcal{R}}
(\mathcal{S},\mathcal{R})
\end{equation}
of the adjoint of $ E_{8} $ under $ G_{(i)} \times H_{(i)} $ $ (i=1,2) $.
The number of families for a certain representation $ \mathcal{R} $ is
given by the dimension of 
$ H^{1} \! \left( \mathbf{Z}, 
         V^{ \textrm{\tiny{($i$)}} }_{\mathbf{Z}}(\mathcal{S}) \right) $,
where 
$ V^{ \textrm{\tiny{($i$)}} }_{\mathbf{Z}}(\mathcal{S}) $ 
is the vector bundle 
$ V^{ \textrm{\tiny{($i$)}} }_{\mathbf{Z}} $ 
in the representation $ \mathcal{S} $. For example, the decomposition of
the adjoint of $ E_{8} $ under $ SU(4) \times SO(10) $ is
\begin{equation}
\label{decomp}
\mathbf{248}_{E_{8}} 
  = (\mathbf{15},\mathbf{1}) \oplus (\mathbf{1},\mathbf{45}) 
    \oplus (\mathbf{6},\mathbf{10}) \oplus (\mathbf{4},\mathbf{16}) 
    \oplus (\mathbf{\overline{4}},\mathbf{\overline{16}}).
\end{equation}
The relevant representations are then 
$ (\mathcal{S},\mathcal{R}) = (\mathbf{4},\mathbf{16}) $ 
for the families,
$ (\mathcal{S},\mathcal{R}) 
  = (\mathbf{\overline{4}},\mathbf{\overline{16}}) $
for the anti-families, and
$ (\mathcal{S},\mathcal{R}) = (\mathbf{6},\mathbf{10}) $
for the Higgs fields.  A chiral matter field on the observable 
and hidden orbifold fixed planes is denoted by
$ C^{ \textrm{\tiny{(1)}} \mathcal{I} p } (\mathcal{R}) $ and
$ C^{ \textrm{\tiny{(2)}} \mathcal{I} p } (\mathcal{R}) $, respectively.
Here,
\begin{equation} 
\mathcal{I}, \mathcal{J}, \mathcal{K}, \ldots
  = 1,\ldots, 
    \textrm{dim} \!
       \left[ 
         H^{1} \! 
               \left(
                 \mathbf{Z}, 
                 V^{ \textrm{\tiny{($i$)}} }_{\mathbf{Z}}(\mathcal{S})
                \right) 
       \right]
\end{equation}
are family indices, and
\begin{equation} 
p,q,\ldots = 1, \ldots, \textrm{dim}(\mathcal{R})
\end{equation}
are representation indices of $ \mathcal{R} $.  The cohomology group
$ H^{1} \! \left( \mathbf{Z}, 
           V^{ \textrm{\tiny{($i$)}} }_{\mathbf{Z}}(\mathcal{S}) \right) $
has basis 
$ \{ u^{ \textrm{\tiny{($i$)}} x }_{\mathcal{I}}(\mathcal{R}) \} $,
where
\begin{equation} 
x,y,\ldots = 1,\ldots,\textrm{dim}(\mathcal{S}) 
\end{equation}
are representation
indices of $ \mathcal{S} $.  The generators $ T_{xp}(\mathcal{R}) $ and
their complex conjugates $ T^{xp}(\mathcal{R}) $ satisfy
\begin{equation}
\textrm{tr} \left( T_{xp}(\mathcal{R}) T^{yq}(\mathcal{R}) \right) 
  = \delta_{x}{}^{y} \delta_{p}{}^{q}.
\end{equation}
Recall that the vector bundle $ V^{ \textrm{\tiny{(2)}} }_{\mathbf{Z}} $
on the hidden orbifold fixed plane is assumed to be trivial so that 
$ H_{(2)}  = E_{8} $.

The $ N $ M5-branes yield gauge groups $ \mathcal{H}_{(n)} $ 
$ (n=1,\ldots,N) $.  Generically, these groups are 
$ \mathcal{H}_{(n)} = U(1)^{g_{n}} $, where $ g_{n} $ is the genus of the
holomorphic curve $ W^{(n)}_{\mathbf{Z}} $ on which the 
$ n^{\textrm{th}} $ M5-brane wraps.  $ \mathcal{H}_{(n)} $ can be enhanced
to non-Abelian groups, typically unitary groups, when M5-branes overlap or
a curve $ W^{(n)}_{Z} $ degenerates.  For example, $ N $ M5-branes,
wrapped on the same holomorphic curve with genus $ g $ and positioned at
different points in the orbifold lead to a group $ U(1)^{gN} $.  Moving
all $ M5 $-branes to the same orbifold point enhances the group to
$ U(N)^{g} $.

By construction, the low energy massless modes fall into 4-dimensional 
$ \mathcal{N} = 1 $ chiral or vector supermultiplets.  The low energy
theory is then characterized by specifying three types of functions.  The
K\"{a}hler potential $ K $ describes the structure of the kinetic energy
terms for the chiral fields while the holomorphic superpotential $ W $
encodes their potential.  The holomorphic gauge kinetic functions
$ f^{(1)} $ and $ f^{(2)} $ represent the coupling of the chiral fields
to the $ H_{(1)} = H $ and $ H_{(2)} = E_{8} $ gauge group fields.  For a
set of chiral fields $ Y^{\imath} $, the relevant terms of the low energy 
action are
\begin{multline}
S_{Y^{\imath}} 
  = - \frac{1}{16 \pi G_{N}} \int_{\mathbf{M}^{4}} d^{4}x \sqrt{g}
      \left[
         R + 2K_{\imath \bar{\jmath} } \,
         \partial_{\mu} Y^{\imath} \partial^{\mu} \bar{Y}^{\bar{\jmath}}
      \right.
\\
      \left.
    + 2 e^{K} 
        \left( 
            K^{\imath \bar{\jmath}} D_{\imath}W \overline{D_{\jmath}W} 
          - 3|W|^{2}
        \right) 
    + \textrm{D-terms} 
      \right]
\\
    - \frac{1}{ 16 \pi \alpha_{ \textrm{GUT} } } \sum_{i=1}^{2}
      \int_{\mathbf{M}^{4}_{ \textrm{\tiny{($i$)}} }} d^{4}x \sqrt{g} \,
      \Bigl\{
         \text{Re} \left[ f^{(i)}(Y^{\imath}) \right]
         \textrm{tr} 
           \left( 
             F^{ \textrm{\tiny{($i$)}} } \wedge 
             F^{ \textrm{\tiny{($i$)}} } 
           \right) 
      \Bigr.
\\ 
      \Bigl. 
        + \text{Im} \left[ f^{(i)}(Y^{\imath}) \right]
          \textrm{tr} \, 
            \bigl( 
              F^{ \textrm{\tiny{($i$)}} } \wedge 
              \tilde{F}^{ \textrm{\tiny{($i$)}} } 
            \bigr)
      \Bigr\}
\end{multline} 
where 
\begin{equation}
K_{\imath \bar{\jmath}} 
  = \frac{ \partial^{2} K } 
         { \partial Y^{ \imath } \partial \bar{Y}^{ \bar{\jmath} } }
\end{equation}
is the K\"{a}hler metric, 
\begin{equation}
D_{\imath} W  
  =   \partial_{\imath} W 
    + \frac{\partial K}{\partial Y^{\imath}} W 
\end{equation}
is the K\"{a}hler covariant derivative acting on the superpotential,
\begin{equation}
\tilde{F}^{ \textrm{\tiny{($i$)}} }_{\mu \nu} 
  = \frac{1}{2} \epsilon_{\mu \nu \sigma \rho} 
    F^{ \textrm{\tiny{($i$)}} }_{\sigma \rho}
\end{equation}
is the dual field strength, and
\begin{gather}
\alpha_{\textrm{GUT}} = \frac{ (4 \pi \kappa^{2})^{2/3} }{2 \mathcal{V}}
\\
G_{N} = \frac{\kappa^{2}}{16 \pi^{2} \mathcal{V} \rho}
\end{gather}
are the 4-dimensional gauge coupling and Newton constant, respectively.

Various moduli arise in the low energy theory.  First of all,
associated with the Calabi-Yau 3-fold are the K\"{a}hler moduli $ a^{i} $
$ (i = 1,\ldots,h^{(1,1)}) $, defined as $ \omega = a^{i} \omega_{i} $,
along with their superpartners.
Here  $ \omega_{a \bar{b}} = -i g^{ \textrm{\tiny{(CY)}} }_{a \bar{b}} $
is the K\"{a}hler form. The modulus
\begin{equation}
V = \frac{1}{2 \pi \rho v} 
    \int_{ \mathbf{Z} \times \mathbf{S}^{1} / \mathbb{Z}_{2} }
    d^{7}x \sqrt{^{7}g}
\end{equation}
is the orbifold average of the Calabi-Yau volume 
$ \mathcal{V} 
  = \int_{\mathbf{Z}} d^{6}x \sqrt{g^{ \textrm{\tiny{(CY)}} }} $ 
in units of the Calabi-Yau coordinate volume
\begin{equation} 
v = \int_{\mathbf{Z}} d^{6}x,
\end{equation}
and the modulus
\begin{equation}
R = \frac{1}{2 \pi \rho \mathcal{V}} 
    \int_{ \mathbf{S}^{1} / \mathbb{Z}_{2} \times \mathbf{Z} }
    d^{7}x \sqrt{^{7}g} 
\end{equation}
is the Calabi-Yau averaged orbifold size in units of $ 2 \pi \rho $.
In terms of the K\"{a}hler moduli, 
$ V $ can be written $ V = \frac{1}{6} d_{ijk} a^{i} a^{j} a^{k} $
where
\begin{equation}
d_{ijk} = \int_{Z} \omega_{i} \wedge \omega_{j} \wedge \omega_{k}
\end{equation}
are the Calabi-Yau intersection numbers. Bundle moduli arise
from the vector bundles 
$ V^{ \textrm{\tiny{(1)}} }_{\mathbf{Z}} $ and
$ V^{ \textrm{\tiny{(2)}} }_{\mathbf{Z}} $. 
The moduli $ z_{n} $ $ (n = 1,\ldots,N) $ specify the position of the 
$ N $ M5-branes.  Finally, there are moduli which parameterize the 
total M5-brane curve $ W_{\mathbf{Z}} $.  

In terms of the above moduli, the real parts of the chiral fields $ S $, $
T^{i} $ and $ Z_{n} $ are written
\begin{equation}
\textrm{Re}(S) = V, \quad \textrm{Re}(T^{i}) = Ra^{i}, \quad
\textrm{Re}(Z_{n}) = z_{n}.
\end{equation}
The metric
\begin{equation}
G^{(i)}_{IJ} (a^{i};\mathcal{R}) = \frac{1}{ \mathcal{V} }
\int_{\mathbf{Z}}
  \sqrt{ g^{(CY)} } g^{ (CY) a \overline{b} }
  u^{ \textrm{\tiny{($i$)}} }_{Iax}(\mathcal{R}) 
  u^{ \textrm{\tiny{($i$)}} x }_{J \overline{b}}(\mathcal{R})
\end{equation}
on the moduli space
$ H^{1} \! \left( \mathbf{Z},
         V^{ \textrm{\tiny{($i$)}} }_{\mathbf{Z}}(\mathcal{S}) \right) $
is a function of the K\"{a}hler moduli $ a^{i} $ as indicated, as well as
the complex structure and bundle moduli.  The matter Yukawa couplings
\begin{equation}
\label{lambdaYuk}
\lambda^{ \textrm{\tiny{($i$)}} }_{\mathcal{I} \mathcal{J} \mathcal{K}} 
 ( \mathcal{R}_{1}, \mathcal{R}_{2}, \mathcal{R}_{3} )
 = \int_{\mathbf{Z}} \Omega \wedge 
   u^{ \textrm{\tiny{($i$)}} x }_{\mathcal{I}}(\mathcal{R}_{1}) \wedge
   u^{ \textrm{\tiny{($i$)}} y }_{\mathcal{J}}(\mathcal{R}_{2}) \wedge
   u^{ \textrm{\tiny{($i$)}} z }_{\mathcal{K}}(\mathcal{R}_{3})
   f^{ (\mathcal{R}_{1} \mathcal{R}_{2} \mathcal{R}_{3} ) }_{xyz},
\end{equation}
where $ \Omega $ is the covariantly constant (3,0) form and
$ f^{ (\mathcal{R}_{1} \mathcal{R}_{2} \mathcal{R}_{3} ) }_{xyz} $
projects out the singlet parts of 
$ \mathcal{R}_{1} \times \mathcal{R}_{2} \times \mathcal{R}_{3} $,
depend on the complex structure moduli and the Dolbeault cohomology
classes of the 
$ u^{ \textrm{\tiny{($i$)}} x }_{\mathcal{I}} (\mathcal{R}) $.
For example, for the $ SO(10) $ case (\ref{decomp}), the relevant products
are $ \mathbf{10} \times \mathbf{16} \times \mathbf{16} $ and 
$ \mathbf{10} \times \mathbf{\overline{16}} 
              \times \mathbf{\overline{16}} $.

For the explicit computation of $ K $, $ W $, 
$ f^{ \textrm{\tiny{(1)}} } $ and $ f^{ \textrm{\tiny{(2)}} } $,
it is convenient to express the metric of the vacuum solution in 
Section \ref{Comp} in terms of the above moduli.  One finds
\begin{align}
g_{\mu \nu} 
  &= V^{-1} R^{-1} \left( 1 + \frac{2 \sqrt{2}}{3} b_{i} a^{i} \right)
     \eta_{\mu \nu}
\\
g_{a \bar{b}} 
  &= i \left( 
         a^{i} - \frac{4 \sqrt{2}}{3} b_{j} a^{j} a^{i} 
               + \sqrt{2} b^{i}
       \right) \omega_{i a \bar{b}}.
\end{align} 
The part of the $ E_{8} $ gauge field strength 
$ F^{ \textrm{\tiny{($i$)}} } $ that gives rise to
chiral matter fields can be written as
\begin{equation}
F^{ \textrm{\tiny{($i$)}} }_{\mu \bar{b}} 
  = \sqrt{ 2 \pi \alpha_{\textrm{GUT}} } \sum_{\mathcal{R}} 
    u^{ \textrm{\tiny{($i$)}} x }_{\mathcal{I} \bar{b}}(\mathcal{R})  
    T_{xp}(\mathcal{R}) 
    D_{\mu} C^{ \textrm{\tiny{($i$)}} \mathcal{I} p }(\mathcal{R}).
\end{equation}  
In the following, the representation $ \mathcal{R} $ and the corresponding
representation indices are suppressed in order to simplify the notation.
One can now think of the indices 
$ \mathcal{I}, \mathcal{J}, \mathcal{K},\ldots $
as running over the various relevant representations as well as the
families.  The gauge kinetic functions are
\begin{align}
f^{ \textrm{\tiny{(1)}} }
  &= S + \epsilon T^{i}
         \left(
             \beta^{(0)}_{i} 
           + \sum_{i=1}^{N} (1-Z_{n})^{2} \beta^{(n)}_{i}
         \right)
\\
f^{ \textrm{\tiny{(2)}} }
  &= S + \epsilon T^{i}
         \left(
             \beta^{(0)}_{i}
           + \sum_{i=1}^{N} (Z_{n})^{2} \beta^{(n)}_{i}
         \right).
\end{align}
The matter K\"{a}hler potential and metric
associated with $ \mathbf{M}^{10}_{ \textrm{\tiny{($i$)}} } $ are
\begin{align}
K^{ \textrm{\tiny{($i$)}} }_{\textrm{matter}}
  &= Z^{ \textrm{\tiny{($i$)}} }_{\mathcal{I} \mathcal{J}}
     \bar{C}^{ \textrm{\tiny{($i$)}} \mathcal{I} }
           C^{ \textrm{\tiny{($i$)}} \mathcal{J} }
\\
\label{matterKmetric}
Z^{ \textrm{\tiny{($i$)}} }_{\mathcal{I} \mathcal{J}}
  &= e^{-K_{T}/3}
    \left[
        K^{ \textrm{\tiny{($i$)}} }_{\textrm{B} \mathcal{I} \mathcal{J}}
      - \frac{\epsilon}{2 \mathcal{V}}
        \tilde{\Gamma}
          ^{\textrm{\tiny{($i$)}}i}_{\textrm{B}\mathcal{I}\mathcal{J}}
        \sum_{n=0}^{N+1} (1 - z_{n})^{2} \beta_{i}^{(n)}
\right],
\end{align}
where the $ (1,1) $ K\"{a}hler potential and metric are
\begin{align}
K_{T} &= - \ln \left[ \frac{1}{6} d_{ijk}
          ( T^{i} + \bar{T}^{i} )
          ( T^{j} + \bar{T}^{j} )
          ( T^{k} + \bar{T}^{k} ) \right]
\\ 
K_{Tij} &= \frac{ \partial^{2} K_{T} }
               { \partial T^{i} \partial \bar{T}^{j} },
\end{align}
and the quantities
\begin{align}
K^{ \textrm{\tiny{($i$)}} }_{\textrm{B} \mathcal{I} \mathcal{J}} 
  &= G^{ \textrm{\tiny{($i$)}} }_{\mathcal{I} \mathcal{J}} (T + \bar{T}) 
\\
\Gamma^{ \textrm{\tiny{($i$)}} i }_{\textrm{B} \mathcal{I} \mathcal{J}}
  &= K^{ij}_{T} 
     \frac{ \partial 
            K^{\textrm{\tiny{($i$)}}}_{\textrm{B}\mathcal{I}\mathcal{J}} }
          {\partial T^{j}}
\\
\tilde{\Gamma}^{\textrm{\tiny{($i$)}}i}_{\textrm{B}\mathcal{I}\mathcal{J}}
  &=    \Gamma^{\textrm{\tiny{($i$)}}i}_{\textrm{B}\mathcal{I}\mathcal{J}}
      - (T^{i} + \bar{T}^{i}) 
        Z^{ \textrm{\tiny{($i$)}} }_{\textrm{B} \mathcal{I} \mathcal{J}}
      - \frac{2}{3} (T^{i} + \bar{T}^{i}) (T^{k} + \bar{T}^{k})
        K_{Tkj} 
        \Gamma^{\textrm{\tiny{($i$)}}j}_{\textrm{B}\mathcal{I}\mathcal{J}}
\end{align}
are associated with the vector bundle 
$ V^{ \textrm{\tiny{($i$)}} }_{\mathbf{Z}} $.
The matter superpotential is
\begin{equation}
W_{\textrm{matter}} 
  = \sum_{i=1}^{2}
    \frac{1}{3}  
    \tilde{Y}^{\textrm{\tiny{($i$)}}}_{\mathcal{I}\mathcal{J}\mathcal{K}}
    C^{ \textrm{\tiny{($i$)}} \mathcal{I} }
    C^{ \textrm{\tiny{($i$)}} \mathcal{J} }
    C^{ \textrm{\tiny{($i$)}} \mathcal{K} }
\end{equation}
where
\begin{equation}
\tilde{Y}^{\textrm{\tiny{($i$)}}}_{\mathcal{I}\mathcal{J}\mathcal{K}}
  = 2 \sqrt{2 \pi \alpha_{\textrm{GUT}} } 
    \lambda^{ \textrm{\tiny{($i$)}} }_{\mathcal{I}\mathcal{J}\mathcal{K}}.
\end{equation}
The total K\"{a}hler potential $ K $ is
\begin{equation}
K =   \kappa^{-2}_{ \textrm{\tiny{(4)}} } K_{ \textrm{\textrm{mod}} }  
    + \textstyle{\sum_{i=1}^{2}} 
      K^{ \textrm{\tiny{($i$)}} }_{ \textrm{matter} } 
\end{equation}
where $ \kappa_{ \textrm{\tiny{(4)}} } $ is the 4-dimensional
gravitational coupling
\begin{equation}
\kappa^{2}_{ \textrm{\tiny{(4)}} } 
  = \frac{\kappa^{2}}{2 \pi \rho \mathcal{V} }
\end{equation}
and $ K_{ \textrm{mod} } $ is the K\"{a}hler potential of the moduli.  
Integrating out the moduli yields the effective Yukawa couplings 
\begin{equation}
Y^{\textrm{\tiny{($i$)}}}_{\mathcal{I}\mathcal{J}\mathcal{K}}
  = e^{ K_{\textrm{mod}} / 2 }
    \tilde{Y}^{\textrm{\tiny{($i$)}}}_{\mathcal{I}\mathcal{J}\mathcal{K}}.
\end{equation}
These results hold at the grand unification scale $ M_{\textrm{GUT}} $,
which coincides with the compactification scale $ \mathcal{V}^{1/6} $.
One then uses the supersymmetry renormalization group equations to
evaluate the Yukawa couplings at low energy.

If the perturbative correction to the background discussed in
Section \ref{Comp} is to make sense, the second term
in (\ref{matterKmetric}) must be a small correction to the
first.  However, setting 
$ \mathcal{V}^{1/6} = {M_{\textrm{GUT}}} = 3 \times 10^{16} $ GeV, one
finds $ \epsilon \simeq 0.93 $. Furthermore, one expects
$ G_{IJ} $ and $ \tilde{\Gamma}_{IJ} $ to be of order 1.  Arnowitt and    
Dutta \cite{ArnDut:Yuk} point out that the second term can still be a
small correction to the first if the instanton charges on the observable
orbifold fixed plane vanish and the M5-branes cluster
near the hidden orbifold fixed plane:
\begin{align}
\beta^{(0)}_{i} &= 0       \\
d_{n}           &\equiv (1 - z_{n}) \ll 1, \quad n = 1, \ldots, N. 
\end{align}
The $ \beta^{(0)}_{i} = 0 $ constraint will be imposed in 
Section \ref{Rules}.

\chapter{\label{HetVac}Heterotic M-theory vacua}

This chapter presents rules for constructing a class of heterotic M-theory
vacua.  Compactification to four dimensions with unbroken 
$ \mathcal{N} = 1 $ supersymmetry is achieved on a torus fibered
Calabi-Yau 3-fold $ \mathbf{Z} = \mathbf{X} / \tau_{\mathbf{X}} $ with 
first homotopy group
$ \pi_{1}(\mathbf{Z}) = \mathbb{Z}_{2} $.  Here $ \mathbf{X} $ is an
elliptically fibered Calabi-Yau 3-fold which admits two global sections
and $ \tau_{\mathbf{X}} $ is a freely acting involution on $ \mathbf{X} $.
The vacua in this class have grand unification groups such as 
$ H = E_{6} $, $ SO(10) $, and $ SU(5) $ with an \emph{arbitrary} net
number of generations $ N_{\textrm{gen}} $ of chiral fermions in the
observable sector, and potentially viable matter Yukawa couplings.
Allowing $ E_{6} $ and $ SO(10) $ grand unification groups with arbitrary
$ N_{\textrm{gen}} $ will prove to be useful in Section \ref{TopQuarkYuk}.
The $ H = E_{6} $, $ SO(10) $, and 
$ SU(5) $ vacua correspond to semistable holomorphic vector bundles 
$ V_{\mathbf{Z}} $ over $ \mathbf{Z} $ having structure group 
$ G_{\mathbb{C}} = SU(n)_{\mathbb{C}} $ with $ n = 3 $, 4, and 5,
respectively. Since $ \pi_{1}(\mathbf{Z}) = \mathbb{Z}_{2} $,
$ H $ can be broken with $ \mathbb{Z}_{2} $ Wilson lines. The 
vacua with nonstandard embeddings generically contain M5-branes
in the bulk space at specific points in the 
$ \mathbf{S}^{1}/ \mathbb{Z}_{2} $ orbifold direction. These 
M5-branes span the 4-dimensional uncompactified space and wrap
holomorphic curves in $ \mathbf{Z} $.

\section{\label{Rules}Rules}

The rules presented below are an adaptation of those presented
in \cite{DonOvrPanWal:Sta,DonOvrPanWal:Non}.  They
describe how to construct $ \mathbf{Z} $
and $ V_{\mathbf{Z}} $, and impose phenomenological constraints to obtain
heterotic M-theory vacua.
The vector bundles 
$ V^{ \textrm{\tiny{(1)}} }_{\mathbf{Z}} $ 
and
$ V^{ \textrm{\tiny{(2)}} }_{\mathbf{Z}} $ 
are located on the observable and hidden orbifold fixed planes, 
$ \mathbf{M}^{10}_{ \textrm{\tiny{(1)}} } $ and 
$ \mathbf{M}^{10}_{ \textrm{\tiny{(2)}} } $, respectively.  
$ V^{ \textrm{\tiny{(2)}} }_{\mathbf{Z}} $ is taken to be a
trivial bundle so that $ E_{8} $ remains unbroken in the hidden sector.

\vskip 10pt
\noindent
{\bf Construct $ \mathbf{Z} $:}  Construct a smooth \emph{torus}
fibered Calabi-Yau 3-fold $ \mathbf{Z} $ with 
$ \pi_{1}(\mathbf{Z}) = \mathbb{Z}_{2} $. To  
do this, first construct a smooth \emph{elliptically} fibered
Calabi-Yau 3-fold $ \mathbf{X} \stackrel{\pi} \longrightarrow \mathbf{B} $
which admits a freely-acting involution $ \tau_{\mathbf{X}} $.  
$ \mathbf{Z} $ is then realized as the quotient manifold 
$ \mathbf{Z} = \mathbf{X}/\tau_{\mathbf{X}} $.  

\begin{itemize}

\item {\bf Construct $ \mathbf{X} $:} To construct a smooth elliptically
fibered Calabi-Yau 3-fold $ \mathbf{X} $ which admits a freely acting
involution $ \tau_{\mathbf{X}} $,

\begin{enumerate}

\item {\bf Choose the base $ \mathbf{B} $:}  The Calabi-Yau condition
$ c_{1}(\mathbf{TX}) = 0 $ restricts the base \cite{Gra,MorVaf:ComII} to
be a del Pezzo $ (\mathbf{dP}_{r}, r = 0,\dots,8) $, rational elliptic 
$ (\mathbf{dP}_{9}) $, Hirzebruch $ (\mathbb{F}_{r}, r \geq 0) $, blown-up
Hirzebruch, or an Enriques surface.

\item {\bf Require two global sections:}  Require 
$ \mathbf{X} $ to admit \emph{two} global sections
$ \sigma $ and $ \xi $ satisfying
\begin{equation}
\label{order2}
\xi + \xi = \sigma.
\end{equation}
This requirement will facilitate the construction of a freely acting
involution $ \tau_{\mathbf{X}} $ on $ \mathbf{X} $.

\vskip 10pt
Elliptically fibered manifolds can be described in terms of a Weierstrass
model.  A general elliptic curve can be embedded via a cubic equation into
$ \mathbb{CP}^{2} $.  Without loss of generality, the equation can be
expressed in the Weierstrass form
\begin{equation}
\label{Wpoly}   
z y^{2} = 4 x^{3} - g_{2} z^{2} x - g_{3} z^{3}
\end{equation}
where $ g_{2} $ and $ g_{3} $ are general coefficients and $ (x,y,z) $ are
homogeneous coordinates on $ \mathbb{CP}^{2} $.  To define an elliptic
fibration over a base $ \mathbf{B} $, one needs to specify how the
coefficients $ g_{2} $ and $ g_{3} $ vary as one moves around the base. In 
order to have a pair of sections $ \sigma $ and $ \xi $, the Weierstrass
polynomial (\ref{Wpoly}) must factorize as
\begin{equation}
\label{Wpolyfac}
z y^{2} = 4 (x - \mathsf{a}z)(x^{2} + \mathsf{a}z x + \mathsf{b}z^{2}).  
\end{equation}
Comparing (\ref{Wpoly}) and (\ref{Wpolyfac}), we see that
\begin{equation}
g_{2} = 4(\mathsf{a}^{2} - \mathsf{b}), \quad 
g_{3} = 4\mathsf{a}\mathsf{b}.
\end{equation}
The zero section $ \sigma $ is given by $ (x,y,z) = (0,1,0) $, and the
second section $ \xi $ by $ (x,y,z) = (\mathsf{a},0,1) $.  The zero
section $ \sigma $ marks the zero points 
$ p_{b} = \sigma(b) $ $ (b \in \mathbf{B}) $ on the elliptic fibers 
$ \mathbf{E}_{b} = \pi^{-1} (b) $.

\item {\bf Blow up singularities:} The elliptic fibers are singular when
two roots of the Weierstrass polynomial (\ref{Wpolyfac}) coincide.  The
set of points in the base over which the fibers are singular is given by
the discriminant locus
\begin{equation}
\Delta = 0
\end{equation}
where
\begin{equation}
\Delta = \Delta_{1} \Delta^{2}_{2}
\end{equation}
and
\begin{equation}
\Delta_{1} = \mathsf{a}^{2} - 4\mathsf{b}, \quad \Delta_{2} =
4(2\mathsf{a}^{2} + \mathsf{b}).
\end{equation}  
There is a curve of singularities over the $ \Delta_{2} = 0 $ component
of the discriminant curve.  The vanishing of $ \Delta_{2} $ corresponds to
one of the roots of the 
\begin{equation*}
4 ( x^{2} + \mathsf{a}zx + \mathsf{b}z^{2} )
\end{equation*}
factor of the Weierstrass equation (\ref{Wpolyfac}) coinciding with the
zero of the 
\begin{equation*}
x - \mathsf{a} z
\end{equation*}
factor when $ z = 1 $.  Consequently, the singular points over the part of
the discriminant curve where $ \Delta_{2} = 0 $ all live in the $ \xi $
section.  These points form a curve $ \mathcal{L} $ in the section $ \xi $
given by
\begin{equation}
(x,y,z) = (\mathsf{a},0,1), \quad
2 \mathsf{a}^{2} + b = 0. 
\end{equation}
To construct the smooth Calabi-Yau 3-fold $ \mathbf{X} $, it is necessary
to blow up this entire curve. This is achieved by replacing the singular
point of each fiber over $ \Delta_{2} = 0 $ by a sphere 
$ \mathbb{CP}^{1} $.  This is a new curve $ N $ in the Calabi-Yau 3-fold.
The general elliptic fiber $ F $ has now split into two
spheres: a sphere in the new fiber class $ N $, plus the proper transform
of the singular fiber, which is in the class $ F - N $.

\vskip 10pt
The union of the new fibers over the curve $ \mathcal{L} $ forms the
\emph{exceptional divisor} $ \mathcal{E} $.  This is a surface in $
\mathbf{X} $
given by \cite{DonOvrPanWal:Sta}
\begin{equation}
\mathcal{E} = 2 \left( \sigma - \xi + \pi^{*} c_{1}(\mathbf{B}) \right). 
\end{equation}
There is an analogous surface $ \mathcal{E}' $ formed by the union of the
$ F - N $ fibers over $ \Delta_{2} $ in the zero section.
The intersection of $ \mathcal{E} $ with the spectral cover $ \mathbf{C} $ 
(defined in Appendix \ref{Spectral}) is 
\begin{equation}
\mathcal{E}|_{\mathbf{C}} 
  = 2 \left( \sigma - \xi + \pi^{*} c_{1}(\mathbf{B}) \right)
    \cdot (n \sigma + \pi^{*} \eta)
  = 4 \eta \cdot c_{1}(\mathbf{B}) N
\end{equation}
Since $ \mathbf{C} $ and $ \mathcal{E} $ are surfaces in $ \mathbf{X} $,
this implies that they intersect in a curve wrapping the new fiber
$ 4 \eta \cdot c_{1}(\mathbf{B}) $ times.  For generic $ \mathbf{C} $, the
curve $ \mathcal{E}|_{\mathbf{C}} $ projects to 
$ 4 \eta \cdot c_{1}(\mathbf{B}) $ points in the base $ \mathbf{B} $.
Thus, generically, $ \mathcal{E}|_{\mathbf{C}} $ splits into 
$ 4 \eta \cdot c_{1}(\mathbf{B}) $ distinct curves each wrapping the new
component of the fiber over a different point in the base.  In 
$ \mathbf{X} $, all of these curves are in the same homology class.
However, in $ \mathbf{C} $, since each curve can be separately blown down,
they must be in distinct homology classes, denoted by $ N_{i} $
$ (i = 1,\ldots, 4 \eta \cdot c_{1}(\mathbf{B}) ) $.  Thus, as a class in
$ \mathbf{C} $,
\begin{equation}
\mathcal{E}|_{\mathbf{C}} = \sum_{i} N_{i} \quad
(i = 1,\ldots, 4 \eta \cdot c_{1}(\mathbf{B}) ) 
\end{equation}

\end{enumerate}

\end{itemize}

\begin{itemize}

\item {\bf Construct $ \tau_{\mathbf{X}} $:}  To construct a freely acting
involution $ \tau_{\mathbf{X}} $ on $ \mathbf{X} $, 

\begin{enumerate}

\item Specify an involution $ \tau_{\mathbf{B}} $ on $ \mathbf{B} $ with
fixed point set $ \mathcal{F}_{\tau_{\mathbf{B}}} $.

\item $ \tau_{\mathbf{B}} $ lifts to an involution 
$ \alpha $ on $ \mathbf{X} $ which preserves the section $ \sigma $ only
if the fibration is invariant under $ \tau_{\mathbf{B}} $.  This means
that the coefficients $ g_{2} $ and $ g_{3} $ in the Weierstrass
equation (\ref{Wpoly}) must be invariant under $ \tau_{\mathbf{B}} $.  In
terms of the parameters $ \mathsf{a} $ and $ \mathsf{b} $ appearing in
(\ref{Wpolyfac}), this translates into
\begin{equation} 
\label{preserve}
\tau^{*}_{\mathbf{B}} (\mathsf{a}) = \mathsf{a}, \quad
\tau^{*}_{\mathbf{B}} (\mathsf{b}) = \mathsf{b}.
\end{equation}
The involution $ \alpha $ is uniquely determined by the additional
requirements that it fix the zero section $ \sigma $ and that it preserve
the holomorphic volume form.  Note that $ \alpha $ leaves fixed the whole
fiber above each point in $ \mathcal{F}_{\tau_{\mathbf{B}}} $.  Thus, 
$ \alpha $ is not by itself a suitable candidate for 
$ \tau_{\mathbf{X}} $.

\item Now, construct an involution 
$ t_{\xi}: \mathbf{X} \rightarrow \mathbf{X} $ as follows. Let $ x \in
\mathbf{X} $. Then $ x $ lies in a fiber $ \pi^{-1} (b) $ for some 
$ b \in \mathbf{B} $.  It is easily seen from (\ref{order2}) that
\begin{equation}
\label{trans}
t_{\xi} (x)  = x + \xi(b)
\end{equation}
satisfies $ t_{\xi} \circ t_{\xi} = \mathrm{id}_{\mathbf{X}} $ and, hence,
is an involution. A translation acts freely on a smooth torus.   
However, $ t_{\xi} $ might not act freely on singular fibers.  Thus,     
$ t_{\xi} $ by itself is not a suitable candidate for 
$ \tau_{\mathbf{X}} $.

\item Recall that $ \alpha $  leaves fixed the whole fiber above each
point in $ \mathcal{F}_{\tau_{\mathbf{B}}} $.  Thus, the composition
\begin{equation}
\tau_{\mathbf{X}} = \alpha \circ t_{\xi}
\end{equation}
is a freely acting involution on $ \mathbf{X} $ provided none of the   
fibers above $ \mathcal{F}_{\tau_{\mathbf{B}}} $ are singular. To ensure
that none
of the fibers above $ \mathcal{F}_{\tau_{\mathbf{B}}} $ are singular,
require
\begin{equation}
\label{freeact}
\mathcal{F}_{\tau_{\mathbf{B}}} \cap \{ \Delta = 0 \} = \emptyset.
\end{equation}

\end{enumerate}

\end{itemize}

\vskip 10pt
\noindent
{\bf Construct $ V_{\mathbf{Z}} $:}  Construct a semistable holomorphic
vector bundle $ V_{\mathbf{Z}} $ over $ \mathbf{Z} $ with structure group 
$ G_{\mathbb{C}} = SU(n)_{\mathbb{C}} $.  To do this, first use the
spectral cover method
\cite{FriMorWit:Vec,Don:Pri,BerJohPanSad:On} 
to construct a semistable holomorphic vector bundle
$ V_{\mathbf{X}} $ over $ \mathbf{X} $ with the same structure group. A
consistent bundle $ V_{\mathbf{X}} $ must satisfy the 
\emph{$ G_{\mathbb{C}} = SU(n)_{\mathbb{C}} $
bundle constraints} (\ref{BCsigma}), (\ref{BCetaChern}) and
(\ref{BCkappa}) given below. To ensure that $ H $ is the largest subgroup
of $ E_{8} $ preserved by the bundle $ V_{\mathbf{X}} $, impose the
\emph{stability constraint} (\ref{stab}). Since $ \tau_{\mathbf{X}} $ is
freely acting, $ V_{\mathbf{X}} $ descends to a bundle $ V_{\mathbf{Z}} $
over $ \mathbf{Z} $ when $ V_{\mathbf{X}} $ is 
$ \tau_{\mathbf{X}}$-equivariant \cite{DOPR:SU(4)}. Every 
$ \tau_{\mathbf{X}} $-equivariant $ V_{\mathbf{X}} $ is also 
$ \tau_{\mathbf{X}} $-invariant. Requiring  
$ \tau^{*}_{\mathbf{X}}(V_{\mathbf{X}}) = V_{\mathbf{X}} $ leads to the
\emph{bundle involution conditions} (\ref{BItau}) and (\ref{BIkappa}).
Note that the spectral cover method requires the presence of a global
section and hence cannot be used to construct $ V_{\mathbf{Z}} $ directly.

\begin{itemize}

\item {\bf $ G_{\mathbb{C}} = SU(n)_{\mathbb{C}} $ bundle constraints:}
A semistable holomorphic vector bundle $ V_{\mathbf{X}} $ over 
$ \mathbf{X} $ can be constructed by the spectral cover method, as
described in Appendix \ref{Spectral}. This
method requires the specification of a divisor $ \mathbf{C} $ of $
\mathbf{X} $
(known as the \emph{spectral cover}) and a line bundle $ \mathcal{N} $
over $ \mathbf{C} $.  The condition that $ c_{1}(V_{\mathbf{X}}) = 0 $
implies that the spectral data 
$ (\mathbf{C},\mathcal{N}) $ 
can be written in terms of
an effective divisor class $ \eta $ in the base $ \mathbf{B} $ and
coefficients $ \lambda $ and $ \kappa_{i} $
$ (i = 1, \ldots, 4 \eta \cdot c_{1}(\mathbf{B})) $.  Constraints are
placed on $ \eta $ , $ \lambda $ , and the $ \kappa_{i} $ by the condition 
that
\begin{equation}
c_{1}(\mathcal{N}) = n \left( \frac{1}{2} + \lambda \right) \sigma
   + \left( \frac{1}{2} - \lambda \right) \pi^{*}_{\mathbf{C}} \eta
   + \left( \frac{1}{2} + n \lambda \right) \pi^{*}_{\mathbf{C}}
c_{1}(\mathbf{B})
   + \textstyle{\sum_{i}} \kappa_{i} N_{i}
\end{equation}
be an integer class.  To ensure that $ c_{1}(\mathcal{N}) $ is an integer
class, one can impose the sufficient (but not necessary) Class A or 
Class B constraints discussed in Chapter \ref{NoYukVac}. More generally,
$ c_{1}(\mathcal{N}) $ will be an integer class if the
constraints
\begin{gather}
\label{BCsigma}
q \equiv n \left( \frac{1}{2} + \lambda \right) \in \mathbb{Z}
\\
\label{BCetaChern}
\left(\frac{1}{2} - \lambda \right) \pi^{*}_{\mathbf{C}} \eta
   + \left( \frac{1}{2} + n \lambda \right) \pi^{*}_{\mathbf{C}}
                                            c_{1}(\mathbf{B})
\quad \textrm{is an integer class}   
\\
\label{BCkappa}
\kappa_{i} - \frac{1}{2} m \in \mathbb{Z}, \quad m \in \mathbb{Z}
\end{gather} 
are simultaneously satisfied.  These results follow from the discussion
at the end of Appendix \ref{Spectral}.

\item{\bf Stability constraint:} The commutant $ H $ of $ G=SU(n) $ in 
$ E_{8} $ will be the largest subgroup of $ E_{8} $ preserved by the
vector bundle $ V_{\mathbf{X}} $ if \cite{BerMay:Sta}
\begin{equation}
\label{stab}
\eta \geq n c_{1}(\mathbf{B}) \quad
(n \geq 2).
\end{equation}
For the models of interest (which have $ n = 3 $, $ 4 $, and $ 5 $), the 
$ \beta^{(0)}_{i} = 0 $ constraint (\ref{eta6c_1}) ensures that
the stability constraint is satisfied.

\item {\bf Bundle involution conditions:}  Necessary conditions for 
$ V_{\mathbf{X}} $ to be $ \tau_{\mathbf{X}} $-invariant are given by
\begin{align}
\tau_{\mathbf{B}}(\eta) &= \eta    \label{BItau}\\
\textstyle{\sum_{i}} \kappa_{i} &= \eta \cdot c_{1}(\mathbf{B}).
\label{BIkappa}
\end{align}
There may be non-invariant bundles satisfying (\ref{BItau}) and
(\ref{BIkappa}).  The details of selecting only the bundles which are both 
invariant and equivariant are beyond the scope of this thesis.

\end{itemize}
{\bf Impose phenomenological constraints:}  The requirements of 
$ N_{\textrm{gen}} $ net chiral generations, anomaly cancellation, and
potentially viable matter Yuk-
\\
awa couplings are discussed below.

\begin{itemize}

\item {\bf $ N_{\textrm{gen}} $ condition:}  In the models of interest
with
$ V^{ \textrm{\tiny{(1)}} }_{\mathbf{Z}} $ having structure group 
$ G_{\mathbb{C}} = SU(n)_{\mathbb{C}} $ (with $ n = 3 $, 4, or 5), the net
number of generations ($ \# $ generations $ - $ $ \# $ antigenerations) 
$ N_{\textrm{gen}} $ of chiral fermions in the observable sector (in the 
$ \mathbf{27} - \mathbf{\overline{27}}$ of $ E_{6} $,
$ \mathbf{16} - \mathbf{\overline{16}} $ of $ SO(10) $, or 
$ \mathbf{ 10 + \overline{5} } - ( \mathbf{ \overline{10} + 5 } ) $ of 
$ SU(5) $) is given by
\begin{equation}
N_{\textrm{gen}} = \frac{1}{2} \int_{\mathbf{Z}} 
          c_{3} \! \left( V^{ \textrm{\tiny{(1)}} }_{\mathbf{Z}} \right).
\end{equation}
Since $ \mathbf{X} $ is a double cover of $ \mathbf{Z} $, it follows that 
\begin{equation}
c_{3} (V_{\mathbf{Z}}) = \frac{1}{2} c_{3}(V_{\mathbf{X}}).
\end{equation}
$ c_{3}(V_{\mathbf{X}}) $ has been computed by Curio \cite{Cur:Chi} and
Andreas \cite{And:On}:
\begin{equation}
c_{3}(V_{\mathbf{X}}) = 2 \lambda \sigma \wedge \eta \wedge ( \eta - n
c_{1}(\mathbf{B}) ).
\end{equation}
Integrating over the fiber yields
\begin{equation}
\label{Ngen}
N_{\textrm{gen}} = \frac{1}{2} \int_{\mathbf{B}} \lambda \eta 
                   \wedge (\eta - n c_{1}(\mathbf{B}))
        = \frac{1}{2} \lambda \eta \cdot ( \eta - n c_{1}(\mathbf{B}) ).
\end{equation}

\item {\bf Effectiveness conditions:}  As described in Section \ref{Comp},
anomaly cancellation requires
\begin{equation}
\label{W_{Z}}
[W_{\mathbf{Z}}] 
  =   c_{2}(\mathbf{TZ}) 
    - c_{2} \! \left( V^{ \textrm{\tiny{(1)}} }_{\mathbf{Z}} \right) 
    - c_{2} \! \left( V^{ \textrm{\tiny{(2)}} }_{\mathbf{Z}} \right).
\end{equation} 
$ V^{ \textrm{\tiny{(2)}} }_{\mathbf{Z}} $ is taken to be a trivial bundle
so that $ E_{8} $ remains unbroken in the hidden sector, 
$ c_{2} \! \left( V^{ \textrm{\tiny{(2)}} }_{\mathbf{Z}} \right) $
vanishes, and (\ref{W_{Z}}) simplifies accordingly.  Condition
(\ref{W_{Z}}) can then be pulled back onto $ \mathbf{X} $ to give
\begin{equation}
\label{W_{X}}
[W_{\mathbf{X}}] 
  =   c_{2}(\mathbf{TX}) 
    - c_{2} \! \left( V^{ \textrm{\tiny{(1)}} }_{\mathbf{X}} \right).
\end{equation}
The Chern classes appearing in (\ref{W_{X}}) have been evaluated to
be \cite{DonOvrPanWal:Sta}
\begin{equation}
\label{c_2TX}
c_{2}(\mathbf{TX}) = 12 \sigma_{*} c_{1} + \left( c_{2} + 11c^{2}_{1}
\right)(F-N)
            + \left( c_{2} - c^{2}_{1} \right) N
\end{equation}  
\begin{equation}
\label{c_2V_X}
c_{2}(V_{\mathbf{X}}) = \sigma_{*} \eta 
       - \left( f(n) - k^{2} \right) (F - N)
       - \left( f(n) - k^{2} + \textstyle{\sum_{i}} \kappa_{i} \right) N
\end{equation}  
where 
\begin{align}
c_{i} & \equiv c_{i}(\mathbf{B}) 
\\
k^{2} & \equiv \textstyle{\sum_{i}} \kappa^{2}_{i} 
\\
f(n)  & \equiv \frac{1}{24} \left( n^{3} - n \right) c^{2}_{1}
            - \frac{1}{2} \left( \lambda^{2} - \frac{1}{4} \right)
              n \eta \cdot ( \eta - n c_{1} ).
\end{align}
Using these expressions for $ c_{2}(\mathbf{TX}) $ and
$ c_{2}(V_{\mathbf{X}}) $, (\ref{W_{X}}) becomes
\begin{equation}
\label{[W_{X}]}
[W_{\mathbf{X}}] = \sigma_{*} W_{\mathbf{B}} + c (F-N) + dN
\end{equation}
where
\begin{align}
\label{ceqn}
c &= c_{2} + f(n) + 11c^{2}_{1} - k^{2}     \\
\label{deqn}
d &= c_{2} + f(n) - c^{2}_{1} - k^{2} + \textstyle{\sum_{i}} \kappa_{i}
\end{align}
and
\begin{equation}
W_{\mathbf{B}} = 12c_{1}(\mathbf{B}) - \eta.
\end{equation}
The class  $ [W_{\mathbf{Z}}] $ must represent a physical holomorphic
curve in the Calabi-Yau 3-fold $ \mathbf{Z} $ since M5-branes are required
to wrap around it. Hence $ [W_{\mathbf{Z}}] $ must be an effective class,
and its pull-back $ [W_{\mathbf{X}}] $ is an effective class in the
covering 3-fold $ \mathbf{X} $.  This leads to the effectiveness
conditions
\begin{gather}
\label{EffW_{B}}
W_{\mathbf{B}} = 12 c_{1} - \eta  \quad 
\textrm{is effective in $ \mathbf{B} $}
\\  
\label{Effc,d}
c \geq 0, \quad  d \geq 0.
\end{gather}

\item {\bf $ \beta^{(0)}_{i} = 0 $ constraints:}  As explained at the end
of Section \ref{4DLow}, to obtain potentially viable matter Yukawa
couplings, require vanishing instanton charges, $ \beta^{(0)}_{i} $, on
the observable orbifold fixed plane.  $ \beta^{(0)}_{i} = 0 $ implies that
\begin{equation}
\Omega 
  \equiv  c_{2} \! 
          \left( V^{ \textrm{\tiny{(1)}} }_{\mathbf{X}} \right) 
        - \frac{1}{2} c_{2}(\mathbf{TX})
  = 0
\end{equation}  
and thus from (\ref{c_2TX}) and (\ref{c_2V_X})
\begin{equation}
\sigma_{*} (6c_{1} - \eta) + \tilde{c} (F-N) + \tilde{d}N = 0
\end{equation}
where
\begin{align}
\tilde{c} &= c -\frac{1}{2} c_{2} - \frac{11}{2} c^{2}_{1},
\label{ctilde}   \\
\tilde{d} &= d -\frac{1}{2} c_{2} + \frac{1}{2} c^{2}_{1}.
\label{dtilde}  
\end{align}
This leads to the $ \beta^{(0)}_{i} = 0 $ constraints
\begin{gather}
\label{eta6c_1}
\eta = 6c_{1}(\mathbf{B}) 
\\
\label{tc=td=0}
\quad \tilde{c} = 0 \quad \quad \tilde{d} = 0.
\end{gather}

\end{itemize}

Applying the above rules amounts to finding a class $ W_{\mathbf{B}} $ and
coefficients $ c $, $ d $ in the expression (\ref{[W_{X}]})
which satisfy all of the constraints.  Each such solution corresponds to a 
heterotic M-theory vacuum with grand unification group $ H $, 
$ N_{\textrm{gen}} $ chiral generations, and potentially viable matter
Yukawa couplings.  Since $ \pi_{1}(\mathbf{Z}) = \mathbb{Z}_{2} $, $ H $
can be broken with $ \mathbb{Z}_{2} $ Wilson lines.  
For example \cite{FarGarIsi:Non}, $ H = SO(10) $ can be
broken to $ SU(5) \times U(1) $ by a $ \mathbb{Z}_{2} $ Wilson line
with generator \cite{CamEllHagNanTic:Fli}
\begin{equation}
-i \mathcal{G} =
\left(
  \begin{array}{ccccc}
     0 & -1 &        &   &      \\
     1 &  0 &        &   &      \\
       &    & \ddots &   &      \\
       &    &        & 0 & -1   \\
       &    &        & 1 &  0
 \end{array}
\right).
\end{equation}

\chapter{\label{NoYukVac}Vacua with nonvanishing $ \beta^{(0)}_{i} $}

The rules discussed in Section \ref{Rules} allow the classification
of the torus fibered Calabi-Yau manifolds 
$ \mathbf{Z} = \mathbf{X} / \tau_{\mathbf{X}} $  according to
phenomenological criteria.  This section considers the cases
where the elliptically fibered Calabi-Yau manifold $ \mathbf{X} $ has a 
Hirzebruch $ (\mathbb{F}_{r}, r \geq 0) $ or a $ \mathbf{dP}_{3} $ del
Pezzo base surface. The $ \beta^{(0)}_{i} = 0 $ constraints
(\ref{eta6c_1}) and (\ref{tc=td=0}) which ensure
potentially viable matter Yukawa couplings \emph{will not} be imposed.
These constraints will be considered in Chapter \ref{YukVac}.

It will be demonstrated that while a restricted class of 
$ N_{\textrm{gen}} = 3 $, 
$ H = SO(10) $ vacua is admitted by the $ \mathbb{F}_{2} $ surface,
the $ \mathbf{dP}_{3} $ surface does not admit such a class.  This class
is restricted in the sense that sufficient (but not necessary) 
$ G_{\mathbb{C}} = SU(4)_{\mathbb{C}} $ bundle constraints are used
instead of the more general ones given by (\ref{BCsigma}),
(\ref{BCetaChern}) and (\ref{BCkappa}) with $ n = 4 $.  Specifically,
\emph{two} classes of sufficient 
$ G_{\mathbb{C}} = SU(4)_{\mathbb{C}} $ bundle constraints, denoted by
Class A and Class B,
will be considered:
\begin{align}
\label{ClassA}
\textrm{Class A:}&  \quad \lambda \in \mathbb{Z}, \quad
\eta = c_{1}(\mathbf{B}) \ \textrm{mod} \ 2, \quad 
\kappa_{i} - \frac{1}{2} m \in \mathbb{Z} 
\\
\label{ClassB}
\textrm{Class B:}&  \quad \lambda - \frac{1}{2} \in \mathbb{Z}, \quad  
c_{1}(\mathbf{B}) \ \textrm{is an even class}, \quad
\kappa_{i} - \frac{1}{2} m \in \mathbb{Z}
\end{align}
where $ m \in \mathbb{Z} $. The analysis will show that when 
$ N_{\textrm{gen}} = 3 $, 
$ \mathbb{F}_{r} $ $ (r \geq 0) $ does not admit Class A, and 
$ \mathbf{dP}_{3} $ does not admit Class A or Class B.  $ \mathbb{F}_{0} $
and $ \mathbb{F}_{r} $ $ (r \ \textrm{odd} \geq 1) $
do not admit Class B (the former is excluded by the stability constraint
(\ref{stab})). $ \mathbb{F}_{2} $ admits Class B, and 
$ \mathbb{F}_{r} $ $ (r \ \textrm{even} \geq 4) $ admits Class B when the
constraints (\ref{preserve}), (\ref{freeact}), and (\ref{BItau}) on 
$ \tau_{\mathbb{F}_{r}} $ are satisfied.  These results have been
published by Faraggi, Garavuso and Isidro \cite{FarGarIsi:Non}.

\section{\label{Hirz}Hirzebruch surfaces}

A Hirzebruch surface $ \mathbb{F}_{r} $ $(r \geq 0)$, is a 2-dimensional
complex manifold constructed as a fibration with base $ \mathbb{CP}^{1} $ and
fiber $ \mathbb{CP}^{1} $.  We denote the class of the base and fiber of
$ \mathbb{F}_{r} $ by $ S $ and $ E $, respectively.  Their intersection
numbers are
\begin{equation}
\label{HirzInt}
S \cdot S = -r  \quad  S \cdot E = 1  \quad   E \cdot E = 0
\end{equation}
$ S $ and $ E $ form a basis of the homology class
$ H_{2}(\mathbb{F}_{r}, \mathbb{Z})$. This pair has the advantage that it
is also the set of generators for the Mori cone.  That is, the class
\begin{equation}
\eta = s S + e E
\end{equation}
is effective on $ \mathbb{F}_{r} $ for integers $ s $ and $ e $ if and
only if
\begin{equation}
s \geq 0,   \quad  e \geq 0.
\end{equation}
The Chern classes of $ \mathbb{F}_{r} $ are
\begin{align}
c_{1}(\mathbb{F}_{r}) &= 2 S + (r+2)E      \\
c_{2}(\mathbb{F}_{r}) &= 4.
\end{align}
We will need the result
\begin{equation}
\label{HirzFCS}
c^{2}_{1}(\mathbb{F}_{r}) = 8.
\end{equation}

\subsection{\label{NoYukHirz}$ H = SO(10) $ vacua with 
$ N_{\textrm{gen}} = 3 $ and
$ \mathbf{B} = \mathbb{F}_{r} $ $ (r \geq 0) $}

$ H = SO(10) $ vacua with $ N_{\textrm{gen}} = 3 $ and 
$ \mathbf{B} = \mathbb{F}_{r} $ $ (r \geq 0) $
will now be searched for.  It is demonstrated in \cite{DonOvrPanWal:Sta}
that there is an elliptically fibered Calabi-Yau 3-fold with 
$ \mathbb{F}_{2} $ base that admits a freely acting involution.  The
projection $ \tau_{\mathbb{F}_{2}} $ of this involution to the base 
$ \mathbb{F}_{2} $ has a finite fixed point set 
$ \mathcal{F}_{\tau_{\mathbb{F}_{2}}} $ and leaves an arbitrary
class $ \eta \in H_{2}(\mathbb{F}_{2},\mathbb{Z}) $ invariant.  It is not
possible to find an involution $ \tau_{\mathbb{F}_{r}} $ with finite fixed
point set $ \mathcal{F}_{\tau_{\mathbb{F}_{r}}} $ for all 
$ \mathbb{F}_{r} $ $ (r \geq 0) $.  The analysis which follows will work
with an arbitrary value of $ r $.  The results are valid when the
contraints (\ref{preserve}), (\ref{freeact}), and (\ref{BItau}) on 
$ \tau_{\mathbb{F}_{r}} $ are satisfied. For these cases, there exists an
elliptically fibered Calabi-Yau 3-fold $ \mathbf{X} $ which admits a
freely acting involution $ \tau_{\mathbf{X}} $.  The torus fibered
Calabi-Yau 3-fold $ \mathbf{Z} = \mathbf{X} / \tau_{\mathbf{X}} $ has 
$ \pi_{1}(\mathbf{Z}) = \mathbb{Z}_{2} $.

The results (\ref{HirzInt}) through (\ref{HirzFCS}) allow the remaining
rules of Section \ref{Rules} to be expressed in terms of $ r $, $ s $, 
$ e $, $ S $, and $ E $.  This will now be done, keeping the parameters 
$ n $  and $ N_{\textrm{gen}} $ for completeness.  These parameters will
later be set to $ n = 4 $ (corresponding to $ H = SO(10) $) and 
$ N_{\textrm{gen}} = 3 $.  
Analysis in this section is restricted to the Class A and Class B 
$ G_{\mathbb{C}} = SU(4)_{\mathbb{C}} $ bundle constraints (\ref{ClassA})
and (\ref{ClassB}), respectively, which can be written
\begin{align}
\textrm{Class A:}&  \quad \lambda \in \mathbb{Z}, \quad
s \ \mathrm{even}, \quad 
e - r \ \mathrm{even}, \quad
\kappa_{i} - \frac{1}{2} m \in \mathbb{Z}
\\
\textrm{Class B:}&  \quad \lambda - \frac{1}{2} \in \mathbb{Z}, \quad
r \ \mathrm{even}, \quad
\kappa_{i} - \frac{1}{2} m \in \mathbb{Z}
\end{align}
where $ m \in \mathbb{Z} $. The most general
$ G_{\mathbb{C}} = SU(n)_{\mathbb{C}} $ bundle constraints are
\begin{gather}
q \equiv n \left( \frac{1}{2} + \lambda \right) \in \mathbb{Z}
\\
\left(\frac{1}{2} - \lambda \right) \pi^{*}_{\mathbf{C}} (sS + eE)
   + \left( \frac{1}{2} + n \lambda \right) \pi^{*}_{\mathbf{C}} 
                                            [2S +(r+2)E]     
\quad \textrm{is an integer class}
\\
\kappa_{i} - \frac{1}{2} m \in \mathbb{Z}, \quad m \in \mathbb{Z}.
\end{gather}
The stability constraint (\ref{stab}) yields
\begin{equation}
\label{stab_nNgen}
2n \leq s, \quad n(r+2) \leq e
\end{equation}
and the second bundle involution condition (\ref{BIkappa}) becomes
\begin{equation}
\sum_{i} \kappa_{i} = (2 - r)s + 2e \quad 
\left( i = 1, \ldots, 4[(2 - r)s + 2e] \right).
\end{equation}
The $ N_{\textrm{gen}} $ condition (\ref{Ngen}) is
\begin{equation}
\label{Ngen=3}
N_{\textrm{gen}} = \frac{1}{2} \lambda 
    \left[ 2(s-n)e  - r s^{2} + n(r-2)s \right].
\end{equation}
The effectiveness conditions (\ref{EffW_{B}}) and (\ref{Effc,d}) imply
\begin{equation}
\label{eff_s_e_nNgen}
s \leq 24, \quad e \leq 12r + 24
\end{equation}
\begin{equation}
k^{2} \equiv \sum_{i} \kappa^{2}_{i} \leq \mathrm{min}
\left \{ 
 \begin{array}{c}
   92 + \frac{1}{3} (n^{3} - n) 
      + n \left( \frac{1}{4 \lambda} - \lambda \right) N_{\textrm{gen}}
\\
   -4 + \frac{1}{3} (n^{3} - n)
      + n \left( \frac{1}{4 \lambda} - \lambda \right) N_{\textrm{gen}}
      + \sum_{i} \kappa_{i},  
 \end{array}
\right. 
\end{equation}
respectively. Each M5-brane homology class
\begin{equation}
[W_{\mathbf{X}}] = \sigma_{*} W_{\mathbb{F}_{r}} + c (F-N) + dN
\end{equation}
with
\begin{align}
W_{\mathbb{F}_{r}} &= (24 - s) S + (12r + 24 - e) E
\\
c &= 92 + \frac{1}{3} (n^{3} - n) 
        + n \left( \frac{1}{4 \lambda} - \lambda \right) N_{\textrm{gen}}
        - k^{2} 
\\
d &= -4 + \frac{1}{3} (n^{3} - n) 
        + n \left( \frac{1}{4 \lambda} - \lambda \right) N_{\textrm{gen}}
        - k^{2}
        + \sum_{i} \kappa_{i}
\end{align}
satisfying  the above constraints corresponds to a heterotic 
M-theory vacuum.
 
$ H = SO(10) $ vacua (corresponding to $ n = 4 $) with 
$ N_{\textrm{gen}} = 3 $ net chiral generations in the observable sector
can be systematically searched for as follows. Rewrite the 
$ N_{\textrm{gen}} = 3 $ condition (\ref{Ngen=3}) as 
\begin{equation}
\label{Ngen=3,n=4}
6 / \lambda = 2(s-4)e  - r s^{2} + 4(r-2)s.
\end{equation}
The right side of this expression is always an even integer.
Thus, the left side, $ 6 / \lambda $, must also be an even integer. 
This in turn implies that the only allowed values of $ \lambda $ for Class
A and Class B are
\begin{align}
\mathrm{Class \ A:}& \quad \lambda = \pm 1, \pm 3
\\    
\mathrm{Class \ B:}& \quad \lambda = \pm 1/2, \pm 3/2.
\end{align}
Next, for each integer $ s $ allowed by (\ref{stab_nNgen}) and
(\ref{eff_s_e_nNgen}), solve (\ref{Ngen=3,n=4}) for $ e $:
\begin{equation}
e(r,\lambda) = \frac{1}{2s - 8} 
    \left( r s^{2} + (8 - 4r) s + \frac{6}{\lambda} \right) \quad
(8 \leq s \leq 24).
\end{equation}
If the resulting $ e(r,\lambda) $ is, for some allowed value of $ \lambda
$, an integer satisfying
\begin{equation}
\label{StabEff:n=4,Ngen=3}
4r + 8 \leq e \leq 12 r + 24
\end{equation}
as required by (\ref{stab_nNgen}) and (\ref{eff_s_e_nNgen}), then the
M5-brane homology class 
\begin{equation}
\label{M5Hom:n=4,Ngen=3}
[W_{\mathbf{X}}] = \sigma_{*} W_{\mathbb{F}_{r}} + c (F-N) + dN
\end{equation}
with
\begin{align}
W_{\mathbb{F}_{r}} &= (24 - s) S + (12r + 24 - e) E
\\
c &= 112 + \frac{3}{\lambda} - 12 \lambda - k^{2}  
\\
d &=  16 + \frac{3}{\lambda} - 12 \lambda - k^{2} 
         + \sum_{i} \kappa_{i}
\end{align}
and
\begin{gather}
\sum_{i} \kappa_{i} = (2 - r)s + 2e \quad
\left( i = 1, \ldots, 4[(2 - r)s + 2e] \right) 
\\
\label{effc,d:n=4,Ngen=3}
k^{2} \equiv \sum_{i} \kappa^{2}_{i} \leq \mathrm{min}
\left \{
 \begin{array}{c}
  112 + \frac{3}{\lambda} - 12 \lambda
\\
   16 + \frac{3}{\lambda} - 12 \lambda + \sum_{i} \kappa_{i}
 \end{array}
\right.
\end{gather}
corresponds to a SO(10) heterotic M-theory
vacuum with $ N_{\textrm{gen}} = 3 $ net chiral generations in the
observable
sector. 

For Class A, the $ e(r, \lambda) $ are 
\begin{equation*}
\begin{array}{ll}
s = 8 \! :  & e(r,\lambda) =  4r + 8 + \frac{3}{4 \lambda} 
\not\in \mathbb{Z}
\\
\\
s = 10 \! : & e(r,\lambda) =  5r + \frac{20}{3} + \frac{1}{2 \lambda}
\not\in \mathbb{Z}
\\
\\
s = 12 \! : & e(r,\lambda) =  6r + 6 + \frac{3}{8 \lambda}
\not\in \mathbb{Z}
\\
\\
s = 14 \! : & e(r,\lambda) =  7r + \frac{28}{5} + \frac{3}{10 \lambda}
\not\in \mathbb{Z}
\\
\\
s = 16 \! : & e(r,\lambda) =  8r + \frac{16}{3} + \frac{1}{4 \lambda}
\not\in \mathbb{Z}
\\
\\ 
s = 18 \! : & e(r,\lambda) =  9r + \frac{36}{7} + \frac{3}{14 \lambda}
\not\in \mathbb{Z}
\\
\\
s = 20 \! : & e(r,\lambda) = 10r + 5 + \frac{3}{16 \lambda}
\not\in \mathbb{Z}
\\
\\
s = 22 \! : & e(r,\lambda) = 11r + \frac{44}{9} + \frac{1}{6 \lambda}
\not\in \mathbb{Z}
\\
\\
s = 24 \! : & e(r,\lambda) = 12r + \frac{24}{5} + \frac{3}{20 \lambda}
\not\in \mathbb{Z}
\end{array}
\end{equation*}
There are no integer solutions for any allowed value of $ \lambda $ in
this this class.  Thus, $ \mathbb{F}_{r} $ $ (r \geq 0) $ does not admit
Class A vacua.

For Class B, the $ e(r,\lambda) $ are
\begin{equation*}
\begin{array}{ll}
s = 8 \! :  & e(r,\lambda) =  4r + 8 + \frac{3}{4 \lambda}
\not\in \mathbb{Z}
\\
\\
s = 9 \! :  & e(r,\lambda = -\frac{1}{2}) = \frac{9}{2} r  + 6
\\
\\
s = 10 \! : & e(r,\lambda = \frac{3}{2}) =  5r + 7
\\
\\
s = 11 \! : & e(r,\lambda = -\frac{3}{2}) = \frac{11}{2} r + 6
\\
\\
s = 12 \! : & e(r,\lambda) =  6r + 6 + \frac{3}{8 \lambda}
\not\in \mathbb{Z}
\\
\\
s = 13 \! : & e(r,\lambda = \frac{3}{2}) = \frac{13}{2} r + 6
\\
\\
s = 14 \! : & e(r,\lambda = - \frac{1}{2}) =  7r + 5
\\
\\
s = 15 \! : & e(r,\lambda = \frac{1}{2}) = \frac{15}{2} r + 6
\\
\\
s = 16 \! : & e(r,\lambda) =  8r + \frac{16}{3} + \frac{1}{4 \lambda}
\not\in \mathbb{Z}
\\
\\
s = 17 \! : & e(r,\lambda) = \frac{17}{2} r + \frac{68}{13} 
                                        + \frac{3}{13 \lambda}
\not\in \mathbb{Z}
\\
\\ 
s = 18 \! : & e(r,\lambda = - \frac{3}{2}) =  9r + 5
\\
\\
s = 19 \! : & e(r,\lambda) = \frac{19}{2} r + \frac{76}{15} 
                                       + \frac{1}{5 \lambda}
\not\in \mathbb{Z}
\\
\\
s = 20 \! : & e(r,\lambda) = 10r + 5 + \frac{3}{16 \lambda}
\not\in \mathbb{Z}
\\
\\
s = 21 \! : & e(r,\lambda) = \frac{21}{2} r + \frac{84}{17} 
                                        + \frac{3}{17 \lambda}
\not\in \mathbb{Z}
\\
\\
s = 22 \! : & e(r,\lambda = \frac{3}{2}) = 11r + 5
\\
\\
s = 23 \! : & e(r,\lambda) = \frac{23}{2} r + \frac{92}{19} 
                                        + \frac{3}{19 \lambda}
\not\in \mathbb{Z}
\\
\\
s = 24 \! : & e(r,\lambda) = 12r + \frac{24}{5} + \frac{3}{20 \lambda}
\not\in \mathbb{Z}.
\end{array}
\end{equation*}
For each $ e(r,\lambda) $ with an integer solution satisfying
(\ref{StabEff:n=4,Ngen=3}), the associated M5-brane homology class
obtained from (\ref{M5Hom:n=4,Ngen=3}) through
(\ref{effc,d:n=4,Ngen=3}) is given below.

\begin{itemize}

\item $ s = 9 $: $ \sum_{i} \kappa_{i} = 30 $, $ k^{2} \leq 46 $, 
$ r \ \mathrm{even} \geq 4 $
\\
\\
$ [W_{\mathbf{X}}] = \sigma_{*} 
            \left[ 15 S + \left( \frac{15}{2} r + 18 \right) E \right]
            + (112 - k^{2})(F - N) + (46 - k^{2}) N $
\\

\item $ s = 10 $: $ \sum_{i} \kappa_{i} = 34 $, $ k^{2} \leq 34 $, 
$ r \ \mathrm{even} \geq 2 $
\\
\\
$ [W_{\mathbf{X}}] = \sigma_{*} 
            \left[ 14 S + \left( 7r + 17 \right) E \right]
            + (96 - k^{2})(F - N) + (34 - k^{2}) N $
\\

\item $ s = 11 $: $ \sum_{i} \kappa_{i} = 34 $, $ k^{2} \leq 66 $,
$ r \ \mathrm{even} \geq 2 $
\\
\\
$ [W_{\mathbf{X}}] = \sigma_{*}
            \left[ 13 S + \left( \frac{13}{2} r + 18 \right) E \right]
            + (128 - k^{2})(F - N) + (66 - k^{2}) N $
\\

\item $ s = 13 $: $ \sum_{i} \kappa_{i} = 38 $, $ k^{2} \leq 38 $,
$ r \ \mathrm{even} \geq 2 $
\\
\\
$ [W_{\mathbf{X}}] = \sigma_{*}
            \left[ 11 S + \left( \frac{11}{2}r + 18 \right) E \right]
            + (96 - k^{2})(F - N) + (38 - k^{2}) N $
\\

\item $ s = 14 $: $ \sum_{i} \kappa_{i} = 38 $, $ k^{2} \leq 54 $,
$ r \ \mathrm{even} \geq 2 $
\\
\\
$ [W_{\mathbf{X}}] = \sigma_{*}
            \left[ 10 S + \left( 5r + 19 \right) E \right]
            + (112 - k^{2})(F - N) + (54 - k^{2}) N $
\\

\item $ s = 15 $: $ \sum_{i} \kappa_{i} = 42 $, $ k^{2} \leq 58 $,
$ r \ \mathrm{even} \geq 2 $
\\
\\
$ [W_{\mathbf{X}}] = \sigma_{*}
            \left[ 9 S + \left( \frac{9}{2}r + 18 \right) E \right]
            + (112 - k^{2})(F - N) + (58 - k^{2}) N $
\\

\item $ s = 18 $: $ \sum_{i} \kappa_{i} = 46 $, $ k^{2} \leq 78 $,
$ r \ \mathrm{even} \geq 2 $
\\
\\
$ [W_{\mathbf{X}}] = \sigma_{*}
            \left[ 6 S + \left( 3r + 19 \right) E \right]
            + (128 - k^{2})(F - N) + (78 - k^{2}) N $
\\

\item $ s = 22 $: $ \sum_{i} \kappa_{i} = 54 $, $ k^{2} \leq 54 $,
$ r \ \mathrm{even} \geq 2 $
\\
\\
$ [W_{\mathbf{X}}] = \sigma_{*}
            \left[ 2 S + \left( r + 19 \right) E \right]
            + (96 - k^{2})(F - N) + (54 - k^{2}) N $

\end{itemize}
Each of the above homology classes corresponds to a SO(10) heterotic
M-theory vacuum with $ N_{\textrm{gen}} = 3 $ net chiral generations in
the observable sector.  These vacua demonstrate that $ \mathbb{F}_{2} $
admits Class B, and $ \mathbb{F}_{r} $ $ (r \ \mathrm{even} \geq 4) $
admits Class B when the constraints (\ref{preserve}), (\ref{freeact}),
and (\ref{BItau}) on $ \tau_{\mathbb{F}_{r}} $ are satisfied.

\section{\label{DPez}Del Pezzo surfaces}

A del Pezzo surface $ \mathbf{dP}_{r} $ $ (r = 0,1,\ldots,8) $, is a
2-dimensional complex manifold constructed from complex projective space
$ \mathbb{CP}^{2} $ by blowing up $ r $ points.  A basis of
$ H_{2}(\mathbf{dP}_{r}, \mathbb{Z}) $ composed of effective classes is
given by the hyperplane class $ l $ and $ r $ exceptional divisors 
$ E_{i} $, $ i = 1,\ldots,r $.  Their intersections are
\begin{equation}
l \cdot l = 1, \quad  E_{i} \cdot E_{j} = -\delta_{ij},
\quad E_{i} \cdot l = 0.
\end{equation}
The Chern classes are given by
\begin{align}
\label{c_1dP_r}
c_{1}(\mathbf{dP}_{r}) &= 3l - \sum_{i=1}^{r} E_{i} 
\\
\label{c_2dP_r}
c_{2}(\mathbf{dP}_{r}) &= 3 + r.
\end{align}
We will need the result
\begin{equation}
\label{c^{2}_{1}dP_r}
c^{2}_{1}(\mathbf{dP}_{r}) = 9 - r.
\end{equation}

\subsection{\label{NoYukDPez}$ H = SO(10) $ vacua with 
$ N_{\textrm{gen}} = 3 $  and $ \mathbf{B} = \mathbf{dP}_{3} $}

$ H = SO(10) $ vacua with $ N_{\textrm{gen}} = 3 $ and 
$ \mathbf{B} = \mathbf{dP}_{3} $ will now be searched for. It is
demonstrated in \cite{DonOvrPanWal:Sta} that there is an elliptically
fibered Calabi-Yau 3-fold $ \mathbf{X} $ with $ \mathbf{dP}_{3} $ base
that admits a freely acting involution $ \tau_{\mathbf{X}} $.  
The torus fibered Calabi-Yau 3-fold 
$ \mathbf{Z} = \mathbf{X}/ \tau_{\mathbf{X}} $ has 
$ \pi_{1}(\mathbf{Z}) = \mathbb{Z}_{2} $.

The projection $ \tau_{\mathbf{dP}_{3}} $ of $ \tau_{\mathbf{X}} $ to the
base $ \mathbf{dP}_{3} $ leaves the class
\begin{equation}
\label{etadP_3}
\eta = m_{1} M_{1} + m_{2} M_{2} + m_{3} M_{3} \quad
\in H_{2}(\mathbf{dP}_{3},\mathbb{Z})
\end{equation}
invariant. The $ m_{j} $ $ (j = 1,2,3) $ are integers and the effective
classes
\begin{align}
M_{1} &= l + E_{1} - E_{2} - E_{3}
\\
M_{2} &= l - E_{1} + E_{2} - E_{3}
\\
M_{3} &= l - E_{1} - E_{2} + E_{3}.
\end{align}
with intersection numbers
\begin{equation}
M_{j} \cdot M_{j} = -2, \quad
M_{j} \cdot M_{k} = 2  \ (j \neq k)
\end{equation}
satisfy $ \tau_{\mathbf{dP}_{3}}(M_{j}) = M_{j} $.  In terms of the 
$ M_{j} $, 
(\ref{c_1dP_r}) with $ r = 3 $ is
\begin{equation}
\label{c_1dP_3M_j}
c_{1} (\mathbf{dP}_{3}) = M_{1} + M_{2} + M_{3}.
\end{equation}
Setting $ r = 3 $ in (\ref{c_2dP_r}) and (\ref{c^{2}_{1}dP_r}) yields 
\begin{align}
c_{2}(\mathbf{dP}_{3}) &= 6
\\
\label{c^{2}_{1}dP_3} 
c^{2}_{1}(\mathbf{dP}_{3})  &= 6.
\end{align}

The results (\ref{etadP_3}) through (\ref{c^{2}_{1}dP_3}) allow the
remaining rules of Section \ref{Rules} to be expressed in terms of the 
$ m_{j} $ and $ M_{j} $ $ (j = 1,2,3) $.  This will now be done, setting 
$ n = 4 $ (corresponding to $ H = SO(10) $) and $ N_{\textrm{gen}} = 3 $.
Analysis in this section is restricted to the Class A and Class B 
$ G_{\mathbb{C}} = SU(4)_{\mathbb{C}} $ bundle constraints (\ref{ClassA})
and (\ref{ClassB}), respectively.  Since Class B requires  
$ c_{1}(\mathbf{dP}_{3}) $ to be an even integer class, (\ref{c_1dP_3M_j})
implies that $ \mathbf{dP}_{3} $ does not admit Class B vacua.  The Class
A constraints can be written
\begin{equation}
\textrm{Class A:} \quad \lambda \in \mathbb{Z}, \quad
m_{j} \ \textrm{odd} \ (j = 1,2,3), \quad
\kappa_{i} - \frac{1}{2} m \in \mathbb{Z}
\end{equation}
where $ m \in \mathbb{Z} $.  The stability constraint (\ref{stab}) yields
\begin{equation}
4 \leq m_{j}  \quad (j=1,2,3)
\end{equation}   
and the second bundle involution condition (\ref{BIkappa}) becomes
\begin{equation}
\sum_{i} \kappa_{i} = 2 ( m_{1} + m_{2} + m_{3} ) \quad
(i = 1,\ldots, 4[ 2 ( m_{1} + m_{2} + m_{3} ) ] ).
\end{equation} 
The $ N_{\textrm{gen}} $ condition (\ref{Ngen}) with 
$ N_{\textrm{gen}} = 3 $ is
\begin{equation}
\label{NgendP3}
\frac{3} {\lambda} = - m^{2}_{1} - m^{2}_{2} - m^{2}_{3} 
                     + 2 ( m_{1} m_{2} + m_{1} m_{3} + m_{2} m_{3} )
                     - 4 ( m_{1} + m_{2} + m_{3} ).
\end{equation}
The effectiveness conditions (\ref{EffW_{B}}) and (\ref{Effc,d}) imply
\begin{gather}
m_{j} \leq 12 \quad (j=1,2,3) \\
k^{2} \equiv \sum_{i} \kappa^{2}_{i} \leq \mathrm{min}
\left \{
 \begin{array}{c}
   87 + \frac{3}{\lambda} - 12 \lambda
\\ 
   15 + \frac{3}{\lambda} - 12 \lambda + \sum_{i} \kappa_{i}
 \end{array}
\right.
\end{gather}
respectively. Each M5-brane homology class
\begin{equation}
[W_{\mathbf{X}}] = \sigma_{*} W_{\mathbf{dP}_{3}} + c (F - N) + d N
\end{equation}
with
\begin{align}
W_{\mathbf{dP}_{3}} &= (12 - m_{1}) M_{1} + (12 - m_{2}) M_{2} + (12 -
m_{3}) M_{3}
\\
c &= c_{2}(\mathbf{dP}_{3}) + \frac{27}{2} c^{2}_{1}(\mathbf{dP}_{3}) 
     + \frac{3}{\lambda} - 12 \lambda - k^{2}
\\
d &= c_{2}(\mathbf{dP}_{3}) + \frac{3}{2} c^{2}_{1}(\mathbf{dP}_{3}) 
     + \frac{3}{\lambda} - 12 \lambda - k^{2} + \sum_{i} \kappa_{i}
\end{align}
satisfying the above constraints corresponds to a $ SO(10) $ heterotic
M-theory vacuum with $ N_{\textrm{gen}} = 3 $ net chiral generations in
the observable sector.

It turns out that $ \mathbf{dP}_{3} $ does not admit Class A vacua with 
$ N_{\textrm{gen}} = 3 $.  This can be seen as follows.  Since Class A
requires the $ m_{j} $ to be odd, the right side of (\ref{NgendP3}) is
always an odd integer.  Thus, the left side, $ 3/ \lambda $ must also be
an odd integer. This, in turn, implies that the only allowed values of 
$ \lambda $ for Class A are $ \lambda = \pm 1, \pm 3 $.  For these values
of $ \lambda $, (\ref{NgendP3}) has no odd integer solutions in the range
$ 4 \leq m_{j} \leq 12 $.  To verify this, first assume that 
$ m_{1} = m_{2} = m_{3} = \mathsf{m} $.  This gives 
$ \mathsf{m} = 2 \pm \sqrt{4 + (1/ \lambda)} $, which for 
$ \lambda = \pm 1, \pm 3 $ is not an integer.  Next, assume that only two
of the $ m_{j} $ are equal, say $ m_{1} = m_{2} = \mathsf{m} $ and 
$ m_{3} \neq \mathsf{m} $.  This gives 
$ \mathsf{m} = \left[ m_{3} ( m_{3} + 4 ) + \frac{3}{\lambda} \right] /  
                     (4m_{3} - 8 ) $, 
which is not an odd integer for $ \lambda = \pm 1, \pm 3 $ when
$ m_{3} = 5,7,9,11 $.  Finally, when no two $ m_{j} \in \{ 5,7,9,11 \} $
are the same, one can check that (\ref{NgendP3}) has no solution for 
$ \lambda =  \pm 1, \pm 3 $.

\chapter{\label{YukVac}Vacua with $ \beta^{(0)}_{i} = 0 $}

The vacua in Chapter \ref{NoYukVac} are not restricted to have vanishing
instanton charges $ \beta^{(0)}_{i} $ $ (i = 1,\ldots,h^{(1,1)}) $ on the
observable fixed plane. The $ \beta^{(0)}_{i} = 0 $ constraints
(\ref{eta6c_1}) and (\ref{tc=td=0}) will now be imposed.  As explained at
the end of Section \ref{4DLow}, the $ \beta^{(0)}_{i} = 0 $
constraints yield potentially viable matter Yukawa couplings.

It turns out that requiring vanishing instanton charges on the observable
fixed plane rules out the vacua found in Chapter \ref{NoYukVac}.  However,
by replacing the sufficient (but not necessary) Class A and Class B
bundle constraints with more general constraints given by 
(\ref{BCsigma}), (\ref{BCetaChern}), and (\ref{BCkappa}) with $ n = 4 $, 
it is shown that a class of 
$ N_{\textrm{gen}} = 3 $, $ H = SO(10) $ vacua with potentially viable
matter Yukawa couplings is admitted by 
$ \mathbf{B} = \mathbf{dP}_{7} $ when the constraints 
(\ref{preserve}), (\ref{freeact}), and (\ref{BItau}) on
$ \tau_{\mathbf{dP}_{7}} $ are satisfied.
The analysis will show that $ \mathbb{F}_{r} $ $ (r \geq 0) $ and 
$ \mathbf{dP}_{r} $ $ (r = 0,\ldots,6,8) $ do not admit such a class.   
These results have been published by 
Faraggi and Garavuso \cite{FarGar:Yuk}.

\section{$ H = SO(10) $, $ N_{\textrm{gen}} = 3 $, 
$ \mathbf{B} = \mathbb{F}_{r} $ $ (r\geq 0) $}

$ H = SO(10) $ vacua with $ N_{\textrm{gen}} = 3 $, 
$ \mathbf{B} = \mathbb{F}_{r} $ $ (r\geq 0) $, 
and potentially viable matter Yukawa couplings will now be searched for.
Upon imposing the $ \beta^{(0)}_{i} = 0 $ constraint (\ref{eta6c_1})
\begin{equation}
\eta = 6c_{1}(\mathbb{F}_{r}),
\end{equation}
the $ N_{\textrm{gen}} $ condition (\ref{Ngen}), with 
$ N_{\textrm{gen}} = 3 $, becomes
\begin{equation}
3 = N_{\textrm{gen}} = \frac{1}{2} \lambda 6 (6 - n)
c^{2}_{1}(\mathbb{F}_{r}).
\end{equation}
For $ n=4 $ (corresponding to $ H = SO(10) $), we obtain
\begin{equation}
\lambda = \frac{1}{16}.
\end{equation}
Plugging this value for $ \lambda $ along with $ n = 4 $
into (\ref{BCsigma}), one sees that the 
$ G_{\mathbb{C}} = SU(4)_{\mathbb{C}} $ bundle constraints cannot be
satisfied. Thus, when $ N_{\textrm{gen}} = 3 $, 
$ \mathbb{F}_{r} $ $ (r \geq 0) $
does not admit the class of $ H = SO(10) $ vacua with potentially viable
matter Yukawa couplings being searched for.

\section{$ H = SO(10) $, $ N_{\textrm{gen}} = 3 $, $ \mathbf{B} =
\mathbf{dP}_{r} $ $ (r = 0,\ldots,8) $}

$ H = SO(10) $ vacua with $ N_{\textrm{gen}} = 3 $, 
$ \mathbf{B} = \mathbf{dP}_{r} $ $ (r = 0,\ldots,8) $, 
and potentially viable matter Yukawa couplings will now be searched for.
Assume that there exists an elliptically fibered Calabi-Yau
3-fold $ \mathbf{X} $ with 
$ \mathbf{B} = \mathbf{dP}_{r} $ $ (r = 0,\ldots,8) $ that admits a freely
acting involution $ \tau_{\mathbf{X}} $.  That is, assume there exists an
involution $ \tau_{\mathbf{dP}_{r}} $ on $ \mathbf{dP}_{r} $ satisfying
(\ref{preserve}) and (\ref{freeact}). The torus fibered Calabi-Yau 3-fold 
$ \mathbf{Z} = \mathbf{X} / \tau_{\mathbf{X}} $ has 
$ \pi_{1}(\mathbf{Z}) = \mathbb{Z}_{2} $. Further assume that 
$ \tau_{\mathbf{dP}_{r}} $ leaves some class 
$ \eta \in H_{2}(\mathbf{dP}_{r},\mathbb{Z}) $
invariant so that (\ref{BItau}) is satisfied.

Now, express the remaining rules of Section \ref{Rules} in terms of 
$ \eta $ and $ \mathbf{dP}_{r} $, setting $ n = 4 $ 
(corresponding to $ H = SO(10) $) and $ N_{\textrm{gen}} = 3 $.  The most
general
$ G_{\mathbb{C}} = SU(4)_{\mathbb{C}} $ bundle constraints are
\begin{gather}
\label{BCq}
q = 4 \left( \frac{1}{2} + \lambda \right) \in \mathbb{Z} 
\\ 
\label{BCetaCherndP_r}
\left(\frac{1}{2} - \lambda \right) \pi^{*}_{C} \eta
   + \left( \frac{1}{2} 
   + 4 \lambda \right) \pi^{*}_{C} c_{1}(\mathbf{dP}_{r})
\quad \textrm{is an integer class}
\\
\kappa_{i} - \frac{1}{2} m \in \mathbb{Z}, \quad m \in \mathbb{Z}
\end{gather} 
The stability constraint is
\begin{equation}
\label{stabdP_r}
\eta \geq 4 c_{1}(\mathbf{dP}_{r})
\end{equation}
and the second bundle involution condition is
\begin{equation}
\label{BIkappadP_r}
\sum_{i} \kappa_{i} = \eta \cdot c_{1}(\mathbf{dP}_{r}) \quad
( i = 1,\ldots, 4 \eta \cdot c_{1}(\mathbf{dP}_{r}) ).
\end{equation}
The $ N_{\textrm{gen}} $ condition is
\begin{equation}
\label{Ngen=3dP_r}
3 = N_{\textrm{gen}} = \frac{1}{2} \lambda \eta \cdot 
              \left( \eta - 4 c_{1}(\mathbf{dP}_{r}) \right).
\end{equation}
The effectiveness conditions are
\begin{gather}
\label{effWdP_r} 
W_{\mathbf{dP}_{r}} = 12 c_{1}(\mathbf{dP}_{r}) - \eta \quad 
\textrm{is effective in $ \mathbf{dP}_{r} $}
\\
\label{effcdP_rn=4Ngen=3}
c = c_{2}(\mathbf{dP}_{r}) + \frac{27}{2} c^{2}_{1}(\mathbf{dP}_{r}) 
    + \frac{3}{\lambda} - 12 \lambda - k^{2} 
    \geq 0
\\
\label{effddP_rn=4Ngen=3}
d = c_{2}(\mathbf{dP}_{r}) + \frac{3}{2} c^{2}_{1}(\mathbf{dP}_{r})
    + \frac{3}{\lambda} - 12 \lambda - k^{2}  + \sum_{i} \kappa_{i}
   \geq 0.
\end{gather}
The $ \beta^{(0)}_{i} = 0 $ constraints are
\begin{gather}
\label{betazeroetadP_r}
\eta = 6 c_{1}(\mathbf{dP}_{r})
\\
\label{ctildedP_rn=4Ngen=3}
\tilde{c} = c - \frac{1}{2} c_{2}(\mathbf{dP}_{r}) 
              - \frac{11}{2} c^{2}_{1}(\mathbf{dP}_{r})
          = 0
\\
\label{dtildedP_rn=4Ngen=3}
\tilde{d} = d - \frac{1}{2} c_{2}(\mathbf{dP}_{r}) 
              + \frac{1}{2} c^{2}_{1}(\mathbf{dP}_{r})
          = 0.
\end{gather}
Each M5-brane homology class
\begin{equation}
[W_{\mathbf{X}}] = \sigma_{*} W_{\mathbf{dP}_{r}} + c (F - N) + d N
\end{equation}
satisfying all of the above constraints corresponds to a $ SO(10) $
heterotic M-theory vacuum with $ N_{\textrm{gen}} = 3 $ net chiral
generations and potentially viable matter Yukawa couplings in the
observable sector.

The $ \beta^{(0)}_{i} = 0 $ constraint (\ref{betazeroetadP_r}) allows 
$ \eta $ to be eliminated from the above expressions.  Upon setting 
$ \eta = 6 c_{1}(\mathbf{dP}_{r}) $, the second 
$ G_{\mathbb{C}} = SU(4)_{\mathbb{C}} $ bundle constraint
(\ref{BCetaCherndP_r}) yields
\begin{equation}
\label{BCp}
p \equiv \frac{7}{2} - 2 \lambda \in \mathbb{Z},
\end{equation}
the stability constraint (\ref{stabdP_r}) is satisfied,
the second bundle involution condition (\ref{BIkappadP_r}) becomes
(after using $ c^{2}_{1}(\mathbf{dP}_{r}) = 9 - r $)
\begin{equation}
\label{BIkappabetazero}
\sum_{i} \kappa_{i} = 6(9-r) \quad
(i = 1,\ldots,4[6(9-r)]),
\end{equation}
the $ N_{\textrm{gen}} $ condition (\ref{Ngen=3dP_r}) yields
\begin{equation}
\label{lambda(r)}
\lambda = \frac{1}{2(9-r)},
\end{equation}
and the first effectivess condition (\ref{effWdP_r}), which becomes
\begin{equation}
\label{effWdP_rbetazero}
W_{\mathbf{dP}_{r}} = 6 c_{1} (\mathbf{dP}_{r}) \quad
\textrm{is effective in $ \mathbf{dP}_{r} $}
\end{equation}
is satisfied.  

The values of $ \lambda $ given by (\ref{lambda(r)}) for 
$ r = 0,\ldots,8 $, and the corresponding values of $ q $ given by 
(\ref{BCq}) are shown in Table \ref{Tablerlambdaq}. 
\begin{table}   
\begin{center}
\begin{tabular}{|c|c|c|c|c|c|c|c|c|c|}
\hline
 $ r $        & 0               & 1               & 2
              & 3               & 4               & 5
              & 6               & 7               & 8             \\
\hline
 $ \lambda $  & 1/18            & 1/16            & 1/14   
              & 1/12            & 1/10            & 1/8
              & 1/6             & 1/4             & 1/2           \\
\hline
$ q $        & 20/9            & 9/4             & 16/7
              & 7/3             & 12/5            & 5/2
              & 8/3             & 3               & 4             \\
\hline
\end{tabular}
\end{center}

\caption{The values of $ \lambda = 1 / [2(9-r)] $ and
$ q = 4 \left( \frac{1}{2} + \lambda \right) $ are shown for 
$ r = 0,\ldots,8 $.}

\label{Tablerlambdaq}

\end{table}
If $ q $ is not an integer, then (\ref{BCq}) is not satisfied.  The table
indicates that $ q \in \mathbb{Z} $ only for $ r = 7 $ and $ r = 8 $.  Thus, 
$ \mathbf{dP}_{r} $ $ (r = 0,\ldots,6) $ does not admit the class of 
$ H = SO(10) $ vacua being searched for.  Requiring (\ref{BCp}) to hold shows 
that $ \mathbf{dP}_{8} $ also does not admit this class.  

Having eliminated $ \mathbf{dP}_{r} $ $ (r = 0,\ldots,6,8) $ from
consideration, the only remaining possibility is $ \mathbf{dP}_{7} $, for
which
\begin{equation}
\lambda = 1 / 4.
\end{equation}  
For this value of $ \lambda $, (\ref{BCp}) is satisfied.  Note that 
$ \lambda = 1 / 4 $ would not be permitted if the sufficient (but not
necessary) Class A and Class B 
$ G_{\mathbb{C}} = SU(4)_{\mathbb{C}} $ bundle constraints in Chapter
\ref{NoYukVac} had been imposed.  

Now, compute the  M5-brane homology class
\begin{equation}
[W_{\mathbf{X}}] = \sigma_{*} W_{\mathbf{dP}_{7}} + c ( F - N) + d N
\end{equation}
From (\ref{effWdP_rbetazero}) and (\ref{c_1dP_r}), one obtains for 
$ W_{\mathbf{dP}_{7}} $
\begin{equation}
W_{\mathbf{dP}_{7}} = 6 c_{1}(\mathbf{dP}_{7}) 
                    = 18 l - 6 \sum_{i}^{7} E_{i}.
\end{equation}
With $ r=7 $, (\ref{BIkappabetazero}) becomes
\begin{equation}
\label{sumkappa=12}
\sum_{i} \kappa_{i} = 12 \quad 
( i = 1,\ldots,48 ).
\end{equation} 
Using this result, along with $ \lambda = 1 / 4 $, 
$ c^{2}_{1}(\mathbf{dP}_{7}) = 2 $, and $ c_{2}(\mathbf{dP}_{7}) = 10 $,
one obtains
from (\ref{effcdP_rn=4Ngen=3}) and (\ref{effddP_rn=4Ngen=3}) 
\begin{gather}
c = 46 - k^{2}
\\
d = 34 - k^{2}.
\end{gather}
Imposing the effectiveness conditions $ c \geq 0 $  and $ d \geq 0 $
gives
\begin{equation}
\label{dP_7ksquared}
k^{2} \equiv \sum_{i} \kappa^{2}_{i} \leq 34.
\end{equation}
The value of $ k^{2} $ is further restricted by the 
$ \beta^{(0)}_{i} = 0 $ constraints (\ref{ctildedP_rn=4Ngen=3}) and
(\ref{dtildedP_rn=4Ngen=3}). With $ c = 46 - k^{2} $ and 
$ d = 34 - k^{2} $, one obtains
\begin{equation}
0 = \tilde{c} = \tilde{d} = 30 - k^{2}
\end{equation}
and hence
\begin{equation}
\label{ksquared=30}
k^{2} \equiv \sum_{i} \kappa^{2}_{i} = 30
\end{equation}
which is consistent with (\ref{dP_7ksquared}).
Thus, for each set of $ \kappa_{i} $ satisfying (\ref{sumkappa=12})
and (\ref{ksquared=30}),
\begin{equation}
[W_{\mathbf{X}}] = \sigma_{*} \left( 18 l - 6 \sum_{i=1}^{7} E_{i} \right)
                   + 16 ( F - N ) + 4 N
\end{equation}
corresponds to a $ SO(10) $ heterotic M-theory vacuum with 
$ N_{\textrm{gen}} = 3 $ net chiral generations and potentially viable
matter Yukawa couplings in the observable sector when the constraints
(\ref{preserve}), (\ref{freeact}) and (\ref{BItau}) on the involution 
$ \tau_{\mathbf{dP}_{7}} $ are satisfied.

\chapter{\label{Free}Realistic free-fermionic models}

Heterotic string theories in four dimensions can be constructed by
compactifying six dimensions of the 10-dimensional heterotic string on an
orbifold \cite{DixHarVafWit:Str,DixHarVafWit:StrII} or a Calabi-Yau
manifold \cite{CanHorStrWit:Vac}.  It is also possible to construct
4-dimensional heterotic string theories without the intermediate state of
a 10-dimensional theory.  To do this, start from the heterotic string
with 10 and 26 dimensions for the superstring left-movers and bosonic
string right-movers, respectively. Next, toroidally compactify on the
left \emph{and} right to four dimensions and either bosonize
\cite{LerLusSch:Chi} each internal fermion or fermionize \cite{free} each
internal boson.  This chapter reviews realistic free-fermionic models and
their
$ \mathbb{Z}_{2} \times \mathbb{Z}_{2} $ correspondence,
explains how this correspondence identifies associated
Calabi-Yau 3-folds which possess the structure of $ \mathbf{Z} $ and
$ \mathbf{X} $ of the previous chapters, and argues how the top quark
Yukawa coupling in these models can be reproduced in the heterotic
M-theory limit.

\section{\label{GenStruc}General structure}
In the free-fermionic formulation \cite{free} of the heterotic string in
four dimensions, the internal degrees of freedom required to cancel the
conformal anom-\\aly are represented as free fermions propagating on
the string worldsheet.  Consider starting from a heterotic string with 10
dimensions for the superstring left-movers and 26 dimensions for the
bosonic string right-movers. Now, toroidally compactify on the left
\emph{and} right to four dimensions and fermionize each internal boson.
In the lightcone gauge, the worldsheet field content is as follows:
Associated with the 4-dimensional spacetime are two transverse left-moving
bosons $ X^{i} $, their Majorana-Weyl fermionic 
superpartners $ \psi^{i} $, 
and two transverse right-moving bosons 
$ \overline{X}^{i} $, $ i = 1,2 $.  
There are 18 left-moving 
$ \chi^{I}, y^{I}, \omega^{I} $ $ (I = 1,\dots,6) $
and 44 right-moving $ \overline{\varphi}^{a} $ $ (a = 1,\ldots,44) $ 
internal Majorana-Weyl fermions. Alternatively, some real fermions may be
paired to form complex fermions.  

The above 64 fermionic fields pick up a phase upon parallel transport
around a noncontractible loop on the worldsheet.  Specification of the
phases for all worldsheet fermions around all noncontractible loops
defines the spin structure of the model. The spin structure can be
encoded in 64-dimensional boundary condition basis vectors.  A model is
defined by a set of such basis vectors together with generalized GSO
projection coefficients, which satisfy \emph{string consistency
constraints} \cite{free} imposed by modular invariance and supersymmetry.

The boundary condition basis vectors associated with the realistic
free-fermionic models are constructed in two stages.  The first stage
constructs the NAHE set \cite{FarNan:Nat} of five basis vectors denoted by 
$ \{ \mathbf{1}, \mathbf{S}, \mathbf{b}_{1}, \mathbf{b}_{2}, 
\mathbf{b}_{2} \} $ and given in Table \ref{NAHE}.
\begin{table}
\label{NAHE}
$ \begin{array}{c|c|ccc|c|ccc|c}
  & \psi^{i} & \chi^{12} & \chi^{34} & \chi^{56} &
    \overline{\psi}^{1,\ldots,5} &
    \overline{\eta}^{1} & \overline{\eta}^{2} & \overline{\eta}^{3} &
    \overline{\phi}^{1,\ldots,8} \\
\hline \hline
\mathbf{1} & 1 & 1 & 1 & 1 & 1,\ldots,1 & 1 & 1 & 1 & 1,\ldots,1 \\
\mathbf{S} & 1 & 1 & 1 & 1 & 0,\ldots,0 & 1 & 1 & 1 & 1,\ldots,1 \\
\hline
\mathbf{b}_{1} & 1 & 1 & 0 & 0 & 1,\ldots,1 & 1 & 0 & 0 & 0,\ldots,0 \\
\mathbf{b}_{2} & 1 & 0 & 1 & 0 & 1,\ldots,1 & 0 & 1 & 0 & 0,\ldots,0 \\
\mathbf{b}_{3} & 1 & 0 & 0 & 1 & 1,\ldots,1 & 0 & 0 & 1 & 0,\ldots,0
\end{array} $
\vskip 10pt
$ \begin{array}{c|cc|cc|cc}
  & y^{3,\ldots,6} & \overline{y}^{3,\ldots,6}
  & y^{1,2}, \omega^{5,6} & \overline{y}^{1,2}, \overline{\omega}^{5,6}
  & \omega^{1,\ldots,4} & \overline{\omega}^{1,\ldots,4}    \\
\hline \hline
\mathbf{1} & 1,\ldots,1 & 1,\ldots,1 & 1,\ldots,1 & 1,\ldots,1
           & 1,\ldots,1 & 1,\ldots,1 \\
\mathbf{S} & 0,\ldots,0 & 0,\ldots,0 & 0,\ldots,0 & 0,\ldots,0
           & 0,\ldots,0 & 0,\ldots,0\\
\hline
\mathbf{b}_{1} & 1,\ldots,1 & 1,\ldots,1 & 0,\ldots,0 & 0,\ldots,0
           & 0,\ldots,0 & 0,\ldots,0 \\
\mathbf{b}_{2} & 0,\ldots,0 & 0,\ldots,0 & 1,\ldots,1 & 1,\ldots,1
           & 0,\ldots,0 & 0,\ldots,0 \\
\mathbf{b}_{3} & 0,\ldots,0 & 0,\ldots,0 & 0,\ldots,0 & 0,\ldots,0
           & 1,\ldots,1 & 1,\ldots,1
\end{array} $

\caption{The NAHE set. Ramond and Neveu-Schwarz
boundary conditions are indicated by `1' and `0', respectively.}

\end{table}
After generalized GSO projections over the NAHE set, the residual gauge
group is 
\begin{equation}
SO(10) \times SO(6)^{3} \times E_{8}. 
\end{equation}
The vector bosons
which generate the $ SO(10) \times SO(6)^{3}
\times E_{8} $ gauge group arise from the Neveu-Schwarz sector and from
the sector 
\begin{equation}
\boldsymbol{\xi}_{2} \equiv \mathbf{1} + \mathbf{b}_{1} +
\mathbf{b}_{2} + \mathbf{b}_{3}.
\end{equation}
The Neveu-Schwarz sector produces the generators of $ SO(10) \times
SO(6)^{3} \times SO(16) $. The $ \boldsymbol{\xi}_{2} $ sector
produces the spinorial $ \mathbf{128} $ of $ SO(16) $ and completes the
hidden gauge group to $ E_{8} $.  NAHE set models have 
$ \mathcal{N} = 1 $ spacetime supersymmetry
and $ 48 $ chiral generations in the $ \mathbf{16} $
representation of $ SO(10) $. The vector $ \mathbf{1} $
has only periodic boundary conditions.  $ \mathbf{S} $ generates 
$ \mathcal{N} = 4 $ spacetime supersymmetry.  The set $ \{\mathbf{1},
\mathbf{S} \} $ yields an $ \mathcal{N} = 4 $ model with $ SO(44) $ gauge
group in the right-moving sector.  The vectors $ \mathbf{b}_{1} $ and  
$ \mathbf{b}_{2} $ reduce the number of supersymmetries in turn to 
$ \mathcal{N} = 2 $ and $ \mathcal{N} = 1 $, respectively.  The choice of
generalized GSO projection coefficients
\begin{equation}
C \left( \begin{array}{c} \mathbf{b}_{i} \\ \mathbf{b}_{j} \end{array}
  \right) 
= C \left( \begin{array}{c} \mathbf{b}_{i} \\ \mathbf{S} \end{array} 
    \right) 
= C \left( \begin{array}{c} \mathbf{1} \\ \mathbf{1} \end{array} 
    \right) 
= -1.
\end{equation}
ensures $ \mathcal{N} = 1 $ spacetime supersymmetry. The remaining
projection coefficients can be determined from those above through the
string consistency constraints. The 
$ \mathbf{b}_{1} $  and $ \mathbf{b}_{2} $ projections break
\begin{equation}
SO(44) \rightarrow SO(10) \times SO(6)^{2} \times SO(22). 
\end{equation}
Including the 
$ \mathbf{b}_{3} $
projections breaks
\begin{equation}
SO(10) \times SO(6)^{2} \times SO(22) \rightarrow SO(10) \times SO(6)^{3}
\times E_{8}.
\end{equation}
Each sector 
$ \mathbf{b}_{1} $, $ \mathbf{b}_{2} $ and $ \mathbf{b}_{3} $
produces 16 multiplets in the spinorial $ \mathbf{16} $ representation of
$ SO(10) $. The states from the sectors $ \mathbf{b}_{j} $ are singlets of
the hidden $ E_{8} $ gauge group and transform under the horizontal 
$ SO(6)_{j} $ $ (j=1,\ldots,3) $ symmetries.
The NAHE set groups the 44 right- and 18 left-moving internal
fermions in the following way: Five complex right-movers 
$ \overline{\psi}^{1,\ldots,5} $ produce the observable 
$ SO(10) $ gauge group.  Eight complex right-movers
$ \overline{\phi}^{1,\ldots,8} $ produce the hidden $ E_{8} $ gauge group.   
Three complex right-movers 
$ \overline{\eta}^{1} $, $ \overline{\eta}^{2} $, $ \overline{\eta}^{3} $
and 12 real right-movers 
$  \overline{y}^{1,\ldots,6} $, $ \overline{\omega}^{1,\ldots,6} $  
form the states
\begin{equation*}
\{ \overline{\eta}^{1}, \overline{y}^{3,\ldots,6} \}, \quad
\{ \overline{\eta}^{2}, \overline{y}^{1,2}, \overline{\omega}^{5,6} \},
\quad
\{ \overline{\eta}^{3}, \overline{\omega}^{1,\ldots,4} \} 
\end{equation*}
which produce the three horizontal $ SO(6) $ symmetries.
The six left-movers $ \chi^{I} $ $ (I = 1,\ldots,6) $ 
are combined into three complex fermions 
$ \chi^{12}, \chi^{34}, \chi^{56} $. 
Twelve real left-movers $ y^{1,\ldots,6} $, $ \omega^{1,\ldots,6} $ 
form the states 
\begin{equation*}
\{ y^{3,\ldots,6} \}, \quad 
\{ y^{1}, y^{2}, \omega^{5}, \omega^{6} \}, \quad
\{ \omega^{1,\ldots,4} \}. 
\end{equation*}

The second stage of the construction adds three (or four) basis vectors,
typically denoted by 
$\{\boldsymbol{\alpha},\boldsymbol{\beta},\boldsymbol{\gamma},\ldots\}$.
These basis vectors break the
$ SO(10) \times SO(6)^{3} \times E_{8} $ gauge group
and reduce the net number of chiral generations from 48 to 3 
(one from each of the sectors 
$ \mathbf{b}_{1} $, $ \mathbf{b}_{2} $, and $ \mathbf{b}_{3} $).
The models which can be constructed are distinguished by the choice of 
$\{\boldsymbol{\alpha},\boldsymbol{\beta},\boldsymbol{\gamma},\ldots\}$. 
The assignment of boundary conditions to $ \overline{\psi}^{1,\ldots,5} $
breaks $ SO(10) $ to one of its subgroups. The 
flipped $ SU(5) $ \cite{(FSU5)},
Pati-Salam \cite{(PS)}, 
Standard-like \cite{(SLM)}, and 
left-right symmetric \cite{(LRS)} 
$ SO(10) $ breaking patterns are shown in Table \ref{SO(10)breaking}.
In the former two cases, an additional 
$ \mathbf{16} $ and $ \mathbf{\overline{16}} $ 
representation of $ SO(10) $ is obtained from the set 
$\{\boldsymbol{\alpha},\boldsymbol{\beta},\boldsymbol{\gamma},\ldots\}$.
Similarly, the assignment of boundary conditions to 
$ \overline{\phi}^{1,\ldots,8} $ breaks the hidden $ E_{8} $ to one of its 
subgroups.  The flavor $ SO(6) $ symmetries are broken to flavor $ U(1) $
symmetries. Three such symmetries $ U(1)_{j} $ arise from 
the worldsheet currents 
\begin{equation*}
\overline{\eta}^{j} \overline{\eta}^{j*} \quad (j=1,\dots,3).
\end{equation*}
Additional $ U(1) $ symmetries arise from the pairing of two real fermions
from the sets 
\begin{equation*}
\{ \overline{y}^{3,\ldots,6} \}, \quad
\{ \overline{y}^{1,2}, \overline{\omega}^{5,6} \}, \quad 
\{ \overline{\omega}^{1,\ldots,4} \}.
\end{equation*}
The final observable gauge group
depends on the number of such pairings.  For every right-moving $ U(1) $
symmetry, there is a corresponding left-moving global $ U(1) $ symmetry
that is obtained by pairing two of the left-moving real fermions
$ \{ y^{1,\ldots,6},\omega^{1,\ldots,6} \} $.  Each of the remaining
worldsheet left-moving real fermions is paired with a right-moving real
fermion from the set 
$ \{ \overline{y}^{1,\ldots,6},\overline{\omega}^{1,\ldots,6} \} $.

\section{\label{NonExt}$ \mathbb{Z}_{2} \times \mathbb{Z}_{2} $
correspondence}

The precise geometrical realization of the full $ N_{\textrm{gen}} = 3 $
realistic free-fermionic models is not yet known.  However, an extended NAHE 
set $ \{ \mathbf{1}, \mathbf{S}, \mathbf{b}_{1}, \mathbf{b}_{2},
     \mathbf{b}_{3}, \\ \boldsymbol{\xi}_{1} \} $, 
where
\begin{equation}
\boldsymbol{\xi}_{1} 
  = ( 0,\ldots,0 |
      \underbrace{1,\ldots,1}_{ \overline{\psi}^{1,\ldots,5}, 
                              \overline{\eta}^{1,2,3} },
      0,\ldots,0 ),
\end{equation}
or equivalently
$ \{ \mathbf{1}, \mathbf{S}, \boldsymbol{\xi}_{1}, \boldsymbol{\xi}_{2},
\mathbf{b}_{1}, \mathbf{b}_{2} \} $, has been shown to yield the same data
as the $ \mathbb{Z}_{2} \times \mathbb{Z}_{2} $ orbifold of a
toroidal Narain model \cite{Nar:New} with nontrivial background fields
\cite{NarSarWit:A_No}. The corresponding Narain model has 
$ \mathcal{N} = 4 $ spacetime supersymmetry and either 
\begin{equation}
SO(12) \times E_{8} \times E_{8} 
\end{equation}
or
\begin{equation}
SO(12) \times SO(16) \times SO(16) 
\end{equation}
gauge group, depending on the
choice of sign for the GSO projection coefficient 
$ C (^{ \boldsymbol{\xi_{1}} }_{ \boldsymbol{\xi_{2}} } ) $.
Let the former and latter Narain models be denoted by 
$ \mathbf{N}_{+} $ and $ \mathbf{N}_{-} $, respectively.
The corresponding $ \mathbb{Z}_{2} \times \mathbb{Z}_{2} $ orbifolds are
\begin{equation}
\mathbf{Z}_{+}
  \equiv \frac{ \mathbf{N}_{+} }
              { \mathbb{Z}_{2} \times \mathbb{Z}_{2} }
\end{equation}
and
\begin{equation}
\mathbf{Z}_{-}
  \equiv \frac{ \mathbf{N}_{-} }
              { \mathbb{Z}_{2} \times \mathbb{Z}_{2} }.
\end{equation}
As the NAHE set is common to all realistic free-fermionic
models, $ \mathbf{Z}_{+} $ and $ \mathbf{Z}_{-} $ are at their core.
Table \ref{NZ} lists the gauge group and spacetime
supersymmetry obtained from $ \mathbf{N}_{+} $, $ \mathbf{N}_{-} $,
$ \mathbf{Z}_{+} $, and $ \mathbf{Z}_{-} $.
\begin{table}
$$ \begin{array}{c|c|c}
               &                      & \textrm{spacetime}
\\
               & \textrm{gauge group} & \textrm{SUSY} 
\\
\hline
\mathbf{N}_{+} & SO(12) \times E_{8} \times E_{8} & \mathcal{N} = 4
\\
\mathbf{N}_{-} & SO(12) \times SO(16) \times SO(16) & \mathcal{N} = 4
\\
\mathbf{Z}_{+} & SO(4)^{3} \times E_{6} \times U(1)^{2} \times E_{8}
               & \mathcal{N} = 1
\\
\mathbf{Z}_{-} & SO(4)^{3} \times SO(10) \times U(1)^{3} \times SO(16)
               & \mathcal{N} = 1
\end{array} $$
\caption{The gauge group and spacetime supersymmetry obtained from the
Narain models $ \mathbf{N}_{+} $, $ \mathbf{N}_{-} $ and the
corresponding $ \mathbb{Z}_{2} \times \mathbb{Z}_{2} $ orbifolds 
$ \mathbf{Z}_{+} $, $ \mathbf{Z}_{-} $.} 
\label{NZ}
\end{table}
In the free-fermionic formulation, the 
$ \mathbf{N}_{+} $ and $ \mathbf{N}_{-} $
data is produced by the set 
$\{ \mathbf{1}, \mathbf{S}, \boldsymbol{\xi}_{1}, \boldsymbol{\xi}_{2} \}$
with appropriate choices for the sign of 
$ C (^{ \boldsymbol{\xi_{1}} }_{ \boldsymbol{\xi_{2}} } ) $.
Adding the basis vectors $ \mathbf{b}_{1} $ and $ \mathbf{b}_{2} $ 
corresponds to $ \mathbb{Z}_{2} \times \mathbb{Z}_{2} $ modding.  These
vectors reduce the spacetime supersymmetry to $ \mathcal{N} = 1 $, break
\begin{equation}
SO(12) \rightarrow SO(4)^{3} 
\end{equation}
and either
\begin{equation} 
E_{8} \times E_{8} \rightarrow E_{6} \times U(1)^{2} \times E_{8} 
\end{equation}
or
\begin{equation}
SO(16) \times SO(16) \rightarrow SO(10) \times U(1)^{3} \times SO(16).
\end{equation}
The three sectors
\begin{equation*} 
\mathbf{b}_{1} \oplus ( \mathbf{b}_{1} + \boldsymbol{\xi}_{1} ), \quad
\mathbf{b}_{2} \oplus ( \mathbf{b}_{2} + \boldsymbol{\xi}_{2} ), \quad
\mathbf{b}_{3} \oplus ( \mathbf{b}_{3} + \boldsymbol{\xi}_{3} )
\end{equation*}
correspond to the three twisted sectors of the 
$ \mathbb{Z}_{2} \times \mathbb{Z}_{2} $ orbifold, each sector producing
eight chiral generations in either the 
$ \mathbf{27} $ representation of $ E_{6} $ or 
$ \mathbf{16} $ representation of $ SO(10) $.  
The 
\begin{equation*}
\textrm{Neveu-Schwarz} \, \oplus \, \boldsymbol{\xi}_{1} 
\end{equation*}
sector corresponds to
the untwisted sector, producing an additional three 
$ \mathbf{27} $ and $ \mathbf{ \overline{27} } $ or
$ \mathbf{16} $ and $ \mathbf{ \overline{16} } $
repesentations of $ E_{6} $ or $ SO(10) $, respectively.  


\section{Associated Calabi-Yau 3-folds}

The $ \mathbf{Z}_{+} $  and $ \mathbf{Z}_{-} $ orbifolds of Section
\ref{NonExt} have $ ( h^{(1,1)}, h^{(2,1)} )  =  (27,3) $.
Elliptically fibered Calabi-Yau 3-folds with structure 
\begin{equation}
\frac{ \mathbf{K3} \times \mathbf{E} }{ \tau_{ \mathbf{K3} } \times
\tau_{\mathbf{E}} },
\end{equation}
where $ \tau_{ \mathbf{K3} } $ is an involution acting by $ -1 $ on 
$ H^{(2,0)} (\mathbf{K3}) $ and $ \mathbf{E} $ is an elliptic curve with
involution
$ \tau_{\mathbf{E}} (z) = - z $, have been analysed by Voisin
\cite{Voi:Mir} and Borcea \cite{Bor:K3}. They are elliptically fibered
over a base 
\begin{equation}
\mathbf{B} = \frac{\mathbf{K3} }{ \tau_{ \mathbf{K3} } } 
\end{equation}
and 
have been further classified by  Nikulin \cite{Nik:Dis} in terms of three
invariants $ (r,a,\delta) $ in which
\begin{equation}
( h^{(1,1)}, h^{(2,1)} )  = ( 5 + 3r - 2a, 65 - 3r - 2a ).  
\end{equation}
Naively, $ \mathbf{Z}_{+} $ and $ \mathbf{Z}_{-} $ would correspond to a
Voisin-Borcea Calabi-Yau
3-fold with $ (r,a,\delta) = (14,10,0) $.  However, such a Voisin-Borcea
Calabi-Yau 3-fold does not exist.  The problem is that in
Voisin-Borcea models, the factorization of the 6-dimensional manifold as a
product of three disjoint manifolds is essential.  $ \mathbf{Z}_{+} $ and 
$ \mathbf{Z}_{-} $, being an orbifolds of a $ SO(12) $ lattice, are
intrinsically $ \mathbf{T}^{6} $ manifolds and hence such a factorization
is not possible. 

To make a connection with the Calabi-Yau 3-folds $ \mathbf{Z} $ and 
$ \mathbf{X} $ of the previous chapters, 
$ \mathbb{Z}_{2} \times \mathbb{Z}_{2} $ orbifolds at a generic point in
the Narain moduli space will now be constructed.  Start with the
10-dimensional space compactified on the torus 
$ \mathbf{T}^{2}_{1} \times \mathbf{T}^{2}_{2} \times \mathbf{T}^{2}_{3} $
parameterized by three complex cooordinates 
$ z_{1} $, $ z_{2} $, and $ z_{3} $, with the identification
\begin{equation}
z_{i} = z_{i} + 1, \quad 
z_{i} = z_{i} + \tau_{i} \quad \quad
(i = 1,2,3)
\end{equation}
where $ \tau_{i} $ is the complex parameter of the torus 
$ \mathbf{T}^{2}_{i} $.  Under a  $ \mathbb{Z}_{2} $ twist 
$ z_{i} \rightarrow - z_{i} $, the torus $ \mathbf{T}^{2}_{i} $
has four fixed points at
\begin{equation}
z_{i} = \{ 0, 1/2, \tau_{i}/2, (1 + \tau_{i})/2 \}.
\end{equation}
With the two $ \mathbb{Z}_{2} $ twists
\begin{align}
\alpha: \ & ( z_{1},z_{2},z_{3} ) \rightarrow ( -z_{1}, - z_{2},  z_{3} ) 
\\
\beta: \ & ( z_{1},z_{2},z_{3} ) \rightarrow (  z_{1}, - z_{2}, - z_{3} )
\end{align}
there are three twisted sectors, $ \alpha $, $ \beta $, and 
$ \alpha \beta  = \alpha \cdot \beta $, each with 16 fixed points 
and producing 16 chiral generations in the $ \mathbf{27} $ representation
of $ E_{6} $ or $ \mathbf{16} $ representation of $ SO(10) $, for a total
of 48. The untwisted sector produces an additional three 
$ \mathbf{27} $ and $ \mathbf{\overline{27}} $ 
representations of $ E_{6}$ or 
$ \mathbf{16} $ and $ \mathbf{\overline{16}} $ 
representations of $ SO(10) $.
These orbifolds, denoted respectively by $ \mathbf{X}_{+} $ and $
\mathbf{X}_{-} $, have $ (h^{(1,1)},h^{(2,1)}) = (51,3) $. They correspond
\cite{DabPar:A_not} to Voisin-Borcea Calabi-Yau 3-folds with 
$ (r,a,\delta)  = (18,4,0) $,
$ \mathbf{K3} = \mathbf{T}^{4} / \mathbb{Z}_{2} $, and 
$ \mathbf{E} = \mathbf{T}^{2} $.
Now, add the shift
\begin{equation}
\gamma : ( z_{1},z_{2},z_{3} ) \rightarrow 
         \left( 
           z_{1} + \frac{1}{2} , z_{2} + \frac{1}{2}, z_{3} + \frac{1}{2}
         \right).
\end{equation}
The product of the $ \gamma $ shift with any of the three twisted sectors
$ \alpha $, $ \beta $, and $ \alpha \beta $ does not produce any
additional fixed points.  Therefore, $ \gamma $ acts freely.  Under the
action of $ \gamma $, the fixed points from the twisted sectors are
identified in pairs.  Hence, each twisted sector now has 8 fixed points
and produces 8 chiral generations in the $ \mathbf{27} $ representation of 
$ E_{6} $ or $ \mathbf{16} $ representation of $ SO(10) $, for a total of
24. Consequently, the resulting orbifolds have 
$ ( h^{(1,1)}, h^{(2,1)} ) = (27,3) $, reproducing
the data of $ \mathbf{Z}_{+} $ and $ \mathbf{Z}_{-} $ , which are
constructed at the free-fermionic point in the Narain moduli space.  Since
$ \gamma $ acts freely, the associated Calabi-Yau 3-folds have 
$ \pi_{1} = \mathbb{Z}_{2} $ \cite{BerEllFarNanQiu:Tow}.
Thus, $ \mathbf{Z}_{+} $ and $ \mathbf{Z}_{-} $ correspond to 
$ \mathbf{Z} $, and $ \mathbf{X}_{+} $ and $ \mathbf{X}_{-} $ correspond
to $ \mathbf{X} $.

\section{\label{TopQuarkYuk}Top quark Yukawa couplings}

The Yukawa coupling of the top quark is obtained at the cubic level of the
superpotential and is a coupling between states from the
twisted-twisted-untwisted sectors.  For example, in the Standard-like
models \cite{(SLM)}, the relevant coupling is
$ t^{c}_{1} Q_{1} \overline{h}_{1} $, where $ t^{c}_{1} $ and $ Q_{1} $
are respectively the quark SU(2) singlet and doublet from the sector
$ \mathbf{b}_{1} $, and $ \overline{h}_{1} $ is the untwisted Higgs.
One can calculate this coupling in the full $ N_{\textrm{gen}} = 3 $
model, or at the level of either the 
$ \mathbf{Z}_{+} $ or $ \mathbf{Z}_{-} $ orbifolds as a 
$ \mathbf{27}^{3} $ $ E_{6} $ 
or 
$ \mathbf{16} \cdot \mathbf{16} \cdot \mathbf{10} $  $ SO(10) $ coupling,
respectively.  This allows the perturbative calculation of the top quark
Yukawa coupling to be reproduced in the heterotic M-theory limit.
The top quark Yukawa coupling at the grand unification scale 
$ M_{\textrm{GUT} } $ is computed, at least in principle, using
(\ref{lambdaYuk}) with $ \mathbf{Z} $ taken to be the
Calabi-Yau manifold associated with $ \mathbf{Z}_{+} $ or 
$ \mathbf{Z}_{-} $.

\chapter{\label{Summary}Summary}

The rules presented in Section \ref{Rules} extend the available formalism
for studying heterotic M-theory vacua.  Compactification to four
dimensions with $ \mathcal{N} = 1 $ supersymmetry is achieved on a torus
fibered
Calabi-Yau 3-fold $ \mathbf{Z} = \mathbf{X} / \tau_{\mathbf{X}} $ with
first homotopy group
$ \pi_{1}(\mathbf{Z}) = \mathbb{Z}_{2} $.  Here $ \mathbf{X} $ is an
elliptically fibered Calabi-Yau 3-fold which admits two global sections   
and $ \tau_{\mathbf{X}} $ is a freely acting involution on $ \mathbf{X} $.
The rules allow the construction of vacua with grand unification groups
such as $ H = E_{6} $, $ SO(10) $, and $ SU(5) $ with an \emph{arbitrary}
net number of generations $ N_{\textrm{gen}} $ of chiral fermions in the
observable sector.  Requiring vanishing instanton charges in the
observable sector yields potentially viable matter Yukawa couplings.
Since $ \pi_{1}(\mathbf{Z}) = \mathbb{Z}_{2} $, $ H $ can be broken with 
$ \mathbb{Z}_{2} $ Wilson lines.  The vacua with nonstandard embeddings
generically contain M5-branes in the bulk space which wrap holomorphic
curves in $ \mathbf{Z} $.  Chapters \ref{NoYukVac} and \ref{YukVac} apply
these rules to the case $ H = SO(10) $, $ N_{\textrm{gen}} = 3 $ with
respectively nonvanishing and vanishing instanton charges 
$ \beta^{(0)}_{i} $ $ (i=1,\ldots,h^{(1,1)}) $ in the observable sector.
The results obtained are summarized in Table \ref{vacuasum}. 
\begin{table}[t]

$$ \begin{array}{ll}
\beta^{(0)}_{i} \neq 0 & \exists \ \mathbf{B} = \mathbb{F}_{2} \
\textrm{Class B vacua} 
\\
                       & \exists \ \mathbf{B} = \mathbb{F}_{r} \ 
(r \ \textrm{even} \geq 4) \ \textrm{Class B vacua*}
\\
                       & \nexists \ \mathbf{B} = \mathbb{F}_{0} \
\textrm{Class A or Class B vacua}
\\                                
                       & \nexists \ \mathbf{B} = \mathbb{F}_{r} \
(r \ \textrm{odd} \geq 1) \ \textrm{Class A or Class B vacua}
\\
                       & \nexists \ \mathbf{B} = \mathbf{dP}_{3} \
\textrm{Class A or Class B vacua}
\\
\\
\beta^{(0)}_{i} = 0   & \exists \ \mathbf{B} = \mathbf{dP}_{7} 
\ \textrm{vacua*}
\\
                      & \nexists \ \mathbf{B} = \mathbf{dP}_{r} \
(r = 0,\ldots,6,8) \ \textrm{vacua}

\\
                      & \nexists \ \mathbf{B} = \mathbb{F}_{r} \
(r \geq 0) \ \textrm{vacua}                
\end{array} $$

\caption{$ H = SO(10) $ heterotic M-theory vacua with 
$ N_{\textrm{gen}} = 3 $.  The `$*$' indicates that these vacua have been 
demonstrated to exist when the constraints (\ref{preserve}),
(\ref{freeact}), and (\ref{BItau}) are satisfied.  The cases with
nonvanishing and vanishing instanton charges  
$ \beta^{(0)}_{i} $ $ (i=1,\ldots,h^{(1,1)}) $ 
in the observable sector are shown separately.  The latter case
corresponds to potentially viable matter Yukawa couplings.}
\label{vacuasum}
\end{table}

$ H = SO(10) $ vacua are of interest because experimental evidence for
neutrino masses supports the $ SO(10) $ embedding of the Standard Model
spectrum.  A class of string models which preserves this embedding are the
realistic free-fermionic models. These models have a stage in their 
construction which corresponds to $ \mathbb{Z}_{2} \times \mathbb{Z}_{2} $
orbifold compactification of the weakly coupled 10-dimensional heterotic
string. This correspondence identifies associated Calabi-Yau 3-folds which
possess the structure of the above $ \mathbf{Z} $ and $ \mathbf{X} $.
This, in turn, allows the above formalism to be used to study heterotic
M-theory vacua associated with realistic free-fermionic models.
Specifically, while the precise geometrical realization of the full 
$ N_{\textrm{gen}} = 3 $ realistic free-fermionic models is not yet known,
the 
$ \mathbb{Z}_{2} \times \mathbb{Z}_{2} $ orbifolds 
$ \mathbf{Z}_{+} $ and $ \mathbf{Z}_{-} $ with 
$ (h^{(1,1)},h^{(2,1)}) = (27,3) $ described in Section
\ref{NonExt} are at their core. A freely
acting shift relates these orbifolds to the 
$ (h^{(1,1)},h^{(2,1)}) = (51,3) $ 
$ \mathbb{Z}_{2} \times \mathbb{Z}_{2} $ orbifolds 
$ \mathbf{X}_{+} $ and $ \mathbf{X}_{-} $, respectively.  The Calabi-Yau
3-folds associated with $ \mathbf{Z}_{+} $ $ ( \mathbf{Z}_{-} ) $  
and $ \mathbf{X}_{+} $ $ ( \mathbf{X}_{-} ) $ possess the structure of 
$ \mathbf{Z} $ and $ \mathbf{X} $, respectively.
The top quark Yukawa coupling can be calculated at the level of either the 
$ \mathbf{Z}_{+} $ or $ \mathbf{Z}_{-} $ orbifolds as a
$ \mathbf{27}^{3} $ $ E_{6} $ or 
$ \mathbf{16} \cdot \mathbf{16} \cdot \mathbf{10} $ $ SO(10) $ coupling,
respectively. This allows the top quark Yukawa coupling in realistic 
free-fermionic models to be reproduced in the heterotic M-theory limit.
The top quark Yukawa coupling at the grand unification scale
$ M_{\textrm{GUT} } $ is computed, at least in principle, using
(\ref{lambdaYuk}) with $ \mathbf{Z} $ taken to be the
Calabi-Yau manifold associated with $ \mathbf{Z}_{+} $ or
$ \mathbf{Z}_{-} $.

\appendix
\chapter{\label{Chern}Chern classes}

Let $ \mathcal{V}_{\mathbf{M}} $ be a rank $ n $ complex vector
bundle\footnote{The rank of a vector bundle is the dimension of the
fiber.} over a manifold $ \mathbf{M} $ of complex dimension $ m $.
Assume that the structure group $ \mathcal{G} $ of 
$ \mathcal{V}_{\mathbf{M}} $ is a subgroup of 
$ GL(n,\mathbb{C}) \equiv U(n)_{\mathbb{C}} $.
The \emph{Chern classes}
\begin{equation}
c_{i}(\mathcal{V}_{\mathbf{M}}) \in H^{2i}(\mathbf{M}), \quad
i = 0,\ldots,\textrm{min}
    \left( 
      \textrm{rank}(\mathcal{V}_{\mathbf{M}}), 
      \textrm{dim}(\mathbf{M}) 
    \right),        
\end{equation}
are given by the expansion
\begin{equation}
c(\mathcal{V}_{\mathbf{M}}) 
  \equiv \textrm{det} 
         \left( 
           \mathbf{1} + \frac{i\mathcal{F}}{2 \pi}
         \right)
  = \sum_{i = 0}^{ \textrm{min}(n,m) } c_{i}(\mathcal{V}_{\mathbf{M}})
\end{equation}
where $ \mathcal{F} $ is the curvature 2-form on 
$ \mathcal{V}_{\mathbf{M}} $.  When $ \mathcal{V}_{\mathbf{M}} =
\mathbf{TM} $, i.e., when $ \mathcal{F} $ is the Riemann curvature 2-form,
$ c_{i}(\mathbf{TM}) $ is sometimes written as $ c_{i}(\mathbf{M}) $ or
simply $ c_{i} $.  

The Chern class 
$ c_{i}(\mathcal{V}_{\mathbf{M}}) $ with $ i > m $ vanishes trivially.
The classes with with $ i = 0,\ldots,n $ can be identified by diagonalizing
\begin{equation}
\mathcal{X} \equiv \frac{i \mathcal{F}}{2 \pi}.
\end{equation}
Let $ x_{i} $ $ (i = 1,\ldots,n) $ be the eigenvalues of $ \mathcal{X} $.
Then
\begin{align}
c(\mathcal{V}_{\mathbf{M}})
  &= \prod_{i=1}^{n} ( 1 + x_{i} )
\nonumber
\\
  &= 1 + \sum_{i=1}^{n} x_{i} + \sum_{i<j} x_{i} x_{j} 
      + \sum_{i<j<k} x_{i} x_{j} x_{k} + \ldots + x_{1} x_{2}\cdots x_{n}.   
\end{align}
Thus,

\begin{align}
c_{0}(\mathcal{V}_{\mathbf{M}}) &= 1
\\
c_{1}(\mathcal{V}_{\mathbf{M}}) &= \frac{i}{2 \pi}
\textrm{tr}(\mathcal{F})
\\
c_{2}(\mathcal{V}_{\mathbf{M}}) 
  &= \left( \frac{i}{2 \pi} \right)^{2}
     \left\{
       \frac{   (\textrm{tr} \mathcal{F}) \wedge (\textrm{tr} \mathcal{F}) 
              - \textrm{tr}(\mathcal{F} \wedge \mathcal{F})}
            {2}
     \right\}     
\\
\vdots \quad & \quad
\nonumber
\\
c_{n}(\mathcal{V}_{\mathbf{M}}) 
  &= \left( \frac{i}{2 \pi} \right)^{n} \textrm{det}(\mathcal{F}).          
\end{align}

  
\chapter{\label{Spectral}Spectral cover method}

This appendix will describe the spectral cover construction of a
semistable\footnote{A vector bundle $ V $ is stable if
$ \frac{ \textrm{deg}(V) }{ \textrm{rank}(V) } >  
  \frac{ \textrm{deg}(U) }{ \textrm{rank}(U) } \,
  \forall U \subsetneq V $.  Replacing `$>$' with `$ \geq $' defines
semistability.  For further discussion of stability and semistability,
refer to \cite{DOPR:SU(4)}.} holomorphic vector bundle $ V_{\mathbf{X}} $ 
with structure group
\begin{equation}
G_{\mathbb{C}} = SU(n)_{\mathbb{C}} = SL(n,\mathbb{C})
\end{equation}
over an elliptically fibered Calabi-Yau 3-fold
\begin{equation}
\mathbf{X} \stackrel{\pi}{\rightarrow} \mathbf{B}
\end{equation}
with zero section
\begin{equation}
\sigma: \mathbf{B} \rightarrow \mathbf{X}.
\end{equation}
The elliptic fiber at point $ b \in B $ is
\begin{equation}
\mathbf{E}_{b} = \pi^{-1}(b).
\end{equation}
$ \mathbf{E}_{b} $ is a Riemman surface of genus one.  The zero section
assigns to every point $ b \in \mathbf{B} $ the zero element
\begin{equation}
\sigma(b) = p_{b} \in \mathbf{E}_{b}.
\end{equation}
The idea of the spectral cover method is to understand the bundle
structure on a given fiber $ \mathbf{E}_{b} $ and then patch these
bundles together over the base.

\section{Fiber description}

Consider a particular elliptic fiber $ \mathbf{E}_{b} $ at some point 
$ b \in \mathbf{B} $.  This section will explain the construction of a
semistable holomorphic vector bundle $ V_{\mathbf{E}_{b}} $ 
over $ \mathbf{E}_{b} $ with structure group 
$ G_{\mathbb{C}} = SU(n)_{\mathbb{C}} $.  To define a bundle on $
\mathbf{E}_{b} $ one must specify the holonomy; that is, how the bundle
twists as one moves around $ \mathbf{E}_{b} $.  The holonomy is a map from
the fundamental group of $ \mathbf{E}_{b} $ into the structure group 
$ G_{\mathbb{C}} $.  Since the fundamental group of $ \mathbf{E}_{b} $ 
is Abelian, the holonomy must map into the maximal torus of 
$ G_{\mathbb{C}} $ \cite{Bor:Sou}. This means that the transition
functions can be diagonalized, so that $ V_{\mathbf{E}_{b}} $ is the
direct sum of holomorphic line bundles
\begin{equation}
\label{V_E_b}
V_{\mathbf{E}_{b}} = \oplus_{i=1}^{n} \mathcal{N}_{ib}.
\end{equation}
Since $ G_{\mathbb{C}} = SU(n)_{\mathbb{C}} $, the determinant of the
transition functions can be taken to be unity.  This implies that
\begin{equation}
\label{trivE_b}
\otimes_{i=1}^{n} \mathcal{N}_{ib} = \mathcal{O}_{\mathbf{E}_{b}}
\end{equation}
where $ \mathcal{O}_{\mathbf{E}_{b}} $ is the trivial line bundle on 
$ \mathbf{E}_{b} $.  The condition of semistability implies that the line
bundles $ \mathcal{N}_{ib} $ $ (i = 1,\ldots,n) $ are of the same degree,
which can be taken to be zero.  This can be understood from the Hermitian
Yang-Mills equations.  On a Riemann surface, these equations imply that
the field strength is zero.  Thus, the first Chern class of each of the
bundles $ \mathcal{N}_{ib} $ must vanish, or equivalently 
each $ \mathcal{N}_{ib} $ must be of degree zero.  

For each degree zero line bundle 
$ \mathcal{N}_{ib} $ over $ \mathbf{E}_{b} $, there is a
unique point $ Q_{ib} \in \mathbf{E}_{b} $ with the following property: 
\begin{quote}
$ \mathcal{N}_{ib} $ has a holomorphic section that vanishes only at 
$ Q_{ib} $ and has a simple pole only at $ p_{b} $.
\end{quote}
This can be written as
\begin{equation}
\mathcal{N}_{ib} = \mathcal{O}(Q_{ib}) \otimes \mathcal{O}(p_{b})^{-1}
\end{equation}
where 
$ \mathcal{O}(D) $ is the line bundle 
associated\footnote{One can associate a line bundle
to any divisor in a complex manifold $ \mathcal{C} $.  A divisor is a
subspace defined locally by the vanishing of a single holomorphic
function or any linear combination of such subspaces 
$ D = \sum_{i} a_{i} S_{i} $ with integer coefficients $ a_{i} $, and so
is a space of one complex dimension lower than $ \mathcal{C} $.  The
associated line bundle has a section corresponding to a holomorphic
function on $ \mathcal{C} $ with poles of order $ \leq a_{i} $ for $ a_{i}
$ positive, and zeroes of order $ \geq - a_{i} $ for $ a_{i} $ negative.}
with the divisor $ D $. 
The property (\ref{trivE_b}) means that
\begin{equation}
\sum_{i=1}^{n} Q_{ib} = 0.
\end{equation}
The bundle $ V_{\mathbf{E}_{b}} $ determines the $ \mathcal{N}_{ib} $
and $ Q_{ib} $ up to permutations.
Thus, $ V_{\mathbf{E}_{b}} $ is
described by the unordered $ n $-tuple of points 
$ Q_{ib} $ $ (i = 1,\ldots,n) $ in $ \mathbf{E}_{b} $ which add to zero.

For each such $ n $-tuple, there exists a unique (up to multiplication by
a complex scalar) holomorphic function $ s $ which vanishes (to first
order) at the $ Q_{ib} $ and has a pole (at most of order $ n $) only at
$ p_{b} $.  Some of the $ Q_{ib} $ may be coincident.  If the pole at $
p_{b} $ is of order less than $ n $, this means some of the
$ Q_{ib} $ coincide with $ p_{b} $. The function $ s $ can be expressed in
terms of the Weierstrass coordinates of $ \mathbf{E}_{b} $.  In affine
coordinates, where $ z = 1 $, the Weierstrass equation
\begin{equation}
zy^{2} = 4 x^{3} - g_{2} x z^{2} - g_{3} z^{3}
\end{equation}
becomes
\begin{equation}
y^{2} = 4 x^{3} - g_{2} x - g_{3}.
\end{equation}
In these coordinates, $ p_{b} $ is the point $ (x,y) = (\infty, \infty) $.
Since $ s $ has a pole of at most order $ n $ at $ p_{b} $, it must be a
polynomial in $ x $ and $ y $.  As $ x $ has a double pole at $ p_{b} $
and $ y $ has a triple pole, $ s $ can be written
\begin{equation}
s = a_{0} + a_{2} x + a_{3} y + a_{4} x^{2} + a_{5} xy + \cdots +
    \left\{
      \begin{array}{ll}
        a_{n} x^{n/2}    & (n \ \textrm{even})
\\
        a_{n}x^{(n-3)/2} & (n \ \textrm{odd})     
      \end{array}.
    \right. 
\end{equation}
Solving the equation $ s = 0 $ gives $ n $ roots, corresponding to the 
$ n $ points $ Q_{ib} $ $ (i = 1,\ldots,n) $.

The dual interpretation of $ \mathbf{E}_{b} $ as a set of points and
parameter space of degree zero line bundles on itself is formalized by
introducing the Poincar\'{e} line bundle
\begin{equation}
\mathcal{P}_{b} 
  = \mathcal{O}_{ \mathbf{E}_{b} \times \mathbf{E}'_{b} } (D_{b})
\end{equation}
where
\begin{equation}
D_{b} = \Delta_{b} - \mathbf{E}_{b} \times p'_{b} - p_{b} 
        \times \mathbf{E}'_{b}
\end{equation}
and $ \Delta_{b} $ is the diagonal divisor in 
$ \mathbf{E}_{b} \times \mathbf{E}'_{b} $. $ \mathcal{P}_{b} $ is a line
bundle over
\begin{equation*}
\mathbf{E}_{b} \times \mathbf{E}'_{b}
\end{equation*}
whose restriction to
\begin{equation*}
Q_{ib} \times \mathbf{E}'_{b} 
\quad \textrm{or} \quad
\mathbf{E}_{b} \times Q'_{ib}
\end{equation*}
is respectively the line bundle
\begin{equation*}
\mathcal{N}_{ib} \ \textrm{over} \ \mathbf{E}'_{b} 
\quad \textrm{or} \quad 
\mathcal{N}'_{ib} \ \textrm{over} \ \mathbf{E}_{b}.
\end{equation*}

\section{Global description}

Given that the bundle $ V_{\mathbf{E}_{b}} $ over $ \mathbf{E}_{b} $ is
described by the $ n $-tuple $ Q_{ib} $ $ (i=1,\ldots,n) $, it seems
reasonable that a bundle $ V_{\mathbf{X}} $ over 
$ \mathbf{X} \stackrel{\pi}{\rightarrow} \mathbf{B} $ determines how the 
$ n $ points vary as one moves around the base $ \mathbf{B} $.  The
variation over $ \mathbf{B} $ of the $ n $ points generates a divisor 
$ \mathbf{C} \subset \mathbf{X} $ called the \emph{spectral cover}.  
$ \mathbf{C} $ is an $ n $-fold cover of $ \mathbf{B} $ with
\begin{equation}
\pi_{\mathbf{C}}: \mathbf{C} \rightarrow \mathbf{B}.
\end{equation}
$ \mathbf{C} $ is specified by the equation
\begin{equation}
s = 0.
\end{equation}
Here, the pole order condition leads to $ s $ being a section of 
$ \mathcal{O}_{\mathbf{X}} (n \sigma) $, but now one can still twist by
a line bundle $ \mathcal{M} $ over $ \mathbf{B} $ with the properties
\begin{align}
\eta &\equiv c_{1}(\mathcal{M}) \in H^{2}(\mathbf{B},\mathbb{Z})
\\
\label{M|E_b}
\mathcal{M}|_{\mathbf{E}_{b}} &= \mathcal{O}_{\mathbf{E}_{b}}.
\end{align}
Thus, $ s $ is a section of
\begin{equation}
\mathcal{O}_{\mathbf{X}}(\mathbf{C}) 
  = \mathcal{O}_{\mathbf{X}}( n \sigma ) \otimes \mathcal{M}.
\end{equation}
the cohomology class of $ \mathbf{C} $ in 
$ H^{2}(\mathbf{X},\mathbb{Z}) $ is
\begin{equation}
\label{cohomC}
[\mathbf{C}] = n \sigma + \pi^{*} \eta
\end{equation}
where $ \pi^{*} \eta $ is the pull-back of $ \eta $ to $ \mathbf{X} $.

The spectral cover alone does not contain enough information to construct
$ V_{\mathbf{X}} $.  To do this, one must also specify a line bundle 
$ \mathcal{N} $ over $ \mathbf{C} $.  It follows from (\ref{V_E_b}) that
the restriction of $ V_{\mathbf{X}} $ to a fiber $ \mathbf{E}_{b} $ is
given by
\begin{equation}
\label{V_X|E_b}
V_{\mathbf{X}} |_{\mathbf{E}_{b}} = \oplus_{i=1}^{n} \mathcal{N}_{ib}.
\end{equation}
In particular, this gives a decomposition of the fiber 
$ V_{\mathbf{X}} |_{\sigma(b)} $ of 
$ V_{\mathbf{X}} $ at $ p_{b} = \sigma (b) $.
Let $ V_{\mathbf{X}} |_{\mathbf{B}} $ be the restriction of 
$ V_{\mathbf{X}} $ to the base $ \mathbf{B} $ embedded in $ \mathbf{X} $
via the section $ \sigma $.  As was just shown, each fiber of 
$ V_{\mathbf{X}} |_{\mathbf{B}} $ is given by a direct sum of $ n $ line
bundles.  As the point $ b $ moves around the base $ \mathbf{B} $, these
$ n $ line bundles move in one-to-one correspondence with the points 
$ Q_{ib} $ above $ b $.  This data specifies a unique 
line bundle\footnote{When $ \mathbf{C} $ is singular, $ \mathcal{N} $ may
be more generally a rank 1 torsion free sheaf over $ \mathbf{C} $.} 
$ \mathcal{N} $ over $ \mathbf{C} $ such that
\begin{equation}
\label{V_X|_B}
V_{\mathbf{X}} |_{\mathbf{B}} = \pi_{\mathbf{C}*} \mathcal{N}
\end{equation}
where $ \pi_{\mathbf{C}*} \mathcal{N} $ is the push-forward of 
$ \mathcal{N} $ to $ \mathbf{B} $.  $ V_{\mathbf{X}} |_{\mathbf{B}} $ is a
vector
bundle over $ \mathbf{B} $ whose fiber at a generic point $ b $ is the
direct sum
\begin{equation}
V_{\mathbf{X}} |_{\sigma(b)}
  = \oplus_{i=1}^{n} \mathcal{N} |_{Q_{ib}}.
\end{equation}

To describe the entire vector bundle $ V_{\mathbf{X}} $, define the
Poincar\'{e} line bundle
\begin{equation}
\mathcal{P} 
  = \mathcal{O}_{ \mathbf{X} \times_{\mathbf{B}} \mathbf{X}' } (D)
    \otimes \mathcal{K}_{\mathbf{B}}
\end{equation}
where
\begin{equation}
D = \Delta - \sigma \times_{\mathbf{B}} \mathbf{X}'
           - \mathbf{X} \times_{\mathbf{B}} \sigma'.
\end{equation}
$ \mathcal{K}_{\mathbf{B}} $ is the \emph{canonical line bundle} of 
$ \mathbf{B} $, which is
\begin{equation}
\mathcal{K}_{\mathbf{B}} 
  \equiv \textrm{det}({\mathbf{TB}^{*}}) 
  = \wedge^{ \textrm{dim}(\mathbf{B}) } \mathbf{TB}^{*}.
\end{equation}
The \emph{fiber product}
\begin{equation*}
\mathbf{X} \times_{\mathbf{B}} \mathbf{X}'
\end{equation*}
consists of pairs
\begin{equation*}
(x,x') \in \mathbf{X} \times \mathbf{X'}
\end{equation*}
such that
\begin{equation*}
\pi(x) = \pi(x').
\end{equation*}
It is fibered over $ \mathbf{B} $, with the fiber over 
$ b \in \mathbf{B} $ being the ordinary product 
$ \mathbf{E}_{b} \otimes \mathbf{E}'_{b} $.
The equation $ x = x' $ defines the diagonal divisor $ \Delta $ in
$ \mathbf{X} \times_{\mathbf{B}} \mathbf{X}' $.
$ \mathcal{P} $ is a line bundle over 
$ \mathbf{X} \times_{\mathbf{B}} \mathbf{X}' $. 
The restriction of $ \mathcal{P} $ to
\begin{equation*}
Q_{ib} \times \mathbf{E}'_{b} 
\quad \textrm{or} \quad
\mathbf{E}_{b} \times Q^{'}_{ib}
\end{equation*}
is respectively the line bundle
\begin{equation*}
\mathcal{N}_{ib} \ \textrm{over} \ \mathbf{E}'_{b}
\quad \textrm{or} \quad
\mathcal{N}'_{ib} \ \textrm{over} \ \mathbf{E}_{b}.
\end{equation*}
Additionally, the restriction of $ \mathcal{P} $ to
\begin{equation*}
\sigma \times_{\mathbf{B}} \mathbf{X}' 
\quad \textrm{or} \quad
\mathbf{X} \times_{\mathbf{B}} \sigma' 
\end{equation*}
is a trivial bundle.

The Poincar\'{e} line bundle $ \mathcal{P} $ allows $ V_{\mathbf{X}} $
to be constructed from the spectral data $ (\mathbf{C}, \mathcal{N}) $:
\begin{equation}
V_{\mathbf{X}} = p_{1*}(p^{*}_{2} \mathcal{N} \otimes \mathcal{P}).
\end{equation}
Here $ p_{1} $, $ p_{2} $ project 
$ \mathbf{X} \times_{\mathbf{B}} \mathbf{C} $ onto
$ \mathbf{X} $, $ \mathbf{C} $ respectively.  The properties of 
$ \mathcal{P} $  described above guarantee that the restrictions of 
$ V_{\mathbf{X}} $ to $ \mathbf{B} $ and $ \mathbf{E}_{b} $ agree with
(\ref{V_X|_B}) and (\ref{V_X|E_b}), respectively.  The condition
(\ref{M|E_b}) guarantees that the $ V_{\mathbf{X}}|_{\mathbf{E}_{b}} $
are $ SU(n)_{\mathbb{C}} $ bundles.  Requiring 
$ V_{\mathbf{X}}|_{\mathbf{B}} $ to be an $ SU(n)_{\mathbb{C}} $ bundle
as well implies that
\begin{equation}
c_{1}(V_{\mathbf{X}}) = \frac{1}{2 \pi} \,\textrm{tr}(F) = 0.
\end{equation}
This condition is clearly true since, for structure group 
$ SU(n)_{\mathbb{C}} $, the trace must vanish.

An expression for $ c_{1}(V_{\mathbf{X}}) $ in terms of the spectral
data $ (\mathbf{C},\mathcal{N}) $ can be obtained from 
\cite{FriMorWit:Vec}. One finds that
\begin{equation}
c_{1}(V_{\mathbf{X}}) 
  = \pi_{\mathbf{C}*} \left( 
                c_{1}(\mathcal{N}) + \frac{1}{2} c_{1}(\mathbf{C})
              - \frac{1}{2} \pi^{*} c_{1}(\mathbf{B})
             \right). 
\end{equation}
The condition that $ c_{1}(V_{\mathbf{X}}) = 0 $ implies that
\begin{equation}
\label{c_1(N)gamma}
c_{1}(\mathcal{N})
  = -\frac{1}{2} c_{1}(\mathbf{C}) 
    + \frac{1}{2} \pi^{*}_{\mathbf{C}} c_{1}(\mathbf{B}) + \gamma
\end{equation}
where $ \gamma $ is some cohomology class satisfying
\begin{equation}
\pi_{\mathbf{C}*} \gamma = 0.
\end{equation}
The general solution for $ \gamma $ when $ \mathbf{X} $ possesses a second
section $ \xi $ is \cite{DonOvrPanWal:Sta} 
\begin{equation}
\label{gamma}
\gamma 
  = \lambda 
    \left( 
      n \sigma - \pi^{*}_{\mathbf{C}} \eta 
               + n \pi^{*}_{\mathbf{C}} c_{1}(\mathbf{B})
    \right)
    + \sum_{i} \kappa_{i} N_{i} 
\end{equation}
where $ \lambda $ is a rational number.  The 
$ N_{i} \in H_{2}(\mathbf{C},\mathbb{Z}) $, which satisfy
\begin{equation}
\frac{1}{2} \sum_{i=1}^{4 \eta \cdot c_{1}(\mathbf{B})} N_{i} 
\quad \textrm{is an integer class},
\end{equation}
are new fiber classes obtained from the blowup procedure described in
Section \ref{Rules}.  From (\ref{cohomC}), one obtains
\begin{equation}
\label{c_1(C)}
c_{1}(\mathbf{C}) = -n \sigma - \pi^{*}_{\mathbf{C}} \eta.
\end{equation}
Using (\ref{gamma}) and (\ref{c_1(C)}) in (\ref{c_1(N)gamma}) yields
\begin{equation}
c_{1}(\mathcal{N}) = n \left( \frac{1}{2} + \lambda \right) \sigma
   + \left( \frac{1}{2} - \lambda \right) \pi^{*}_{\mathbf{C}} \eta
   + \left( \frac{1}{2} + n \lambda \right) \pi^{*}_{\mathbf{C}}
c_{1}(\mathbf{B})
   + \textstyle{\sum_{i}} \kappa_{i} N_{i}.
\end{equation}
The condition that
\begin{equation}
c_{1}(\mathcal{N}) \quad \textrm{is an integer class}
\end{equation}
constrains the allowed vector bundles.
The section $ \sigma $ is a horizontal divisor, having unit intersection
with the elliptic fibers.  On the other hand, the quantities 
$ \pi^{*}_{\mathbf{C}} c_{1}(\mathbf{B}) $ and
$ \pi^{*}_{\mathbf{C}} \eta $ are vertical, corresponding to curves in the
base lifted to the fibers, and so have zero intersection number with the
fibers.  Thus, a noninteger coefficient of $ \sigma $ cannot be offset by
the coefficients of 
$ \pi^{*}_{\mathbf{C}} c_{1}(\mathbf{B}) $ or
$ \pi^{*}_{\mathbf{C}} \eta $.  The coefficient of $ \sigma $ must itself
be an integer.


\end{document}